\documentclass[aps,twocolumn,showpacs,floatfix,groupedaddress,amsmath,amssymb,prb]{revtex4}
\usepackage{graphicx}
\usepackage{mathrsfs} 
\usepackage{txfonts} 
\usepackage{hyperref} 
\usepackage{array}   
\newcommand{\be}{\begin{equation}}
\newcommand{\ee}{\end{equation}}

\newcommand{\PreserveBackslash}[1]{\let\temp=\\#1\let\\=\temp}

\newcommand{\ket} [1] {| #1 \rangle}
\newcommand{\bra} [1] {\langle #1 |}
\newcommand{\braket}[2]{\langle #1 | #2 \rangle}
\newcommand{\ketbra}[2]{|#1\rangle\langle#2|}

\newcommand{\splitt}[3]{\Upsilon^{\mbox{\tiny \,split}}_{#1\rightarrow #2,#3}}
\newcommand{\fuse}[3]{\Upsilon^{\mbox{\tiny \,fuse}}_{#1, #2 \rightarrow #3}}
\newcommand{\fuser}{\Upsilon^{\mbox{\tiny \,fuse}}}
\newcommand{\splitter}{\Upsilon^{\mbox{\tiny \,split}}}

\newcommand{\SU}{\textit{SU}}
\newcommand{\eref}[1]{(\ref{#1})}
\newcommand{\rmi}{\mathrm{i}}
\newcommand{\sref}[1]{Sec.~\ref{#1}}
\newcommand{\fref}[1]{Fig.~\ref{#1}}
\newcommand{\mc}[1]{\mathcal{#1}}
\newcommand{\mbb}[1]{\mathbb{#1}}

\begin{document}
\title{Tensor network states and algorithms in the presence of a global U(1) symmetry}
\author{Sukhwinder Singh}
\author{Robert N. C. Pfeifer}
\author{Guifre Vidal}
\affiliation{The University of Queensland, Department of Physics,
Brisbane, QLD 4072, Australia}

\begin{abstract}
Tensor network decompositions offer an efficient description of certain many-body states of a lattice system and are the basis of a wealth of numerical simulation algorithms. In a recent paper [arXiv:0907.2994v1] we discussed how to incorporate a global internal symmetry, given by a compact, completely reducible group $\mathcal{G}$, into tensor network decompositions and algorithms. Here we specialize to the case of Abelian groups and, for concreteness, to a $U(1)$ symmetry, often associated with particle number conservation. We consider tensor networks made of tensors that are invariant (or covariant) under the symmetry, and explain how to decompose and manipulate such tensors in order to exploit their symmetry. In numerical calculations, the use of $U(1)$ symmetric tensors allows selection of a specific number of particles, ensures the exact preservation of particle number, and significantly reduces computational costs. We illustrate all these points in the context of the multi-scale entanglement renormalization ansatz.
\end{abstract}


\maketitle

\section{INTRODUCTION\label{sec:intro}}
  
Tensor networks are becoming increasingly popular as a tool to represent wave-functions of quantum many-body systems. Their success is based on the ability to 
{efficiently} describe the ground state of a broad class of local Hamiltonians on the lattice. Tensor network states are used both as a variational ansatz to numerically approximate ground states and as a theoretical framework to characterize and classify quantum phases of matter.
  
Examples of tensor network states for one dimensional systems include the matrix product state\cite{Fannes92,Ostlund95,Perez-Garcia07} (MPS), which results naturally from both Wilson's numerical renormalization group\cite{Wilson75} and White's density matrix renormalization group\cite{White92, White93, Schollwoeck05, McCulloch08} (DMRG) and is also used as a basis for simulation of time evolution;\cite{Vidal03,Vidal04,Daley04,White04,Schollwoeck05b,Vidal07} the tree tensor network\cite{Shi06} (TTN), which follows from coarse-graining schemes where the spins are blocked hierarchically; and the multi-scale entanglement renormalization ansatz\cite{Vidal07b, Vidal08, Evenbly09, Giovannetti08, Pfeifer09, Vidal10} (MERA), which results from a renormalization group procedure known as entanglement renormalization.\cite{Vidal07b,Vidal10} For two dimensional lattices, there are generalizations of these three tensor network states, namely projected entangled pair  states\cite{Verstraete04, Sierra98, Nishino98, Nishio04, Murg07, Jordan08, Gu08, Jiang08, Xie09, Murg09} (PEPS), 2D TTN\cite{Tagliacozzo09, Murg10} and 2D MERA,\cite{Evenbly10f, Evenbly10b, Aguado08, Cincio08, Evenbly09b, Koenig09, Evenbly10} respectively. As variational ans\"atze, PEPS and 2D MERA are particularly interesting since they can be used to address large two-dimensional lattices, including systems of frustrated spins\cite{Murg09, Evenbly10} and interacting fermions,\cite{Corboz09, Kraus09, Pineda09, Corboz09b, Barthel09, Shi09, Corboz10b, Pizorn10, Gu10} where Monte Carlo techniques fail due to the sign problem. 
 
A many-body Hamiltonian $\hat H$ may be invariant under certain transformations, which form a group of symmetries.\cite{Cornwell97} The symmetry group divides the Hilbert space of the theory into symmetry sectors labeled by quantum numbers or conserved charges. On a lattice one can distinguish between \textit{space} symmetries, which correspond to some permutation of the sites of the lattice, and \textit{internal} symmetries, which act on the vector space of each site. An example of space symmetry is invariance under translations by some unit cell, which leads to conservation of momentum. An example of internal symmetry is \SU(2) invariance, e.g. spin isotropy in a quantum spin model. An internal symmetry can in turn be \textit{global}, if it transforms the space of each of the lattice sites according to the same transformation (e.g. a spin independent rotation); or \textit{local}, if each lattice site is transformed according to a different transformation (e.g. a spin-dependent rotation), as it is in the case of gauge symmetric models. A global internal \SU(2) symmetry gives rise to conservation of total spin. By targetting a specific symmetry sector during a calculation, computational costs can often be significantly reduced while explicitly preserving the symmetry. It is therefore not surprising that symmetries play an important role in numerical approaches.

In tensor network approaches, the exploitation of global internal symmetries has a long history, especially in the context of MPS.\cite{White92, Ostlund95, Daley04, Ramasesha96, Sierra97, Tatsuaki00, McCulloch02, Daley05, Bergkvist06, Pittel06, McCulloch07, Danshita07, Perez-Garcia08, Sanz09, Muth09, Mishmash09, Singh10, Cai10} Both Abelian and non-Abelian symmetries have been thoroughly incorporated into DMRG code and have been exploited to obtain computational gains. Symmetries have also been used in more recent proposals to simulate time evolution with MPS, e.g. with the time evolving block decimation (TEBD) algorithm and variations thereof, often collectively referred to as time-dependent DMRG. 

When considering symmetries, it is important to notice that an MPS is a trivalent tensor network. That is, in an MPS each tensor has at most three indices. The Clebsch--Gordan coefficients\cite{Cornwell97} (or coupling coefficients) of a symmetry group are also trivalent, and this makes incorporating the symmetry into a MPS by considering symmetric tensors particularly simple. In contrast, tensor network states with a more elaborated network of tensors, such as MERA or PEPS, consist of tensors having a larger number of indices. In this case a more general formalism is required in order to exploit the symmetry. As explained in Ref.~\onlinecite{Singh09}, a generic symmetric tensor can be decomposed into a \textit{degeneracy} part, which contains all degrees of freedom not determined by symmetry, and a \textit{structural} part, which is completely determined by symmetry and can be further decomposed as a trivalent network of Clebsch--Gordan coefficients.

The use of symmetric tensors in more complex tensor networks has also been discussed in Refs.~\onlinecite{Perez-Garcia10, Zhao10}. In particular, Ref.~\onlinecite{Perez-Garcia10} has shown that under convenient conditions (injectivity), a PEPS that represents a symmetric state can be represented with symmetric tensors, generalizing similar results for MPS obtained in Ref.~\onlinecite{Perez-Garcia08}. Notice that these studies are not concerned with how to decompose symmetric tensors so as to computationally exploit the symmetry. On the other hand, exploitation of $U(1)$ symmetry for computational gain in the context of PEPS was reported in Ref.~\onlinecite{Zhao10}, although no implementation details were provided. Finally, several aspects of \textit{local} internal symmetries in tensor networks algorithms have been addressed in Refs.~\onlinecite{Schuch10, Swingle10, Chen10, Tagliacozzo10}.

The purpose of this paper is to address, in considerable detail and at a pedagogical level, several practical aspects of the exploitation of global internal symmetries not covered in Ref.~\onlinecite{Singh09}. For concreteness we will concentrate on the U(1) symmetry, but extending our results to any Abelian group is straightfoward. A similar analysis of non-abelian groups will be considered in Ref.~\onlinecite{Singh11}.


The paper is organized in sections as follows.
Section \ref{sec:tensor} contains a review of the tensor network formalism and introduces the nomenclature and diagrammatical representation of tensors used in the rest of the paper. It also describes a set $\mathcal{P}$ of primitives for manipulating tensor networks, consisting of manipulations that involves a single tensor (permutation, fusion and splitting of the indices of a tensor) and matrix operations (multiplication and factorization). 

Section \ref{sec:symmetry} reviews basic notions of representation theory of the Abelian group $U(1)$. The action of the group is analysed first on a single system, where $U(1)$ symmetric states and $U(1)$ invariant operators are decomposed in a compact, canonical manner. 
This canonical form allows us to identify the degrees of freedom which are not constrained by the symmetry. The action of the group is then also analysed on the tensor product of two Hilbert spaces and, finally, on the tensor product of a finite number of spaces.
 
Section \ref{sec:symTN} explains how to incorporate the $U(1)$ symmetry into a generic tensor network algorithm, by considering $U(1)$ invariant tensors in a canonical form, and by adapting the set $\mathcal{P}$ of primitives for manipulating tensor networks. These include the multiplication of two $U(1)$ invariant matrices in their canonical form, which is at the core of the computational savings obtained by exploiting the symmetry in tensor network algorithms.

Section \ref{sec:MERA} illustrates the practical exploitation of the $U(1)$ symmetry in a tensor network algorithm by presenting 
MERA calculations of the ground state and low energy states of two quantum spin chain models. Section \ref{sec:conclusions} contain some conclusions.

The canonical form offers a more compact description of $U(1)$ invariant tensors, and leads to faster matrix multiplications and factorizations. However, there is also an additional cost associated with mantaining an invariant tensor in its canonical form while reshaping (fusing and/or splitting) its indices. In some situations, this cost may offset the benefits of using the canonical form. In the appendix we discuss a scheme to lower this additional cost in tensor network algorithms that are based on iterating a sequence of transformations. This is achieved by identifying, in the manipulation of a tensor, operations which only depend on the symmetry. Such operations can be \textit{precomputed} once at the beginning of a simulation. Their result, stored in memory, can be re-used at each iteration of the simulation. The appendix describes two such specific precomputation schemes.


\section{REVIEW: TENSOR NETWORK FORMALISM\label{sec:tensor}}

In this section we review background material concerning the formalism of tensor networks, without reference to symmetry. We introduce basic definitions and concepts, as well as the nomenclature and graphical representation for tensors, tensor networks, and their manipulations, that will be used throughout the paper.

\begin{figure}[t]
\includegraphics[width=8cm]{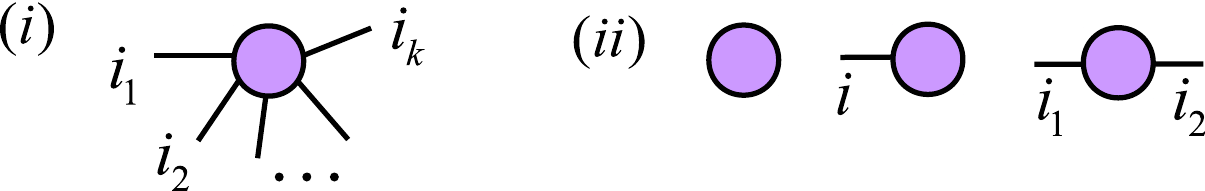}
\caption{(i) Graphical representation of a tensor $\hat{T}$ of rank $k$ and components $\hat{T}_{i_1 i_2 \cdots i_k}$. The tensor is represented by a shape (circle) with $k$ emerging lines corresponding to the $k$ indices $i_1, i_2, \cdots, i_k$. Notice that the indices emerge in counterclockwise order. (ii) Graphical representation of tensors with rank $k=0,1$ and $2$, corresponding to a complex number $c \in \mathbb{C}$, a vector $\ket{v} \in \mathbb{C}^{|i|}$ and a matrix $\hat{M} \in \mathbb{C}^{|i_1|\times |i_2|}$, respectively.\label{fig:tensor}} 
\end{figure}

\subsection{Tensors\label{sec:tensor:tensor}}

A tensor $\hat{T}$ is a multidimensional array of complex numbers $\hat{T}_{i_{1}i_{2}\cdots i_{k}} \in \mathbb{C}$. The \textit{rank} of tensor $\hat{T}$ is the number $k$ of indices. For instance, a rank-zero tensor ($k=0$) is a complex number. Similarly, rank-one ($k=1$) and rank-two ($k=2$) tensors 
represent vectors and matrices, respectively. The \textit{size} of an index $i$, denoted $|i|$, is the number of values that the index takes, $i \in \left\{1, 2, \cdots, |i| \right\}$. The size of a tensor $\hat{T}$, denoted $|\hat{T}|$, is the number of complex numbers it contains, namely $|\hat{T}| = |i_1|\times |i_2| \times \cdots \times |i_k|$.

It is convenient to use a graphical representation of tensors, as introduced in Fig.~\ref{fig:tensor}, where a tensor $\hat{T}$ is depicted as a circle (more generally some shape, e.g. a square) and each of its indices is represented by a line emerging from it. In order to specify which index corresponds to which emerging line, we follow the prescription that the lines corresponding to indices $\{i_1, i_2, \cdots, i_k\}$ emerge in counterclockwise order. Unless stated otherwise, the first index will correspond to the line emerging at nine o'clock (or the first line encoutered while proceeding counterclockwise from nine o'clock).

\begin{figure}[t]
  \includegraphics[width=8cm]{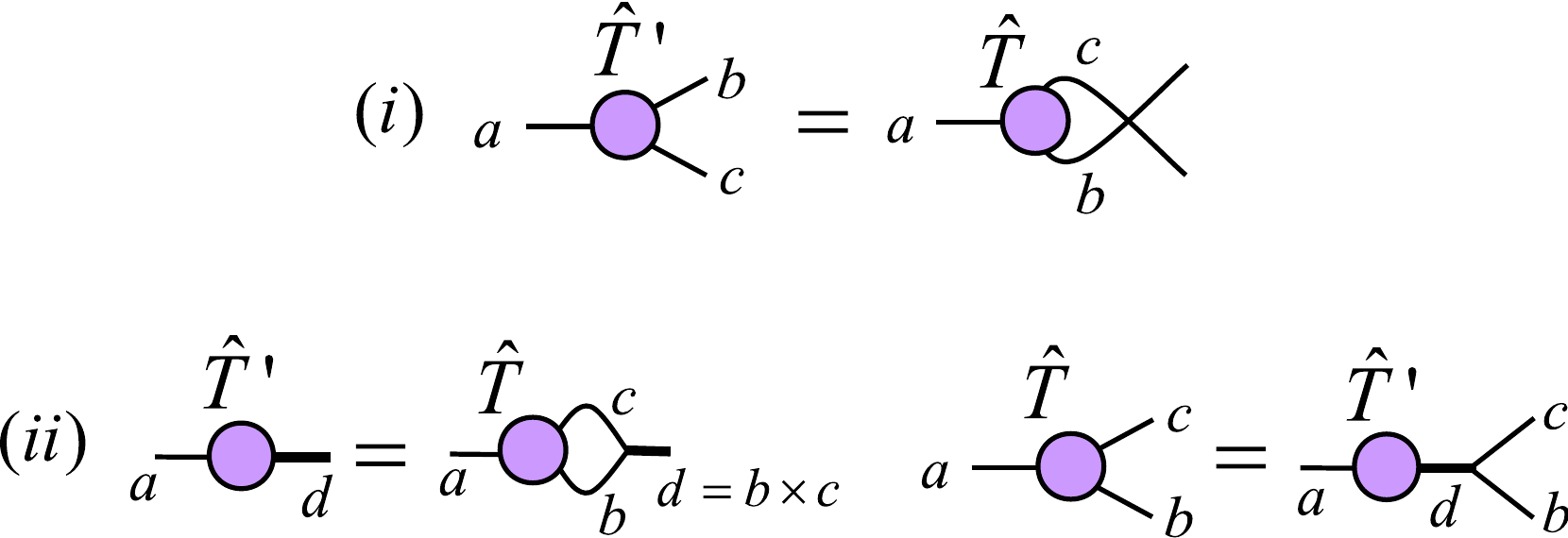}
\caption{
Transformations of a tensor: $(i)$ Permutation of indices $b$ and $c$. $(ii)$ Fusion of indices $b$ and $c$ into $d = b \times c$; splitting of index $d=b \times c$ into $b$ and $c$.\label{fig:tensorman}} 
\end{figure}

Two elementary ways in which a tensor $\hat{T}$ can be transformed are by \textit{permuting} and \textit{reshaping} its indices. A \textit{permutation} of indices corresponds to creating a new tensor $\hat{T}'$ from $\hat{T}$ by simply changing the order in which the indices appear, e.g.
\begin{equation}
	(\hat{T}')_{acb} = \hat{T}_{abc}
	\label{eq:permute}
\end{equation}
On the other hand, a tensor $\hat{T}$ can be \textit{reshaped} into a new tensor $\hat{T}'$ by `fusing' and/or `splitting' some of its indices. For instance, in 
\begin{eqnarray}
	(\hat{T}')_{ad} = \hat{T}_{abc},~~~~~~~d = b\times c,
	\label{eq:fuse}
\end{eqnarray}
tensor $\hat{T}'$ is obtained from tensor $\hat{T}$ by fusing indices $b \in \left\{1, \cdots, |b|\right\}$ and $c \in \left\{1, \cdots, |c|\right\}$ together into a single index $d$ of size $|d| = |b| \cdot |d|$ that runs over all pair of values of $b$ and $c$, i.e.  $ d \in \left\{ (1,1), (1,2), \cdots, (|b|, |c|-1), (|b|,|c|) \right\}$, whereas in
\begin{eqnarray}
	\hat{T}_{abc} = (\hat{T}')_{ad},~~~~~~~d = b\times c,
	\label{eq:split}
\end{eqnarray}
tensor $\hat{T}$ is recovered from $\hat{T}'$ by splitting index $d$ of $\hat{T}'$ back into indices $b$ and $c$. The permutation and reshaping of the indices of a tensor have a straighforward graphical representation; see Fig.~\ref{fig:tensorman}. 

\begin{figure}[t]
  \includegraphics[width=6cm]{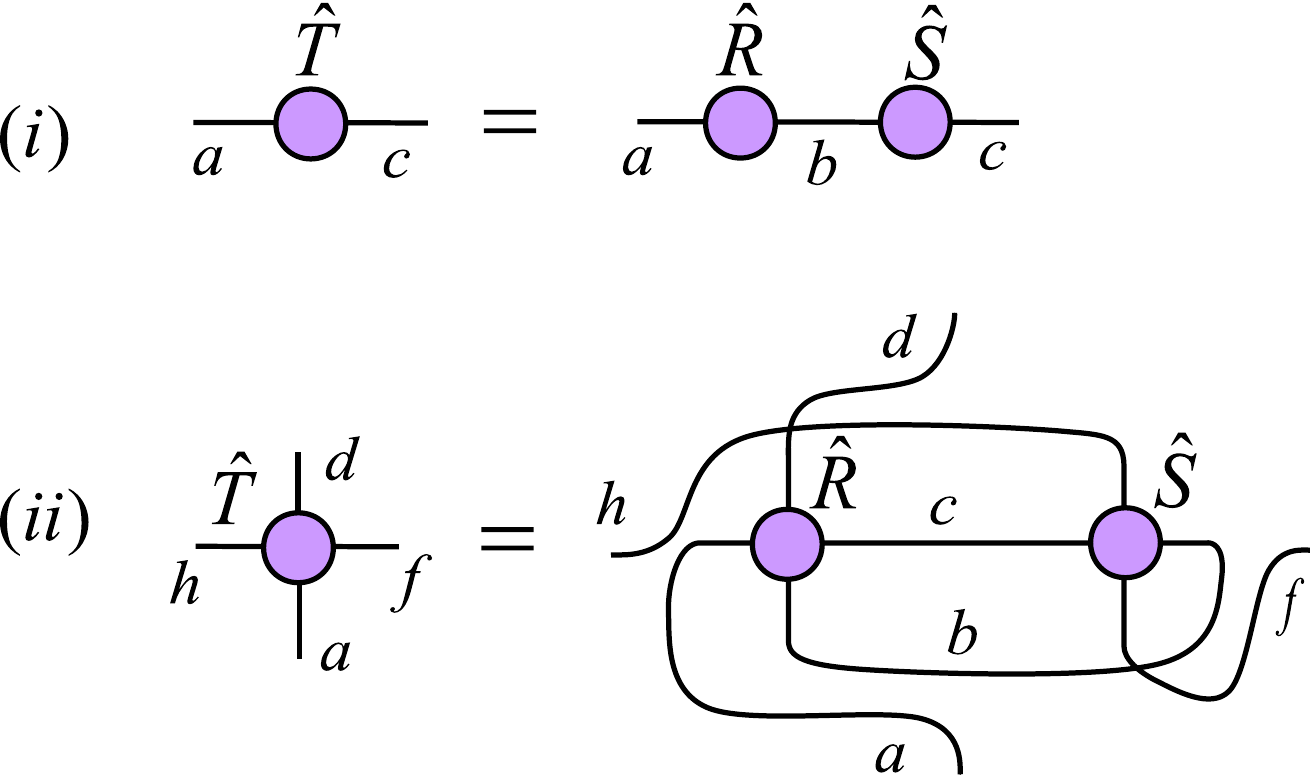}
\caption{
(i) Graphical representation of the matrix multiplication of two matrices $\hat{R}$ and $\hat{S}$ into a new matrix $\hat{T}$ \protect{\eref{eq:Mmultiply}} (ii) Graphical representation of an example of the contraction of two tensors $\hat{R}$ and $\hat{S}$ into a new tensor $\hat{T}$ \protect{\eref{eq:multiply}}. \label{fig:multiply1}} 
\end{figure}

\subsection{Multiplication of two tensors\label{sec:tensor:multiply}}

Given two matrices $\hat{R}$ and $\hat{S}$ with components $\hat{R}_{ab}$ and $\hat{S}_{bc}$, we can multiply them together to obtain a new matrix $\hat{T}$, $\hat{T} = \hat{R}\cdot \hat{S}$ with components
\begin{equation}
	\hat{T}_{ac} = \sum_{b} \hat{R}_{ab}\hat{S}_{bc},
	\label{eq:Mmultiply}
\end{equation}
by summing over or \textit{contracting} index $b$. The multiplication of matrices $\hat{R}$ and $\hat{S}$ is represented graphically by connecting together the emerging lines of $\hat{R}$ and $\hat{S}$ corresponding to the contracted index, as shown in Fig.~\ref{fig:multiply1}(i).

Matrix multiplication can be generalized to tensors. For instance, given tensors $\hat{R}$ and $\hat{S}$ with components $\hat{R}_{abcd}$ and $\hat{S}_{cfbh}$, we can define a tensor $\hat{T}$ with components $\hat{T}_{hafd}$ given by
\begin{equation}
	\hat{T}_{hafd} = \sum_{bc} \hat{R}_{abcd}\hat{S}_{cfbh}.
\label{eq:multiply}
\end{equation}
Again the multiplication of two tensors can be graphically represented by connecting together the lines corresponding to indices that are being contracted (indices $b$ and $c$ in Eq.~\ref{eq:multiply}); see Fig.~\ref{fig:multiply1}(ii).

\begin{figure}[t]
  \includegraphics[width=8cm]{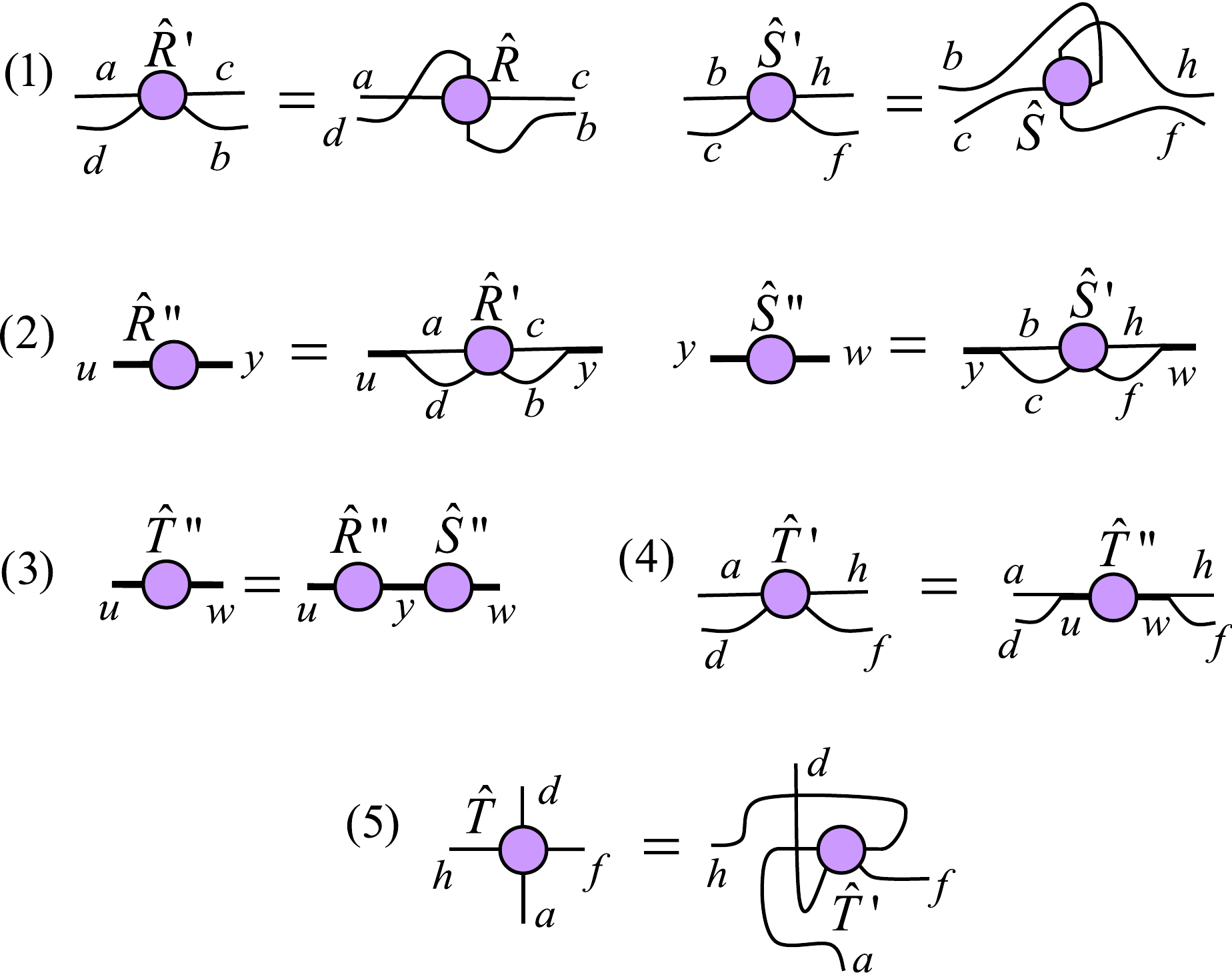}
\caption{
Graphical representations of the five elementary steps 1-5 into which one can decompose the contraction of the tensors of Eq.~\ref{eq:multiply}.\label{fig:multiply2}} 
\end{figure}

The multiplication of two tensors can be broken down into a sequence of elementary steps to transform the tensors into matrices, multiply the matrices, and transform the resulting matrix into a tensor. Next we describe these steps for the contraction given in Eq.~\ref{eq:multiply}. They are illustrated in Fig.~\ref{fig:multiply2}.

\begin{enumerate}
	\item \textit{Permute} the indices of tensor $\hat{R}$ in such a way that the indices to be contracted, $b$ and $c$, appear in the last positions and in a given order, e.g. $bc$; similarly, permute the indices of $\hat{S}$ so that the indices to be contracted, again $b$ and $c$, appear in the first positions and in the same order $bc$: 
	\begin{align}
	(\hat{R}')_{ad ~bc} &= \hat{R}_{abc d}   \nonumber \\
	(\hat{S}')_{bc ~fh} &= \hat{S}_{c f b h} 
	\end{align}
	
	\item \textit{Reshape} tensor $\hat{R}'$ into a matrix $\hat{R}''$ by fusing into a single index $u$ all the indices that are not going to be contracted, $u = a\times d$, and into a single index $y$ all indices to be contracted, $y = b \times c$. Similarly, reshape tensor $\hat{S}'$ into a matrix $\hat{S}''$ with indices $y = b\times c$ and $w = f\times h$,
		\begin{align}\label{eg1}
		(\hat{R}'')_{uy} &= (\hat{R}')_{adbc} \nonumber \\
		(\hat{S}'')_{y w} &= (\hat{S}')_{b c fh}.
	\end{align}
	
	\item \textit{Multiply} matrices $\hat{R}''$ and $\hat{S}''$ to obtain a matrix $\hat{T}''$, with components
	\begin{equation}
	(\hat{T}'')_{uw} = \sum_{y} (\hat{R}'')_{uy} ~~(\hat{S}'')_{y w}
	\end{equation}
	
	\item \textit{Reshape} matrix $\hat{T}''$ into a tensor $\hat{T}'$ by splitting indices $u = a\times d$ and $w = f\times h$,
		\begin{equation}
	(\hat{T}')_{adfh} = (\hat{T}'')_{uw}
	\end{equation}

	\item \textit{Permute} the indices of $\hat{T}'$ into the order in which they appear in $\hat{T}$,
	\begin{equation}
	\hat{T}_{hafd} = (\hat{T}')_{adfh}.\label{eq:endmultiply}
	\end{equation}
\end{enumerate}

We note that breaking down a multiplication of two tensors into elementary steps is not necessary -- one can simply implement the contraction of Eq.~\ref{eq:multiply} as a single process. However, it is often more convenient to compose the above elementary steps since, for instance, in this way one can use existing linear algebra libraries for matrix multiplication. In addition, it can be seen that the leading computational cost in multiplying two large tensors is not changed when decomposing the contraction in the above steps. In Sec.~\ref{sec:symTN:discussion} this subject will be discussed in more detail for $U(1)$ invariant tensors. 

\subsection{Factorization of a tensor\label{sec:tensor:factorize}}


\begin{figure}[t]
  \includegraphics[width=6cm]{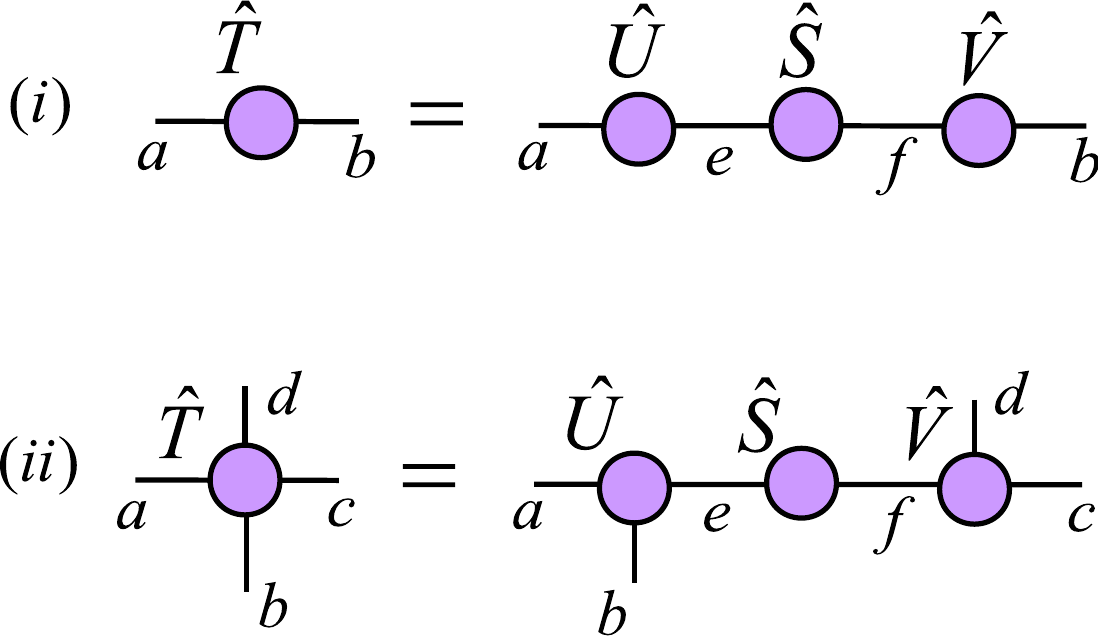}
\caption{
(i) Factorization of a matrix $\hat{T}$ according to a singular value decomposition \protect{\eref{eq:singular}}. (ii) Factorization of a rank-4 tensor $\hat{T}$ according to one of several possible singular value decompositions. \label{fig:decompose}} 
\end{figure}

A matrix $\hat{T}$ can be factorized into the product of two (or more) matrices in one of several canonical forms. For instance, the \textit{singular value decomposition}
\begin{equation}
	\hat{T}_{ab} = \sum_{c,d} \hat{U}_{ac}\hat{S}_{cd}\hat{V}_{db} 
	= \sum_{c} \hat{U}_{ac}s_{c}\hat{V}_{cb}
	\label{eq:singular}
\end{equation}
factorizes $\hat{T}$ into the product of two unitary matrices $\hat{U}$ and $\hat{V}$, and a diagonal matrix $\hat{S}$ with non-negative diagonal elements $s_c = \hat{S}_{cc}$ known as the \textit{singular values} of $\hat{T}$, see Fig.~\ref{fig:decompose}(i). On the other hand, the \textit{eigenvalue} or \textit{spectral decomposition} of a square matrix $\hat{T}$ is of the form
\begin{equation}
	\hat{T}_{ab} = \sum_{c,d} \hat{M}_{ac}D_{cd}(\hat{M}^{-1})_{db} 
	= \sum_{c} \hat{M}_{ac}\lambda_{c}(\hat{M}^{-1})_{cb}
	\label{eq:spectral}
\end{equation}
where $\hat{M}$ is an invertible matrix whose columns encode the eigenvectors $\ket{\lambda_c}$ of $\hat{T}$, 
\begin{equation}
	\hat{T} \ket{\lambda_{c}} = \lambda_c \ket{\lambda_c},
\end{equation}
$\hat{M}^{-1}$ is the inverse of $\hat{M}$, and $\hat{D}$ is a diagonal matrix, with the eigenvalues $\lambda_c=\hat{D}_{cc}$ on its diagonal. Other useful factorizations include the LU decomposition, the QR decomposition, etc. We refer to any such decomposition generically as a \textit{matrix factorization}.

A tensor $\hat{T}$ with more than two indices can be converted into a matrix in several ways, by specifying how two join its indices into two subsets. After specifying how tensor $\hat{T}$ is to be regarded as a matrix, we can factorize $\hat{T}$ according to any of the above matrix factorizations, as illustrated in Fig.~\ref{fig:decompose}(ii) for a singular value decomposition. This requires first permuting and reshaping the indices of $\hat{T}$ to form a matrix, then decomposing the later, and finally restoring the open indices of the resulting matrices into their original form by undoing the reshapes and permutations.

\subsection{Tensor networks and their manipulation\label{sec:tensor:TN}}

\begin{figure}[t]
  \includegraphics[width=7cm]{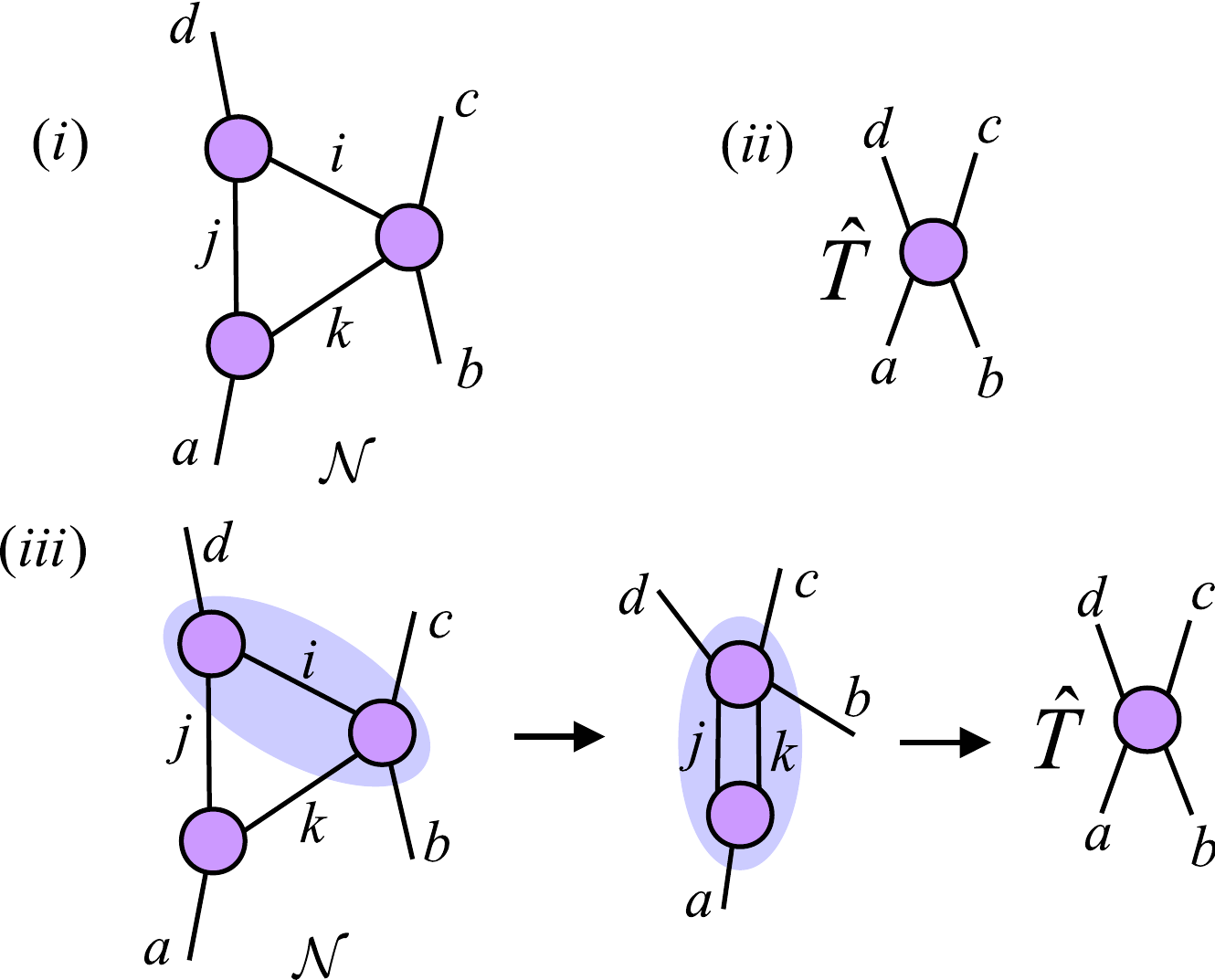}
\caption{
(i) Example of a tensor network $\mathcal{N}$. (ii) Tensor $\hat{T}$ of which the tensor network $\mathcal{N}$ could be a representation. (iii) Tensor $\hat{T}$ can be obtained from $\mathcal{N}$ through a sequence of contractions of pairs of tensors. Shading indicates the two tensors to be multiplied together at each step.\label{fig:TN}} 
\end{figure}

A \textit{tensor network} $\mathcal{N}$ is a set of tensors whose indices are connected according to a network pattern, e.g. Fig.~\ref{fig:TN}. 

Given a tensor network $\mathcal{N}$, a single tensor $\hat{T}$ can be obtained by contracting all the indices that connect the tensors in $\mathcal{N}$. Here, the indices of tensor $\hat{T}$ correspond to the open indices of the tensor network $\mathcal{N}$
. We then say that the 
network $\mathcal{N}$ is a tensor network decomposition of $\hat{T}$. One way to obtain $\hat{T}$ from $\mathcal{N}$ is through a sequence of contractions involving two tensors at a time, Fig.~\ref{fig:TN}.

From a tensor network decomposition $\mathcal{N}$ for a tensor $\hat{T}$, another tensor network decomposition for the same tensor $\hat{T}$ can be obtained in many ways. One possibility is to replace two tensors in $\mathcal{N}$ with the tensor resulting from contracting them together, as is done in each step of Fig.~\ref{fig:TN}(ii). Another way is to replace a tensor in $\mathcal{N}$ with a decomposion of that tensor (e.g. with a singular value decomposition). In this paper, we will be concerned with manipulations of a tensor network that, as in the case of multiplying two tensors or decomposing a tensor, can be broken down into a sequence of operations from the following list:
\begin{enumerate}
	\item Permutation of the indices of a tensor, Eq.~\ref{eq:permute}.
	\item Reshape of the indices of a tensor, Eqs.~\ref{eq:fuse}-\ref{eq:split}.
	\item Multiplication of two matrices, Eq.~\ref{eq:Mmultiply}.
	\item Decomposition of a matrix (e.g. singular value decomposition \protect{\eref{eq:singular}} or spectral decomposition \protect{\eref{eq:spectral}}).
\end{enumerate}

These operations constitute a set $\mathcal{P}$ of \textit{primitive} operations for tensor network manipulations (or, at least, for the type of manipulations we will be concerned with). 

In Section \ref{sec:symTN} we will discuss how this set $\mathcal{P}$ of primitive operations can be generalized to tensors that are symmetric under the action of the group $U(1)$.

\subsection{Tensor network states for quantum many-body systems\label{sec:tensor:TNstates}}


As mentioned in the introduction, tensor networks are used as a means to represent the wave-function of certain quantum many-body systems on a lattice. Let us consider a lattice $\mathcal{L}$ made of $L$ sites, each described by a complex vector space $\mathbb{V}$ of dimension $d$. A generic pure state $\ket{\Psi} \in \mathbb{V}^{\otimes L}$ of $\mathcal{L}$ can always be expanded as
\begin{equation}
\label{eq:purePsi}
\ket{\Psi} = \sum_{i_{1}, i_{2},\cdots, i_{L}} \hat{\Psi}_{i_{1} i_{2}\cdots i_{L}} \ket{i_{1}}\ket{ i_{2}} \cdots \ket{i_{L}},
\end{equation}
where $i_{s} = 1, \cdots, d$ labels a basis $\ket{i_s}$ of $\mathbb{V}$ for site $s \in \mathcal{L}$. Tensor $\hat{\Psi}$, with components $\Psi_{i_{1} i_{2}\cdots i_{L}}$, contains $d^L$ complex coefficients. This is a number that grows exponentially with the size $L$ of the lattice. Thus, the representation of a \textit{generic} pure state $\ket{\Psi} \in \mathbb{V}^{\otimes L}$ is \textit{inefficient}. However, it turns out that an \textit{efficient} representation of \textit{certain} pure states can be obtained by expressing tensor $\hat{\Psi}$ in terms of a tensor network.

\begin{figure}[t]
  \includegraphics[width=8cm]{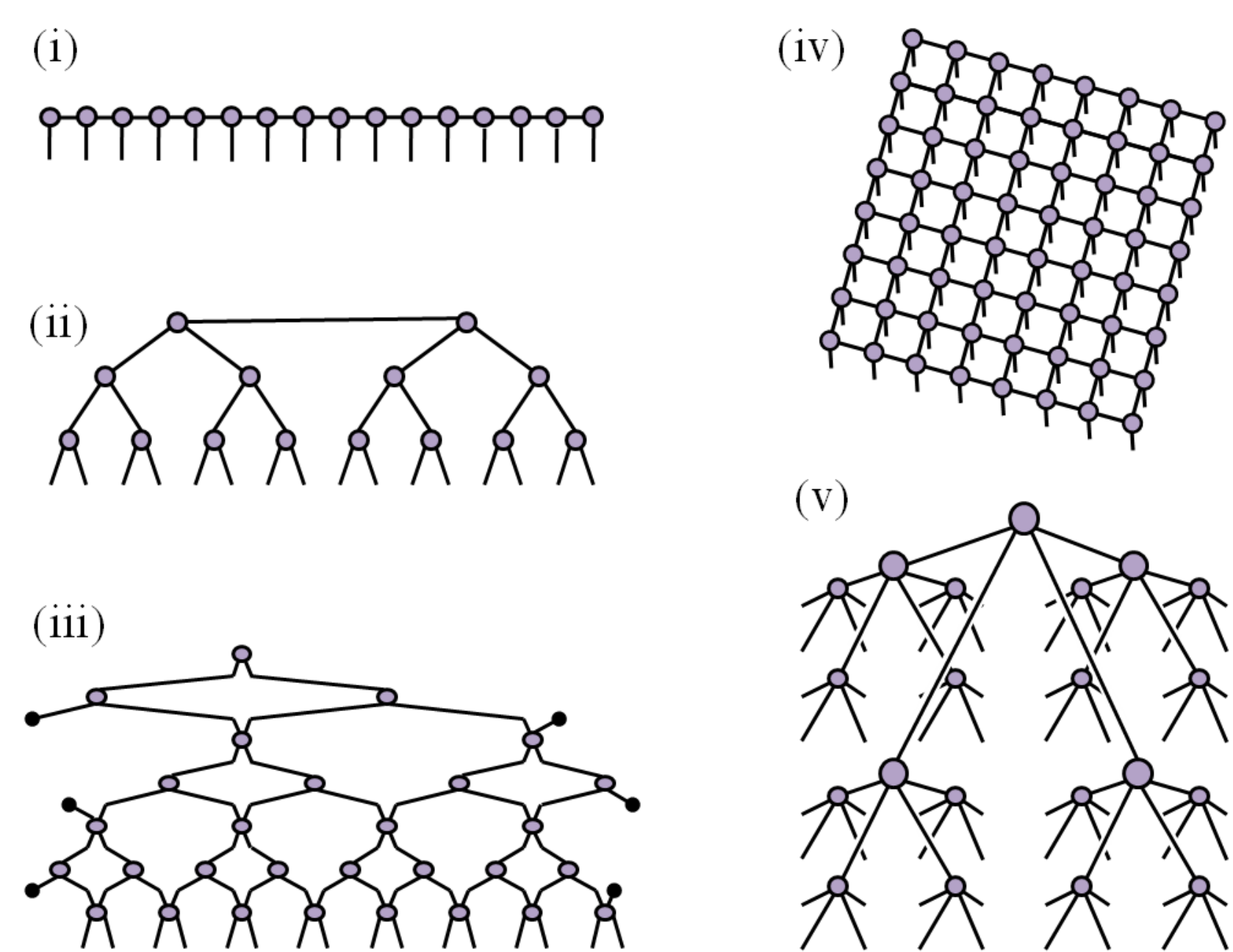}
\caption{
Examples of tensor network states for 1D systems: $(i)$ matrix product state (MPS), $(ii)$ tree tensor network (TTN), $(iii)$ multi-scale entanglement renormalization ansatz (MERA). Examples of tensor network states for 2D systems: $(iv)$ projected entangled-pair state PEPS, $(v)$ 2D TTN. (2D MERA not depicted).\label{fig:TNs}}
\end{figure}

Fig.~\ref{fig:TNs} shows several popular tensor network decompositions used to approximately describe the ground states of local Hamiltonians $H$ of lattice models in one or two spatial dimensions. The open indices of each of these tensor networks correspond to the indices $i_1, i_2, \cdots, i_L$ of tensor $\hat{\Psi}$. Notice that all the tensor networks of Fig.~\ref{fig:TNs} contain $O(L)$ tensors. If $p$ is the rank of the tensors in one of these tensor networks, and $\chi$ is the size of their indices, then the tensor network depends on $O(L\chi^p)$ complex coefficients. For a fixed value of $\chi$ this number grows linearly in $L$, and not exponentially. It therefore does indeed offer an efficient description of the pure state $\ket{\Psi} \in \mathbb{V}^{\otimes L}$ that it represents. Of course only a subset of pure states can be decomposed in this way. Such states, often referred to as tensor network states, are used as variational ans\"atze, with the $O(L\chi^p)$ complex coefficients as the variational parameters.

Given a tensor network state, a variety of algorithms (see e.g. Refs.~\onlinecite{Wilson75}-\onlinecite{Gu10}) are used for tasks such as: ($i$) computation of the expectation value $\bra{\Psi}\hat o\ket{\Psi}$ of a local observable $\hat o$, ($ii$) optimization of the variational parameters so as to minimize the expectation value of the energy $\bra{\Psi}\hat{H}\ket{\Psi}$, or ($iii$) simulation of time evolution, e.g. $e^{-\rmi\hat H t}\ket{\Psi}$. These tasks are accomplished by manipulating tensor networks. 

On most occasions, all required manipulations can be reduced to a sequence of primitive operations in the set $\mathcal{P}$ introduced in Sec.~\ref{sec:tensor:TN}. Thus, in order to adapt the tensor network algorithms of e.g. Refs.~\onlinecite{Wilson75}-\onlinecite{Gu10} to the presence of a symmetry, we only need to modify the set $\mathcal{P}$ of primitive tensor network operations. This will be done in Sec.~\ref{sec:symTN}.

\subsection{Tensors as linear maps\label{sec:tensor:linear}}

\begin{figure}[t]
  \includegraphics[width=7cm]{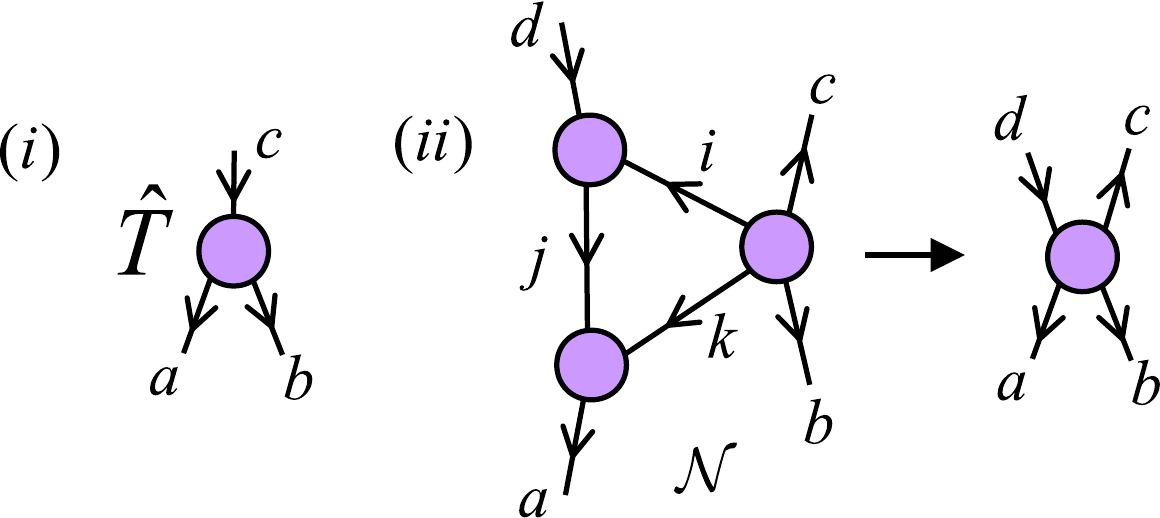}
\caption{
(i) Tensor $\hat{T}$ with one incoming index and two outgoing indices, denoted by incoming and outgoing arrows respectively \protect{\eref{eq:Tabc}}. (ii) A tensor network $\mathcal{N}$ with directed links can be interpreted as a linear map between incoming and outgoing spaces (of the incoming and outgoing indices) obtained by composing the linear maps associated with each of the tensors in $\mathcal{N}$.\label{fig:arrow}} 
\end{figure}

A tensor can be used to define a linear map between vector spaces in the following way. 
First, notice that an index $i$ can be used to label a basis $\{\ket{i}\}$ of a complex vector space $\mathbb{V}^{[i]} \cong \mathbb{C}^{|i|}$ of dimension $|i|$. On the other hand, given a tensor $\hat{T}$ of rank $k$, we can attach a direction `in' or 'out' to each index $i_{1}, i_{2}, \cdots, i_k$. This direction divides the indices of $\hat{T}$ into a subset $I$ of \textit{incoming} indices and the subset $O$ of \textit{outgoing} indices. We can then build input and output vector spaces given by the tensor product of the spaces of incoming and outgoing indices,
\begin{equation}
	\mathbb{V}^{[\text{in}]} = \bigotimes_{i_l \in I} \mathbb{V}^{[i_l]},~~~~~~~
	\mathbb{V}^{[\text{out}]} = \bigotimes_{i_l \in O} \mathbb{V}^{[i_l]},
	\label{eq:inout}
\end{equation}
and use tensor $\hat{T}$ to define a linear map between $\mathbb{V}^{[\text{in}]}$ and $\mathbb{V}^{[\text{out}]}$. For instance, if a rank-3 tensor $\hat{T}_{abc}$ has one incoming index $c \in I$ and two outgoing indices $a,b \in O$, then it defines a linear map $\hat{T} : \mathbb{V}^{[c]} \rightarrow \mathbb{V}^{[a]}\otimes \mathbb{V}^{[b]}$ given by
\begin{equation}
	\hat{T} = \sum_{a,b,c} \hat{T}_{abc}  \ket{a}\ket{b}  \bra{c} 
	\label{eq:Tabc}
\end{equation}
Graphically, we denote the direction of an index by means of an arrow; see Fig.~\ref{fig:arrow}(i).

By decorating the lines of a tensor network $\mathcal{N}$ with arrows (Fig.~\ref{fig:arrow}(ii)), this can be regarded as a composition of linear maps---namely one linear map for each tensor in $\mathcal{N}$. While arrows might be of limited relevance in the absence of a symmetry, they will play an important role when we consider symmetric tensors since they specify how the group acts on each index of a given tensor.

\section {REVIEW: Representation theory of the group U(1)\label{sec:symmetry}}


In this section we review basic background material concerning the representation theory of the group $U(1)$. We first consider the action of $U(1)$ on a vector space $\mathbb{V}$, which decomposes into the direct sum of (possibly degenerate) irreducible representations. We then consider vectors of $\mathbb{V}$ that are symmetric (invariant or covariant) under the action of $U(1)$, as well as linear operators that are $U(1)$ invariant. Then we consider the action of $U(1)$ on the tensor product of two vector spaces, and its generalization to the tensor product of an arbitrary number of vector spaces.

\subsection{Decomposition into direct sum of irreducible representations\label{sec:symmetry:irreps}}


Let $\mathbb{V}$ be a finite dimensional space and let $\varphi \in [0,2\pi)$ label a set of linear transformations $\hat{W}_\varphi$, 
\begin{equation}
\hat{W}_{\varphi}:\mathbb{V}\rightarrow \mathbb{V}, 
\end{equation}
that are a unitary representation of the group $U(1)$. That is
\begin{eqnarray}
&&\hat{W}_{\varphi}^{~\dagger} \hat{W}_{\varphi} = \hat{W}_{\varphi} \hat{W}_{ \varphi}^{~\dagger} = \mathbb{I}, ~~~~~~~~~~~~~~~~~~~~\forall\ \varphi \in [0,2\pi),\\
&&\hat{W}_{\varphi_1}\hat{W}_{\varphi_2} = \hat{W}_{\varphi_2}\hat{W}_{\varphi_1} = \hat{W}_{\varphi_1+\varphi_2|_{2\pi}} ~~~~~~\forall\ \varphi_1,\varphi_2 \in [0,2\pi).~~
\end{eqnarray}
Then $\mathbb{V}$ decomposes as the direct sum of (possibly degenerate) one-dimensional irreducible representations (or \textit{irreps}) of $U(1)$,
\begin{equation}
\mathbb{V} \cong \bigoplus_n \mathbb{V}_{n},
\label{eq:decoV}
\end{equation}
where $\mathbb{V}_{n}$ is a subspace of dimension $d_{n}$, made of $d_n$ copies of an irrep of U(1) with charge $n\in \mathbb{Z}$. We say that irrep $n$ is $d_n$-fold degenerate and that $\mathbb{V}_n$ is the degeneracy space. For concreteness, in this paper we identify the integer charge $n$ as labelling the number of particles (another frequent identification is with the $z$ component of the spin, in which case semi-integer numbers may be considered). 
The representation of group $U(1)$ is generated by the particle number operator $\hat{n}$,
\begin{equation}
\hat{n} \equiv \sum_n n \hat{P}_n,~~~~~\hat{P}_n \equiv \sum_{t_{n}=1}^{d_{n}} \ketbra{nt_{n}}{nt_{n}},
\end{equation}
where $\hat{P}_n$ is a projector onto the subspace $\mathbb{V}_n$ of particle number $n$, and the vectors $\ket{n t_n}$, 
\begin{equation}
	\hat{n}\ket{nt_{n}} = n\ket{nt_{n}},~~~~~~t_n=1,\cdots, d_n,
\label{eq:eigen}
\end{equation}
are an orthonormal basis of $\mathbb{V}_n$. 
In terms of $\hat{n}$, the transformations $\hat{W}_{\varphi}$ read
\begin{equation}
	\hat{W}_{\varphi} = e^{-\rmi\hat{n}\varphi}.
\end{equation}
It then follows from Eq.~\ref{eq:eigen} that
\begin{equation}
	\hat{W}_{\varphi}\ket{nt_{n}} = e^{-\rmi n\varphi}\ket{nt_{n}}.
\end{equation}
The dual basis $\left\{\bra{nt_{n}}\right\}$ is transformed by the \textit{dual representation} of $U(1)$, with elements $\hat{W}_{\varphi}^{~\dagger}$, as
\begin{equation}	
	\bra{nt_{n}} \hat{W}_{\varphi}^{~\dagger} = e^{\rmi n\varphi} \bra{nt_{n}}.
\end{equation}

\textbf{Example 1:} Consider a two-dimensional space $\mathbb{V}$ that decomposes as $\mathbb{V} \cong \mathbb{V}_{0} \oplus \mathbb{V}_{1}$, where the irreps $n = 0$ and $n= 1$ are non-degenerate (i.e. $d_0=d_1=1$). Then the orthogonal vectors $\left\{\ket{n=0, t_0 = 1}, \ket{n=1, t_1 = 1}\right\}$ form a basis of $\mathbb{V}$. In column vector notation,
\begin{equation}
\begin{pmatrix} 1 \\ 0 \end{pmatrix}  \equiv \;  \ket{n=0, t_0 = 1},~~~~
\begin{pmatrix} 0 \\ 1 \end{pmatrix}  \equiv \; \ket{n=1, t_1 = 1},
\end{equation}
the particle number operator $\hat{n}$ and transformation $\hat{W}_{\varphi}$ read
\begin{equation}
\hat{n} \equiv \; \begin{pmatrix} 0 & 0 \\ 0 & 1 \end{pmatrix},~~~~~~
\hat{W}_{\varphi} \equiv \; \begin{pmatrix} 1 & 0 \\ 0 & e^{-\rmi\varphi} \end{pmatrix}.
\end{equation}

\textbf{Example 2:} Consider a four-dimensional space $\mathbb{V}$ that decomposes as $\mathbb{V} \cong \mathbb{V}_{0} \oplus \mathbb{V}_{1} \oplus \mathbb{V}_{2}$, where $d_0=d_2=1$ and $d_1=2$, so that now irrep $n = 1$ is two-fold degenerate. Let $\left\{\ket{n=1, t_1 = 1}, \ket{n=1, t_1 = 2}\right\}$ form a basis of $\mathbb{V}_1$. In column vector notation,
\begin{eqnarray}
\begin{pmatrix} 1 \\ 0 \\ 0 \\ 0 \end{pmatrix}  \equiv \;  \ket{n=0, t_0 = 1},~~~~
\begin{pmatrix} 0 \\ 1 \\ 0 \\ 0 \end{pmatrix}  \equiv \; \ket{n=1, t_1 = 1},\\
\begin{pmatrix} 0 \\ 0 \\ 1 \\ 0 \end{pmatrix}  \equiv \;  \ket{n=1, t_1 = 2},~~~~
\begin{pmatrix} 0 \\ 0 \\ 0 \\ 1 \end{pmatrix}  \equiv \; \ket{n=2, t_2 = 1},
\end{eqnarray}
the particle number operator $\hat{n}$ and transformation $\hat{W}_{\varphi}$ read
\begin{equation}
\hat{n} \equiv \; \begin{pmatrix} 0 & 0 & 0 & 0 \\ 0 & 1 & 0 & 0 \\ 0 & 0 & 1 & 0 \\ 0 & 0 & 0 & 2  \end{pmatrix},~~~
\hat{W} \equiv \; \begin{pmatrix} 1 & 0 & 0 & 0 \\ 0 & e^{-\rmi\varphi} & 0 & 0 \\ 0 & 0 & e^{-\rmi\varphi} & 0 \\ 0 & 0 & 0 & e^{-\rmi2\varphi}  \end{pmatrix}. 
\end{equation}

\subsection{Symmetric states and operators\label{sec:symmetry:states}}


In this work we are interested in states and operators that have a simple transformation rule under the action of $U(1)$. A pure state $\ket{\Psi} \in \mathbb{V}$ is \textit{symmetric} if it transforms as
\begin{equation}
	\hat{W}_{\varphi}\ket{\Psi} = e^{-\rmi n \varphi} \ket{\Psi}.
	\label{eq:nPsi1}
\end{equation}
The case $n=0$ corresponds to an \textit{invariant} state, $\hat{W}_{\varphi}\ket{\Psi} = \ket{\Psi}$, which transforms trivially under $U(1)$, whereas for $n\neq 0$ the state is \textit{covariant}, with $\ket{\Psi}$ being multiplied by a non-trivial phase $e^{-\rmi n\varphi}$. Notice that a symmetric state $\ket{\Psi}$ is an eigenstate of $\hat{n}$: that is, it has a well-defined particle number $n$. $\ket{\Psi}$ can thus be expanded in terms of a basis of the relevant subspace $\mathbb{V}_n$,
\begin{equation}
	\hat{n}\ket{\Psi} = n \ket{\Psi_n},~~~~~~~\ket{\Psi} = \sum_{t_n=1}^{d_n} (\Psi_n)_{t_n} \ket{n t_n}.
	\label{eq:nPsi2}
\end{equation}
 
A linear operator $\hat{T}: \mathbb{V} \rightarrow \mathbb{V}$ is invariant if it commutes with the generator $\hat{n}$ ,
\begin{equation}
	[\hat{T}, \hat{n}] = 0,
	\label{eq:commutator}
\end{equation}
or equivalently if it commutes with the action of the group,
\begin{equation}
	\hat{W}_{\varphi} \hat{T} \hat{W}_{\varphi}^{~\dagger} = \hat{T}~~~~~~~~~\forall \varphi \in [0,2\pi).
	\label{eq:commutator2}
\end{equation}
It follows that $\hat{T}$ decomposes as (\textit{Schur's lemma})
\begin{equation}
\hat{T} = \bigoplus_{n} \hat{T}_{n}
\label{eq:Schur}
\end{equation}
where $\hat{T}_{n}$ is a $d_n\times d_n$ matrix that acts on the subspace $\mathbb{V}_n$ in Eq.~\ref{eq:decoV}.

Notice that the operator $\hat{T}$ in Eq.~\ref{eq:Schur} transforms vectors with a well defined particle number $n$ into vectors with the same particle number. That is, $U(1)$ invariant operators \textit{conserve particle number}. 

\textbf{Example 1 revisited:} In Example 1 above, symmetric vectors must be proportional to either $\ket{n=0, t_0=1}$ or $\ket{n=1,t_1=1}$. An invariant operator $\hat{T} = \hat{T}_0 \oplus \hat{T}_1$ is of the form
\begin{equation}
\hat{T} = \; \begin{pmatrix} \alpha_0 & 0 \\ 0 & \alpha_1 \end{pmatrix},~~~ \alpha_0,\alpha_1 \in \mathbb{C}.
\label{eq:ex1rev}
\end{equation}

\textbf{Example 2 revisited:} In Example 2 above, a symmetric vector $\ket{\Psi}$ must be of the form
\begin{equation}
	\ket{\Psi} = \begin{pmatrix} \alpha_0 \\ 0 \\ 0 \\ 0 \end{pmatrix}, ~~~~~
\ket{\Psi} = \begin{pmatrix} 0 \\ \alpha_1 \\ \beta_1 \\ 0 \end{pmatrix}, ~~~\mbox{or}~~~
\ket{\Psi} = \begin{pmatrix} 0 \\ 0 \\ 0 \\ \alpha_2 \end{pmatrix}, 
	\label{eq:ex2rev}
\end{equation}
where $\alpha_0, \alpha_1, \beta_1, \alpha_2 \in \mathbb{C}$. An invariant operator $\hat{T} = \hat{T}_0 \oplus \hat{T}_1 \oplus \hat{T}_2$ is of the form
\begin{equation}
	\hat{T} = \; \begin{pmatrix} \alpha_0 & 0 & 0 & 0 \\ 0 & \alpha_1 & \beta_1 & 0 \\ 0 & \gamma_1 & \delta_1 & 0 \\ 0 & 0 & 0 & \alpha_2 \end{pmatrix}
	\label{eq:ex2rev2}
\end{equation} 
where $\hat{T}_1$ corresponds to the $2\times 2$ central block 
and $\alpha_0, \alpha_1, \beta_1, \gamma_1, \delta_1, \alpha_2 \in \mathbb{C}$. 

The above examples illustrate that the symmetry imposes constraints on vectors and operators. By using an eigenbasis $\{\ket{n t_n}\}$ of the particle number operator $\hat{n}$, these constraints imply the presence of the zeros in Eqs.~\ref{eq:ex1rev}-\ref{eq:ex2rev2}. Thus, a reduced number of complex coefficients is required in order to describe $U(1)$ symmetric vectors and operators. As we will discuss in Sec.~\ref{sec:symTN}, performing manipulations on symmetric tensors can also result in a significant reduction in computational costs.

\subsection{Tensor product of two representations\label{sec:symmetry:tp}}


Let $\mathbb{V}^{(A)}$ and $\mathbb{V}^{(B)}$ be two spaces that carry representations of $U(1)$, as generated by particle number operators $\hat{n}^{(A)}$ and $\hat{n}^{(B)}$, and let 
\begin{equation}
\mathbb{V}^{(A)} \cong \bigoplus_{n_{A}} \mathbb{V}^{(A)}_{n_A},~~~~~~\mathbb{V}^{(B)} \cong \bigoplus_{n_{B}} \mathbb{V}^{(B)}_{n_B}
\label{eq:AandB}
\end{equation}
be their decompositions as a direct sum of (possibly degenerate) irreps. Let us also consider the action of $U(1)$ on the tensor product $\mathbb{V}^{(AB)} \cong \mathbb{V}^{(A)}\otimes \mathbb{V}^{(B)}$ as generated by the \textit{total particle number operator}  
\begin{equation}
	\hat{n}^{(AB)} \equiv \hat{n}^{(A)}\otimes \mathbb{I} + \mathbb{I} \otimes \hat{n}^{(B)},
\end{equation}
that is, implemented by unitary transformations 
\begin{equation}
	\hat{W}_{\varphi}^{(AB)} \equiv e^{-\rmi\hat{n}^{(AB)}\varphi}.
\end{equation}

The space $\mathbb{V}^{(AB)}$ also decomposes as the direct sum of (possibly degenerate) irreps,
\begin{equation}
\mathbb{V}^{(AB)} \cong \bigoplus_{n_{AB}} \mathbb{V}^{(AB)}_{n_{AB}}. 
\label{eq:decoVAB}
\end{equation}
Here the subspace $\mathbb{V}^{(AB)}_{n_{AB}}$, with total particle number $n_{AB}$, corresponds to the direct sum of all products of subspaces $\mathbb{V}^{(A)}_{n_A}$ and $\mathbb{V}^{(B)}_{n_B}$ such that $n_A + n_B = n_{AB}$,
\begin{equation}
	\mathbb{V}^{(AB)}_{n_{AB}} \cong \bigoplus_{n_A,n_B |_{n_A+n_B = n_{AB}}} \mathbb{V}^{(A)}_{n_A} \otimes \mathbb{V}^{(B)}_{n_B}.
\end{equation}

For each subspace $\mathbb{V}^{(AB)}_{n_{AB}}$ in Eq.~\ref{eq:decoVAB} we introduce a \textit{coupled} basis $\{\ket{n_{AB} t_{n_{AB}}}\}$,  
\begin{equation}
	\hat{n}^{(AB)}\ket{n_{AB} t_{n_{AB}}} = n_{AB} \ket{n_{AB} t_{n_{AB}}},
\end{equation}
where each vector $\ket{n_{AB} t_{n_{AB}}}$ corresponds to the tensor product $\ket{n_{A}t_{n_A};n_B t_{n_B}} \equiv \ket{n_{A}t_{n_A}} \otimes \ket{n_B t_{n_B}}$ of a unique pair of vectors $\ket{n_{A}t_{n_A}}$ and $\ket{n_B t_{n_B}}$, with $n_A+n_B = n_{AB}$. Let table $\fuser$, with components
\begin{equation}
\fuse{n_{A}t_{n_A}}{ n_{B}t_{n_B}}{n_{AB}t_{n_{AB}}} \equiv \braket{n_{AB}t_{n_{AB}}}{n_{A}t_{n_A};n_{B}t_{n_B}},
\label{eq:u1fuse}
\end{equation}
encode this one-to-one correspondence. Notice that each component of $\fuser$ is either a zero or a one. Then
\begin{equation}
	\ket{n_{AB}t_{n_{AB}}} = \sum_{n_A t_{n_A} n_B t_{n_B}}\fuse{n_{A}t_{n_A}}{ n_{B}t_{n_B}}{n_{AB}t_{n_{AB}}} \ket{n_{A}t_{n_A};n_{B}t_{n_B}}.
\label{eq:u1fuse2}
\end{equation}
For later reference, we notice that $\fuser$ can be decomposed into two pieces. The first piece expresses a basis $\{\ket{n_{A}t_{n_{A}}; n_{B}t_{n_{B}}}\}$ of $\mathbb{V}^{(AB)}$ in terms of the basis $\{\ket{n_{A}t_{n_A}}\}$ of $\mathbb{V}^{(A)}$ and the basis $\{\ket{n_{B}t_{n_B}}\}$ of $\mathbb{V}^{(B)}$. This assignment occurs as in the absence of the symmetry, where one creates a composed index $d = b \times c$ by running fast over index $c$, as for example in Eq.~\ref{eq:fuse}. The second piece is a permutation of basis elements that reorganizes them according to their total particle number $n_{AB}$. 
Finally, the product basis can be expressed in terms of the coupled basis
\begin{equation}
	\ket{n_{A}t_{n_A};n_{B}t_{n_B}} = \sum_{n_A t_{n_A} n_B t_{n_B}}\splitt{n_{AB} t_{n_{AB}}}{n_{A}t_{n_{A}}}{n_{B}t_{n_{B}}} \ket{n_{AB}t_{n_{AB}}}, 
\label{eq:u1split2}
\end{equation}
with
\begin{equation}
\splitt{n_{AB} t_{n_{AB}}}{n_{A}t_{n_{A}}}{n_{B}t_{n_{B}}} = \fuse{n_{A}t_{n_A}}{ n_{B}t_{n_B}}{n_{AB}t_{n_{AB}}}.
\label{eq:u1split}
\end{equation}

\textbf{Example 3: } Consider the case where both $\mathbb{V}^{(A)}$ and $\mathbb{V}^{(B)}$ correspond to the space of Example 1, that is
 $\mathbb{V}^{(A)} \cong \mathbb{V}^{(A)}_{0} \oplus \mathbb{V}^{(A)}_{1}$ and $\mathbb{V}^{(B)} \cong \mathbb{V}^{(B)}_{0} \oplus \mathbb{V}^{(B)}_{1}$, where $\mathbb{V}^{(A)}_{0}$, $\mathbb{V}^{(A)}_{1}$, $\mathbb{V}^{(B)}_{0}$, and $\mathbb{V}^{(B)}_{1}$ all have dimension one. Then $\mathbb{V}^{(AB)}$ corresponds to the space in Example 2, namely
\begin{align}
\mathbb{V}^{(AB)} &\cong \mathbb{V}^{(A)} \otimes \mathbb{V}^{(B)} \nonumber \\
 &\cong \left(\mathbb{V}^{(A)}_{0} \oplus \mathbb{V}^{(A)}_{1}\right) \otimes \left(\mathbb{V}^{(B)}_{0} \oplus \mathbb{V}^{(B)}_{1}\right) \nonumber \\
 &\cong \mathbb{V}^{(AB)}_{0} \oplus \mathbb{V}^{(AB)}_{1} \oplus \mathbb{V}^{(AB)}_{2},
\end{align}
where 
\begin{align}
\mathbb{V}^{(AB)}_{0} &\cong \mathbb{V}^{(A)}_{0} \otimes \mathbb{V}^{(B)}_{0}\\
\mathbb{V}^{(AB)}_{1} &\cong 
\left( \mathbb{V}^{(A)}_{0} \otimes \mathbb{V}^{(B)}_{1} \right) 
\oplus \left( \mathbb{V}^{(A)}_{1} \otimes \mathbb{V}^{(B)}_{0} \right)\\
\mathbb{V}^{(AB)}_{2} &\cong \mathbb{V}^{(A)}_{1} \otimes \mathbb{V}^{(B)}_{1}.
\end{align}
The coupled basis $\left\{\ket{n_{AB}t_{n_{AB}}}\right\}$ reads,
\begin{eqnarray}
\ket{n_{AB} = 0, t_0 = 1} ~=~ \ket{n_A=0, t_0 = 1} \otimes \ket{n_B=0, t_0 = 1} \label{eg00}\\
\ket{n_{AB} = 1, t_1 = 1} ~=~ \ket{n_A=0, t_0 = 1} \otimes \ket{n_B=1, t_1 = 1} \label{eg01}\\
\ket{n_{AB} = 1, t_1 = 2} ~=~ \ket{n_A=1, t_1 = 1} \otimes \ket{n_B=0, t_0 = 1} \label{eg02}\\
\ket{n_{AB} = 2, t_2 = 1} ~=~ \ket{n_A=1, t_1 = 1} \otimes \ket{n_B=1, t_1 = 1}, \label{eg03} 
\end{eqnarray}
where we emphasize that the degeneracy index $t_{n_{AB}}$ takes two possible values for $n_{AB} = 1$, i.e. $t_1\in \{1,2\}$, since there are two states $\ket{n_{A}t_{n_A}} \otimes \ket{n_B t_{n_B}}$ with $n_A + n_B = 1$. The components $\fuse{n_{A}t_{A}}{ n_{B}t_{B}}{n_{AB}t_{AB}}$ of the tensor $\fuser$ that encodes this change of basis all zero except for
\begin{align}
	\fuse{01}{01}{01} = \fuse{01}{11}{11} = \fuse{11}{01}{12} = \fuse{11}{11}{21} \;\; &= 1 \nonumber.
	\label{eq:Ex3fuser}
\end{align}

\subsection{Lattice models with $U(1)$ symmetry\label{sec:symmetry:lattice}}


The action of $U(1)$ on the three-fold tensor product 
\begin{equation}
	\mathbb{V}^{(ABC)} \cong \mathbb{V}^{(A)}\otimes \mathbb{V}^{(B)} \otimes \mathbb{V}^{(C)},
\end{equation}
as generated by the total particle number operator
\begin{equation}
	\hat{n}^{(ABC)} = \hat{n}^{(A)} \otimes \mathbb{I} \otimes \mathbb{I} + \mathbb{I} \otimes \hat{n}^{(B)}\otimes \mathbb{I} + \mathbb{I} \otimes \mathbb{I} \otimes \hat{n}^{(C)},
\end{equation}
induces a decomposition
\begin{equation}
\mathbb{V}^{(ABC)} \cong \bigoplus_{n_{ABC}} \mathbb{V}^{(ABC)}_{n_{ABC}} 
\label{eq:decoVABC}
\end{equation}
in terms of irreps $\mathbb{V}^{(ABC)}_{n_{ABC}}$ which we can now relate to $\mathbb{V}^{(A)}_{n_{A}}$, $\mathbb{V}^{(B)}_{n_{B}}$ and $\mathbb{V}^{(C)}_{n_{C}}$. For example, we can first consider the product $\mathbb{V}^{(AB)}_{n_{AB}} \cong \mathbb{V}^{(A)}_{n_{A}}\otimes \mathbb{V}^{(B)}_{n_{B}}$ and then the product $\mathbb{V}^{(ABC)}_{n_{ABC}} \cong \mathbb{V}^{(AB)}_{n_{AB}}\otimes \mathbb{V}^{(C)}_{n_{C}}$, and use two tables $\fuser$ to relate at each step the coupled basis with the product basis, as discussed in the previous section. Similarly we could consider the action of $U(1)$ on four tensor products, and so on.

In particular we will be interested in a lattice $\mathcal{L}$ made of $L$ sites with vector space $\mathbb{V}^{\otimes L}$, where for simplicity we assumed that each site $s\in \mathcal{L}$ is described by the same finite dimensional vector space $\mathbb{V}$ (see Sec.~\ref{sec:tensor:TNstates}). Given a particle number operator $\hat{n}$ defined on each site, we can consider the action of $U(1)$ generated by the total particle number operator
\begin{equation}
	\hat{N} \equiv \sum_{s=1}^{L} \hat{n}^{(s)}
\label{eq:hatN}
\end{equation}
which corresponds to unitary transformations
\begin{equation}
	W^{[L]}_{\varphi} \equiv e^{-\rmi\hat{N}\varphi} = (e^{-\rmi\hat{n}\varphi})^{\otimes L} = \left( \hat{W}_{\varphi} \right)^{\otimes L}.
\end{equation}
The tensor product space $\mathbb{V}^{\otimes L}$ decomposes as
\begin{equation}
\mathbb{V}^{\otimes L} \cong \bigoplus_{N} \mathbb{V}_{N}
\end{equation}
and we denote by $\left\{\ket{Nt_N}\right\}$ the particle number basis in $\mathbb{V}^{\otimes L}$.

We say that a lattice model is $U(1)$ symmetric if its Hamiltonian $\hat{H}: \mathbb{V} \rightarrow \mathbb{V}$ commutes with the action of the group. That is,
\begin{equation}
	[\hat{H}, \hat{N}] = 0
\label{eq:ham0}
\end{equation}
or equivalently, 
\begin{equation}
	\left(\hat{W}_{\varphi}\right)^{\otimes L} \hat{H} \left(\hat{W}_{\varphi}^{~\dagger}\right)^{\otimes L} = \hat{H} ~~~~\forall \varphi \in [0, 2\pi).
\label{eq:ham}
\end{equation}

One example of a $U(1)$ symmetric model is the Hardcore Bose Hubbard Model, with Hamiltonian
\begin{equation}\label{hcbh}
\hat{H}_{HCBH} \equiv \sum_{s=1}^{L}\left(\hat{a}_{s}^{\dagger}\hat{a}_{s+1} + \hat{a}_{s}\hat{a}_{s+1}^{\dagger} + \gamma \hat{n}_{s}\hat{n}_{s+1}\right) -\mu \sum_{s=1}^L \hat{n}_s,
\end{equation}
where we consider periodic boundary conditions (by identifying sites $L+1$ and $1$) and $\hat{a}_{s}^{\dagger}, \hat{a}_{s}$ are hardcore bosonic creation and annihilation operators respectively. In terms of the basis introduced in Example 1, these operators are defined as
\begin{equation}
\hat{a} \equiv \begin{pmatrix} 0 &1 \\ 0 &0 \end{pmatrix}, ~~~~~~~ \hat{n} \equiv \hat{a}^{\dagger}\hat{a} = \begin{pmatrix} 0 &0 \\ 0 &1 \end{pmatrix}. \nonumber
\end{equation}
To see that $\hat{H}_{HCBH}$ commutes with the action of the group we first observe that for two sites 
\begin{equation}
	\left[\hat{a}_{1}^{\dagger} \hat{a}_{2} + \hat{a}_{2}^{\dagger} \hat{a}_{1} \;,\; \hat{n}_{1} + \hat{n}_{2} \right] = 0,
\end{equation}
from which it readily follows that $\left[\hat{H}_{HCBH}, \hat{N}\right ] = 0$. 

Notice that the chemical potential term $-\mu\sum_s \hat{n}_s = -\mu \hat{N}$ also commutes with the rest of the Hamiltonian. The ground state $\ket{\Psi_N^{\tiny\mbox{GS}}}$ of $\hat{H}_{HCBH}$ in a particular subspace $\mathbb{V}_{N}$ or particle number sector can be turned into the absolute ground state by tuning the chemical potential $\mu$. This fact can be used to find the ground state $\ket{\Psi_N^{\tiny\mbox{GS}}}$ of any particle number sector through an algorithm that can only minimize the expectation value of $\hat{H}_{HCBH}$. However, we will later see that the use of symmetric tensors in the context of tensor network states will allow us to directly minimize the expectation value of $\hat{H}_{HCBH}$ in a given particle number sector by restricting the search to states
\begin{equation}
	\ket{\Psi_N} = \sum_{t_N=1}^{d_N} (\Psi_N)_{t_N} \ket{N t_N}
\end{equation}
with the desired particle number $N$.

Finally, by making the identifications 
\begin{equation}
\hat{n} = \frac{\mathbb{I} - \hat{\sigma}_{z}}{2},~~~~~~~ \hat{a} = \frac{\hat{\sigma}_{x} + i\hat{\sigma}_{y}}{2} \nonumber
\end{equation}
where $\hat{\sigma}_{x}, \hat{\sigma}_{y}, \hat{\sigma}_{z}$ are the Pauli matrices 
\begin{equation}
\hat{\sigma}_x \equiv \begin{pmatrix} 0 & 1 \\ 1 & 0 \end{pmatrix}, ~~~~~
\hat{\sigma}_y \equiv \begin{pmatrix} 0 & -i \\ i &0 \end{pmatrix},~~~~~
\hat{\sigma}_z \equiv \begin{pmatrix} 1 & 0 \\ 0 & -1 \end{pmatrix},
\end{equation}
one can map $\hat{H}_{HCBH}$ to the spin-$\frac{1}{2}$ XXZ quantum spin chain
\begin{equation}\label{eq:XXZ}
\hat{H}_{XXZ} \equiv  \sum_{s=1}^{L} 
\left( \hat{\sigma}_{x}^{(s)}\hat{\sigma}_{x}^{(s+1)} 
+ \hat{\sigma}_{y}^{(s)} \hat{\sigma}_{y}^{(s+1)} 
+ \Delta \hat{\sigma}_{z}^{(s)} \hat{\sigma}_{z}^{(s+1)}\right),
\end{equation}
where we have ignored terms proportional to $\hat{N}$ and $\Delta \equiv \gamma/4$. In particular, for $\Delta = 0$ we obtain the quantum XX spin chain 
\begin{equation}\label{eq:XX}
\hat{H}_{XX} \equiv  \sum_{s=1}^{L} 
\left(\hat{\sigma}_{x}^{(s)} \hat{\sigma}_{x}^{(s+1)} 
+ \hat{\sigma}_{y}^{(s)} \hat{\sigma}_{y}^{(s+1)}\right),
\end{equation}
and for $\gamma = 1$, the quantum Heisenberg spin chain
\begin{equation}\label{eq:XXX}
\hat{H}_{XXX} \equiv  \sum_{s=1}^{L} 
\left( \hat{\sigma}_{x}^{(s)}\hat{\sigma}_{x}^{(s+1)} 
+ \hat{\sigma}_{y}^{(s)} \hat{\sigma}_{y}^{(s+1)} 
+ \hat{\sigma}_{z}^{(s)} \hat{\sigma}_{z}^{(s+1)}\right).
\end{equation}
In Sec.~\ref{sec:MERA}, the quantum spin models \eref{eq:XX} and \eref{eq:XXX} will be used to benchmark the performance increase resulting from use of symmetries in tensor networks algorithms.




\section{TENSOR NETWORKS with $U(1)$ symmetry\label{sec:symTN}}


In this section we consider $U(1)$ symmetric tensors and tensor networks. We explain how to decompose $U(1)$ symmetric tensors in a compact, canonical form that exploits their symmetry. We then discuss how to adapt the set $\mathcal{P}$ of primitives for tensor network manipulations in order to work in this form. We also analyse how working in the canonical form affects computational costs.

\subsection{U(1) symmetric tensors\label{sec:symTN:tensor}}


Let $\hat{T}$ be a rank-$k$ tensor with components $\hat{T}_{i_1 i_2 \cdots i_k}$. As in Sec.~\ref{sec:tensor:linear}, we regard tensor $\hat{T}$ as a linear map between the vector spaces $\mathbb{V}^{[\text{in}]}$ and $\mathbb{V}^{[\text{out}]}$ \eref{eq:inout}. This implies that each index is either an incoming or outgoing index. On each space $\mathbb{V}^{[i_l]}$, associated with index $i_l$, we introduce a particle number operator $\hat{n}^{(l)}$ that generates a unitary representation of $U(1)$ given by matrices $\hat{W}_{\varphi}^{(l)} \equiv e^{-\rmi\hat{n}^{(l)}\varphi}$, $\varphi \in [0,2\pi)$. In the following, we use $\hat{W}_{\varphi}^{(l)~*}$ to denote the complex conjugate of $\hat{W}_{\varphi}^{(l)}$. 

\begin{figure}[t]
  \includegraphics[width=8cm]{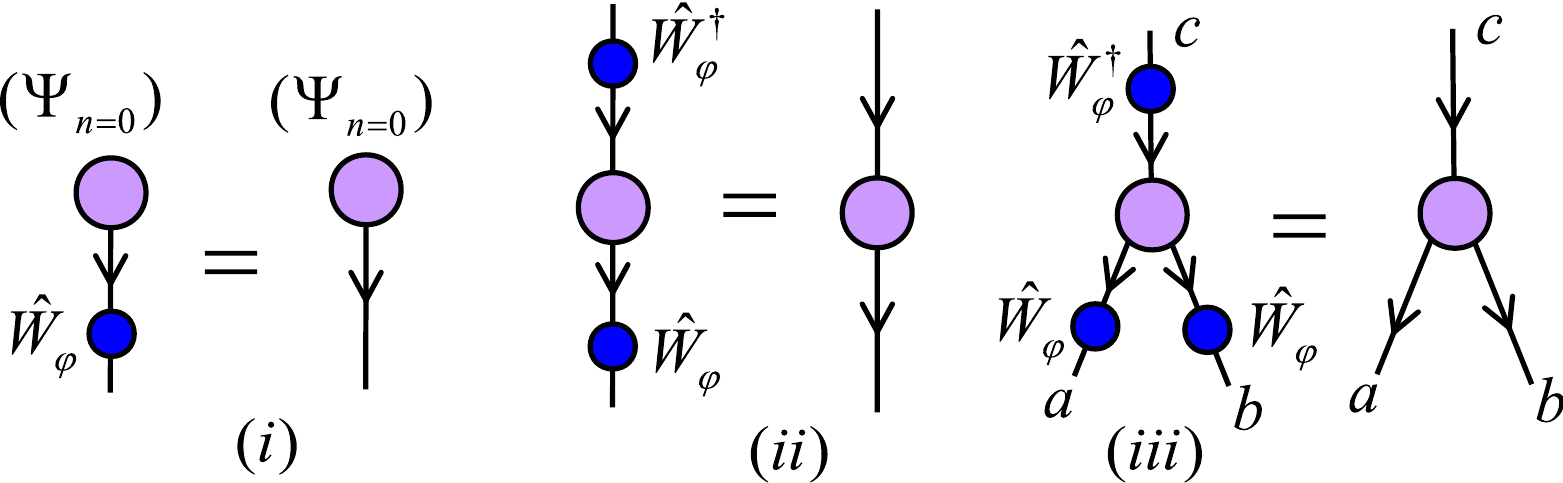}\\
\caption{
(i) Constraint fulfilled by a \textit{U}(1) invariant vector. The only allowed particle number on the single index is $n=0$. (ii) Constraint fulfilled by a \textit{U}(1) invariant matrix. It follows from Schur's lemma that the matrix is block-diagonal in particle number. (iii) Constraint fulfilled by a rank-three tensor with one incoming index and two outgoing indices.\label{fig:invariant}}
\end{figure}

Let us consider the action of $U(1)$ on the space
\begin{equation}
	\mathbb{V}^{[i_1]} \otimes \mathbb{V}^{[i_2]} \otimes \cdots \otimes  \mathbb{V}^{[i_k]}
	\label{eq:prodspace}
\end{equation}
given by
\begin{equation}
	\hat{X}^{(1)}_{\varphi}\otimes \hat{X}^{(2)}_{\varphi}\otimes \cdots \otimes \hat{X}^{(k)}_{\varphi},
\label{eq:Xtrans}
\end{equation}
where 
\begin{equation}
\hat{X}^{(l)}_{\varphi} = \left\{ 
	\begin{array}{cc} \hat{W}^{(l)~*}_{\varphi}& ~~~\mbox{ if } i_l \in I,\\ 
	 									\hat{W}^{(l)}_{\varphi}& ~~~~\mbox{ if } i_l \in O,
	\end{array} \right.
\end{equation}
That is, $\hat{X}^{(l)}_{\varphi}$ acts differently depending on whether index $i_l$ is an incoming or outgoing index of $\hat{T}$.
We then say that tensor $\hat{T}$, with components $T_{i_1 i_2 \cdots i_k}$, is $U(1)$ \textit{invariant} if it is invariant under the transformation of Eq.~\ref{eq:Xtrans}, 
\begin{equation}
	\sum_{i_1, i_2, \cdots, i_k} 	\left(\hat{X}^{(1)}_{\varphi}\right)_{i_1'i_1} \left(\hat{X}^{(2)}_{\varphi} \right)_{i_2' i_2} \cdots \left( \hat{X}^{(k)}_{\varphi} \right)_{i_k' i_k} \hat{T}_{i_1 i_2 \cdots i_k} = \hat{T}_{i_1' i_2' \cdots i_k'},
\end{equation}
for all $\varphi \in [0,2\pi)$. This is depicted in Fig.~\ref{fig:invariant}.

\textbf{Example 4:} A $U(1)$ invariant vector $\ket{\Psi}$---that is, a vector with $\hat{n}\ket{\Psi} = 0$ and components $(\Psi_{n=0})_{t_0}$ in the subspace $\mathbb{V}_{n=0}$ corresponding to vanishing particle number $n=0$ (cf. Eq.~\ref{eq:nPsi2})---fulfills
\begin{equation}
	(\Psi_{n=0})_{{t_0}'} = \sum_{t_0} \left(\hat{W}_{\varphi}\right)_{{t_0}'t_0} (\Psi_{n=0})_{t_0} ~~~~~~\forall\ \varphi \in [0,2\pi),
\end{equation}
in accordance with Eq.~\ref{eq:nPsi1}, as shown in Fig.~\ref{fig:invariant}.

\textbf{Example 5:} A $U(1)$ invariant matrix $\hat{T}$,  Eq.~\ref{eq:Schur}, fulfills
\begin{eqnarray}
	\hat{T}_{a'b'} &=& \sum_{a,b} \left(\hat{W}_{\varphi}\right)_{a'a} \left( \hat{W}^{~*}_{\varphi} \right)_{b'b} \hat{T}_{ab} \\
	&=& \sum_{a,b} \left(\hat{W}_{\varphi}\right)_{a'a} \hat{T}_{ab} \left(\hat{W}^{~\dagger}_{\varphi}\right)_{bb'} ~~~~~~\forall\ \varphi \in [0,2\pi),
\end{eqnarray}
in accordance with Eq.~\ref{eq:commutator2}, see Fig.~\ref{fig:invariant}.
 
\textbf{Example 6:} Tensor $\hat{T}$ in Eq.~\ref{eq:Tabc}, with components $\hat{T}_{abc}$ where $a$ and $b$ are outgoing indices and $c$ is an incoming index, is $U(1)$ invariant if
\begin{eqnarray}
	\hat{T}_{a' b' c'} &=& \sum_{a,b,c} \left(\hat{W}^{(1)}_{\varphi}\right)_{a'a}\left(\hat{W}^{(2)}_{\varphi} \right)_{b'b} \left( \hat{W}^{(3)~*}_{\varphi} \right)_{c'c} \hat{T}_{abc}\\
&=& \sum_{a,b,c} \left(\hat{W}^{(1)}_{\varphi}\right)_{a'a}\left(\hat{W}^{(2)}_{\varphi} \right)_{b'b} \hat{T}_{abc} \left(\hat{W}^{(3)~\dagger}_{\varphi}\right)_{cc'} 
\end{eqnarray}
for all $\varphi \in [0,2\pi)$, see Fig.~\ref{fig:invariant}.

Further, we say that a tensor $\hat{Q}$, with components $\hat{Q}_{i_1 i_2 \cdots i_k}$, is $U(1)$ \textit{covariant} if under the transformation of Eq.~\ref{eq:Xtrans} it simply aquires a non-trivial phase $e^{-\rmi n\varphi}$,
\begin{equation}
	\sum_{i_1, i_2, \cdots, i_k} 	\left(\hat{X}^{(1)}_{\varphi}\right)_{i_1'i_1} \left(\hat{X}^{(2)}_{\varphi} \right)_{i_2 i_2'} \cdots \left( \hat{X}^{(k)}_{\varphi} \right)_{i_k' i_k} \hat{Q}_{i_1 i_2 \cdots i_k} = e^{-\rmi n\varphi} \hat{Q}_{i_1' i_2' \cdots i_k'},
\nonumber	\label{eq:Tcov}
\end{equation}
for all $\varphi \in [0,2\pi)$. 

\textbf{Example 7:} A $U(1)$ covariant vector $\ket{\Psi}$---that is, one which satisfies $\hat{n}\ket{\Psi} = n\ket{\Psi}$ for some $n\neq 0$, and has nonzero components $(\Psi_{n})_{t_n}$ only in the relevant subspace $\mathbb{V}_{n}$ (cf. Eq.~\ref{eq:nPsi2})---fulfills
\begin{equation}
	 \sum_{t_n} \left(\hat{W}_{\varphi}\right)_{{t_n}'t_n} (\Psi_{n})_{t_n} = e^{-\rmi n\varphi}(\Psi_{n})_{{t_n}'} ~~~~~~\forall\ \varphi \in [0,2\pi),
	 \label{eq:coPsi}
\end{equation}
in accordance with Eq.~\ref{eq:nPsi1}.

\begin{figure}[t]
  \includegraphics[width=8cm]{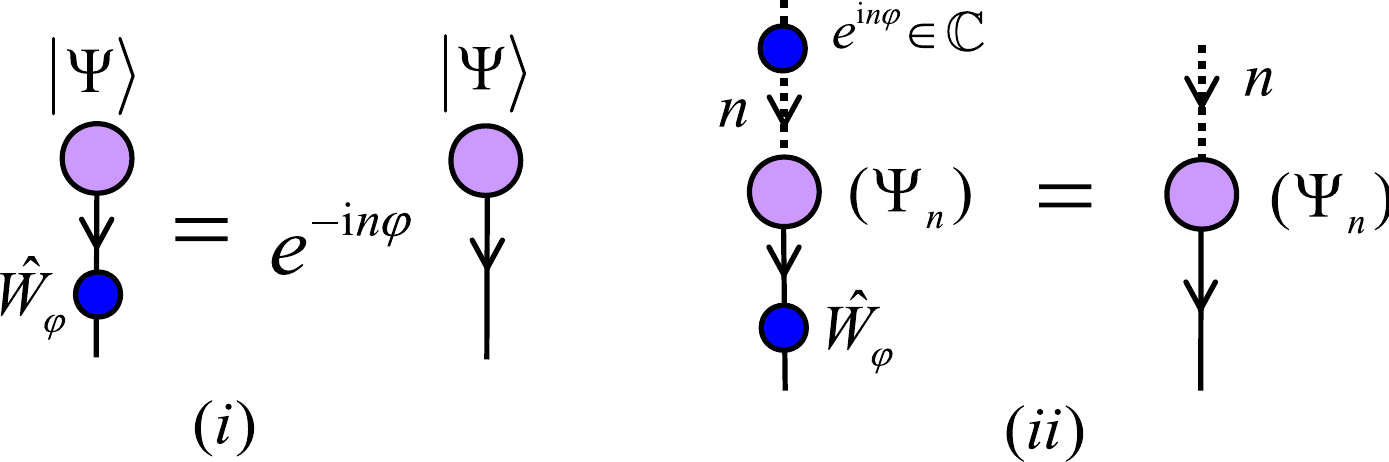}\\
\caption{
(i) $U(1)$ covariant vector $\hat{Q}$, with some non-vanishing particle number $n\neq 0$. Under the action of $U(1)$ on its index, the covariant vector $\hat{Q}$ acquires a phase $e^{-\rmi n\varphi}$ \protect{\eref{eq:coPsi}}. (ii) The $U(1)$ covariant vector $\hat{Q}$, with components $\hat{Q}_{i_1}$, can be represented by a $U(1)$ invariant matrix $\hat{T}$ with components $\hat{T}_{i_1 i} = \hat{Q}_{i_1}$, where $i$ is a trivial index ($|i|=1$) with charge $n$. \label{fig:covariant}}
\end{figure}

Notice that we can describe the rank-$k$ covariant tensor $\hat{Q}$ above by a rank-$(k+1)$ invariant tensor $\hat{T}$ with components
\begin{equation}
	\hat{T}_{i_1i_2\cdots i_k i} \equiv \hat{Q}_{i_1i_2\cdots i_k}.
	\label{eq:TQ}
\end{equation}
This is built from $\hat{Q}$ by just adding an extra incoming index $i$, where index $i$ has fixed particle number $n$ and no degeneracy (i.e., $i$ is associated to a trivial space $\mathbb{V}^{[i]} \cong \mathbb{C}$). We refer to both \textit{invariant} and \textit{covariant} tensors as \textit{symmetric} tensors. By using the above construction, in this work we will represent all $U(1)$ symmetric tensors by means of $U(1)$ invariant tensors. In particular, we represent the non-trivial components $(\Psi_{n})_{t_n}$ of the covariant vector $\ket{\Psi_n}$ in Eqs.~\ref{eq:nPsi1}-\ref{eq:nPsi2} as an  invariant matrix $\hat{T}$ of size $|t_n|\times 1$ with components $\hat{T}_{t_n 1} = (\Psi_n)_{t_n}$.
Consequently, from now on, we will mostly consider only invariant tensors. 

\subsection{Canonical form for U(1) invariant tensors\label{sec:symTN:canonical}}


Let us now write a tensor $\hat{T}$ in a particle number basis on each factor space in Eq.~\ref{eq:prodspace}. That is, each index $i_1$, $i_2$, $\cdots$, $i_k$ is decomposed into a particle number index $n$ and a degeneracy index $t_n$, $i_1 = (n_1, t_{n_1})$,  $i_2 = (n_2, t_{n_2})$, $\cdots$, $i_k = (n_k, t_{n_k})$, and
\begin{equation}
	\hat{T}_{i_1 i_2 \cdots i_k} \equiv \left(\hat{T}_{n_1 n_2 \cdots n_k}\right)_{t_{n_1} t_{n_2} \cdots t_{n_k}}. 
\end{equation}
Here, for each set of particle numbers $n_1, n_2, \cdots, n_k$ we regard $\hat{T}_{n_1n_2\cdots n_k}$ as a tensor with components  $\left(\hat{T}_{n_1 n_2 \cdots n_k}\right)_{t_{n_1} t_{n_2} \cdots t_{n_k}}$. Let $N_{\text{in}}$ and $N_{\text{out}}$ denote the sum of particle numbers corresponding to incoming and outgoing indices,
\begin{equation}
	N_{\text{in}} \equiv \sum_{n_l\in I} n_l,~~~~~~~~N_{\text{out}}\equiv \sum_{n_l\in O} n_l.
\end{equation}
The condition for a non-vanishing tensor of the form $\hat{T}_{n_1n_2\cdots n_k}$ to be invariant under $U(1)$, Eq.~\ref{eq:Xtrans}, is simply that the sum of incoming particle numbers equals the sum of outgoing particle numbers. Therefore, a $U(1)$ invariant tensor $\hat{T}$ satisfies
\begin{equation}
	\hat{T} = \bigoplus_{n_1, n_2, \cdots, n_k} \hat{T}_{n_1 n_2 \cdots n_k} \delta_{N_{\text{in}},N_{\text{out}}}.
	\label{eq:Tcanon}
\end{equation}
(We use the direct sum symbol $\bigoplus$ to denote that the different tensors $\hat{T}_{n_1n_2 \cdots n_k}$ are supported on orthonormal subspaces of the tensor product space of Eq.~\ref{eq:prodspace}.)
In components, the above expression reads,
\begin{equation}
	\hat{T}_{i_1 i_2 \cdots i_k} \equiv \left(\hat{T}_{n_1 n_2 \cdots n_k}\right)_{t_{n_1} t_{n_2} \cdots t_{n_k}}\delta_{N_{\text{in}},N_{\text{out}}}.
	\label{eq:Tcanon2}
\end{equation}
Here, $\delta_{N_{\text{in}},N_{\text{out}}}$ implements particle number conservation: if $N_{\text{in}} \neq N_{\text{out}}$, then all components of $\hat{T}_{n_1n_2\cdots n_{k}}$ must vanish. This generalizes the block structure of $U(1)$ invariant matrices in Eq.~\ref{eq:Schur} (where $\hat{T}_{nn}$ is denoted $\hat{T}_n$) to tensors of arbitrary rank $k$. The canonical decomposition in Eq.~\ref{eq:Tcanon} is important, in that it allows us to identify the degrees of freedom of tensor $\hat{T}$ that are not determined by the symmetry. Expressing tensor $\hat{T}$ in terms of the tensors $\hat{T}_{n_1 n_2 \cdots n_k}$ with $N_{\text{in}} = N_{\text{out}}$ ensures that we store $\hat{T}$ in the most compact possible way.

Notice that the canonical form of Eq.~\ref{eq:Tcanon} is a particular case of the canonical form presented in Eq.~15 of Ref.~\onlinecite{Singh09} for more general (possibly non-Abelian) symmetry groups. There, a symmetric tensor was decomposed into \textit{degeneracy} tensors (analogous to tensors $\hat{T}_{n_1n_2 \cdots n_k}$ in Eq.~\ref{eq:Tcanon}) and structural tensors (generalizing the term $\delta_{N_{\text{in}},N_{\text{out}}}$ in Eq.~\ref{eq:Tcanon}) which can in general be expanded as a trivalent network of Clebsch--Gordan (or coupling) coefficients of the symmetry group. In the case of non-Abelian groups, where some irreps have dimension larger than one, the structural tensors are highly non-trivial. However, for the group $U(1)$ discussed in this paper (as for any other Abelian group) all irreps are one-dimensional and the structural tensors are always reduced to a simple expression such as $\delta_{N_{\text{in}},N_{\text{out}}}$ in Eq.~\ref{eq:Tcanon}. (Nevertheless, in the appendix we will resort to a more elaborate decomposition of the structural tensors in order to further exploit the symmetry during tensor network manipulations of iterative algorithms.)

\begin{figure}[t]
  \includegraphics[width=8cm]{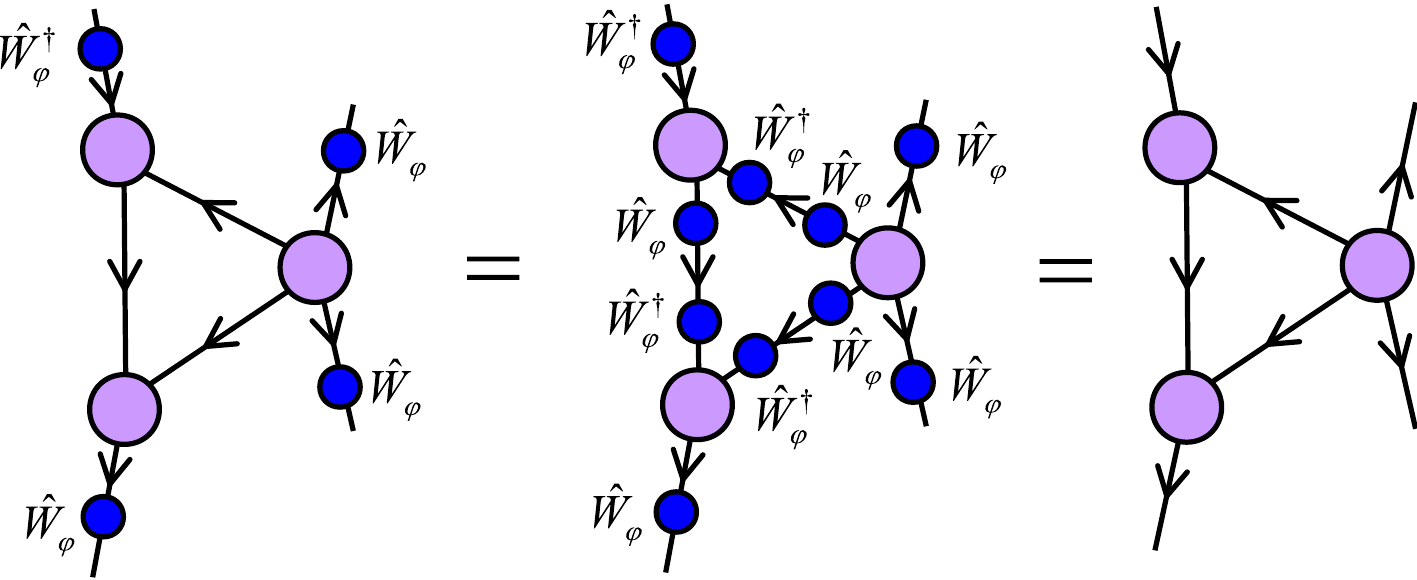}\\
\caption{
A tensor network $\mathcal{N}$ made of $U(1)$ invariant tensors represents a $U(1)$ invariant tensor $\hat{T}$. This is seen by means of two equalities. The first equality is obtained by inserting resolutions of the identity $\mathbb{I} = \hat W_{\varphi} \hat W^{\dagger}_{\varphi}$ on each index connecting two tensors in $\mathcal{N}$. The second equality follows from the fact that each tensor in $\mathcal{N}$ is $U(1)$ invariant. \label{fig:symTN}}
\end{figure}


\subsection{U(1) symmetric tensor networks\label{sec:symTN:TN}}

 
In Sec.~\ref{sec:tensor:linear} we saw that a tensor network $\mathcal{N}$ where each line has a direction (represented with an arrow) can be interpreted as a collection of linear maps composed into a single linear map $\hat{T}$ of which $\mathcal{N}$ is a tensor network decomposition. By introducing a particle number operator on the vector space associated to each line of $\mathcal{N}$, we can define a unitary representation of $U(1)$ on each index of each tensor in $\mathcal{N}$. Then we say that $\mathcal{N}$ is a $U(1)$ invariant tensor network if all its tensors are $U(1)$ invariant. Notice that, by construction, if $\mathcal{N}$ is a $U(1)$ invariant tensor network, then the resulting linear map $\hat{T}$ is also $U(1)$ invariant. This is illustrated in Fig.~\ref{fig:symTN}. 

More generally, we can consider a $U(1)$ symmetric tensor network, made of tensors that are $U(1)$ symmetric (that is, either invariant or covariant). Recall, however, that any covariant tensor can be represented as an invariant tensor by adding an extra index \eref{eq:TQ}. Therefore without loss of generality we can restrict our attention to invariant tensor networks.

\subsection{Tensor network states and algorithms with $U(1)$ symmetry\label{sec:symTN:TNstate}}


As discussed in Sec.~\ref{sec:tensor:TNstates}, a tensor network $\mathcal{N}$ can be used to describe certain pure states $\ket{\Psi}\in \mathbb{V}^{\otimes L}$ of a lattice $\mathcal{L}$. If $\mathcal{N}$ is a $U(1)$ symmetric tensor network then it will describe a pure state $\ket{\Psi}$ that has a well-defined total particle number $N$. That is, a $U(1)$ symmetric pure state
\begin{equation}
 \hat{N}\ket{\Psi} = N \ket{\Psi}, ~~~~~~~~~e^{-\rmi \hat{N}\varphi}\ket{\Psi} = e^{-\rmi N\varphi} \ket{\Psi}.	
\end{equation}
In this way we can obtain a more refined version of popular tensor network states such as MPS, TTN, MERA, PEPS, etc. 
As a variational ansatz, a symmetric tensor network state is more constrained than a regular tensor network state, and consequently it can represent less states $\ket{\Psi} \in \mathbb{V}^{\otimes L}$. However, it also depends on less parameters. This implies a more economical description, as well as the possibility of reducing computational costs during its manipulation.

The rest of this section is devoted to explaining how one can achieve a reduction in computational costs. This is based on storing and manipulating $U(1)$ invariant tensors expressed in the canonical form of Eqs.~\ref{eq:Tcanon}-\ref{eq:Tcanon2}. We next explain how to adapt the set $\mathcal{P}$ of four primitive operations for tensor network manipulation discussed in Sect \ref{sec:tensor:TN}, namely permutation and reshaping of indices, matrix multiplication, and factorization.

\subsection{Permutation of indices\label{sec:symTN:permutation}}


Given a $U(1)$ invariant tensor $\hat{T}$ expressed in the canonical form of Eqs.~\ref{eq:Tcanon}-\ref{eq:Tcanon2}, permuting two of its indices is straightfoward. It is achieved by 
swapping the position of the two particle numbers of $\hat{T}_{n_1n_2 \cdots n_k}$ involved, and also the corresponding degeneracy indices. For instance, if the rank-$3$ tensor $\hat{T}$ of Eq.~\ref{eq:Tabc} is $U(1)$ invariant and has components
\begin{equation}
	\hat{T}_{abc} = \left(\hat{T}_{n_A n_B n_C}\right)_{t_{n_A}t_{n_B}t_{n_C}} \delta_{n_A+n_B, n_C}
	\label{eq:TabcSym}
\end{equation}
when expressed in the particles number basis $a = (n_{A}, t_{n_A})$, $b = (n_B, t_{n_B})$, $c = (n_C, t_{n_C})$, then tensor $\hat{T}'$ of Eq.~\ref{eq:permute}, obtained from $\hat{T}$ by permuting the last two indices, has components
\begin{equation}
	(\hat{T}')_{acb} = \left(\hat{T}_{n_A n_C n_B}'\right)_{t_{n_A}t_{n_C}t_{n_B}} \delta_{n_A+n_B, n_C}.
\end{equation}

Notice that since we only need to permute the components of those $\hat{T}_{n_A n_B n_C}$ such that $n_A+n_B = n_C$, implementing the permutation of indices requires les computational time than a regular index permutation. This is shown in Fig.~\ref{fig:permFuse}, corresponding to a permutation of indices using MATLAB.

\subsection{Reshaping of indices\label{sec:symTN:reshape}}


The indices of a $U(1)$ invariant tensor can be reshaped (fused or split) in a similar manner to those of a regular tensor. However, maintaining the convenient canonical form of Eqs.~\ref{eq:Tcanon}-\ref{eq:Tcanon2} requires additional steps. Two adjacent indices can be fused together using the table $\fuser$ of Eq.~\ref{eq:u1fuse}, which is a sparse tensor made of ones and zeros. Similarly an index can be split into two adjacent indices by using its inverse, the sparse tensor $\splitter$ of Eq.~\ref{eq:u1split}.

\textbf{Example 8 :} Let us consider again the rank-$3$ tensor $\hat{T}$ of Eq.~\ref{eq:Tabc} with components given by Eq.~\ref{eq:TabcSym}, where $a$ and $b$ are outgoing indices and $c$ is an incoming index. We can fuse outgoing index $b$ and incoming index $c$ into an (e.g. incoming) index $d$, obtaining a new tensor $\hat{T}'$ with components
\begin{equation}
	(\hat{T}')_{ad} = \left(\hat{T}'_{n_An_D}\right)_{t_{n_A}t_{t_{n_D}}} \delta_{n_A,n_D},
\end{equation}
where $n_{D}= - n_B + n_C$. [The sign in front of $n_B$ comes from the fact that $d$ is an incoming index and $b$ an outgoing index.] The components of $\hat{T}'$ are in one-to-one correspondence with those of $\hat{T}$ and follow from the transformation
\begin{equation}
	\left(\hat{T}'_{n_{A} n_D}\right)_{t_{n_A} t_{n_D}} = \sum_{n_B,t_{n_B},n_C,t_{n_C}} \left(\hat{T}_{n_A n_B n_C}\right)_{t_{n_A} t_{n_B}t_{n_C}} \fuse{n_B t_{n_B}}{n_C t_{n_C}}{n_D t_{n_D}},
\label{eq:TabcFuse}
\end{equation}
where only the case $n_A=n_D$ needs to be considered.
To complete the example, let us assume that index $a$ is described by the vector space $\mathbb{V}^{(A)}\cong \mathbb{V}_0 \oplus \mathbb{V}_1 \oplus \mathbb{V}_2$ with degeneracies $d_0 = 1$, $d_{1}=2$ and $d_{2}=1$; index $b$ is described by a vector space $\mathbb{V}^{(B)} \cong \mathbb{V}_{-1} \oplus \mathbb{V}_0$ without degeneracies, that is $d_{-1}=d_{0}=1$; and index $c$ is described by a vector space $\mathbb{V}^{(C)} \cong \mathbb{V}_{0} \oplus \mathbb{V}_1$ also without degeneracies, $d_{-1}=d_{0}=1$. Then $\mathbb{V}^{(D)} \cong \mathbb{V}^{(A)}$ and Eq.~\ref{eq:TabcFuse} amounts to 
\begin{eqnarray}
\left(\hat{T}'_{00}\right)_{11} &=& \left(\hat{T}_{000}\right)_{111}, \nonumber\\
\left(\hat{T}'_{11}\right)_{11} &=& \left(\hat{T}_{101}\right)_{111}, \nonumber\\
\left(\hat{T}'_{11}\right)_{12} &=& \left(\hat{T}_{101}\right)_{211}, \nonumber\\
\left(\hat{T}'_{11}\right)_{21} &=& \left(\hat{T}_{1-10}\right)_{111}, \nonumber\\
\left(\hat{T}'_{11}\right)_{22} &=& \left(\hat{T}_{1-10}\right)_{211}, \nonumber\\
\left(\hat{T}'_{22}\right)_{11} &=& \left(\hat{T}_{2-1 1}\right)_{111}, \nonumber
\end{eqnarray}
where we notice that tensor $\hat{T'}$ is a matrix as in Eq.~\ref{eq:ex2rev2}. Similarly, we can split incoming index $d$ of tensor $\hat{T}'$ back into outgoing index $b$ and incoming index $c$ of tensor $\hat{T}$ according to
\begin{equation}
	\left(\hat{T}_{n_A n_B n_C}\right)_{t_{n_A} t_{n_B} t_{n_C}} = \sum_{n_D,t_{n_D}} \left(\hat{T}_{n_A n_D} ' \right)_{t_{n_A} t_{n_D}} \splitt{n_D t_{n_D}}{n_B t_{n_B}}{n_C t_{n_C}}
\label{eq:TadSplit}
\end{equation}
which, again, is non-trivial only for $-n_{B}+n_{C}=n_{D}$ and $n_A+n_B=n_C$.

\begin{figure}
  \includegraphics[width=8cm]{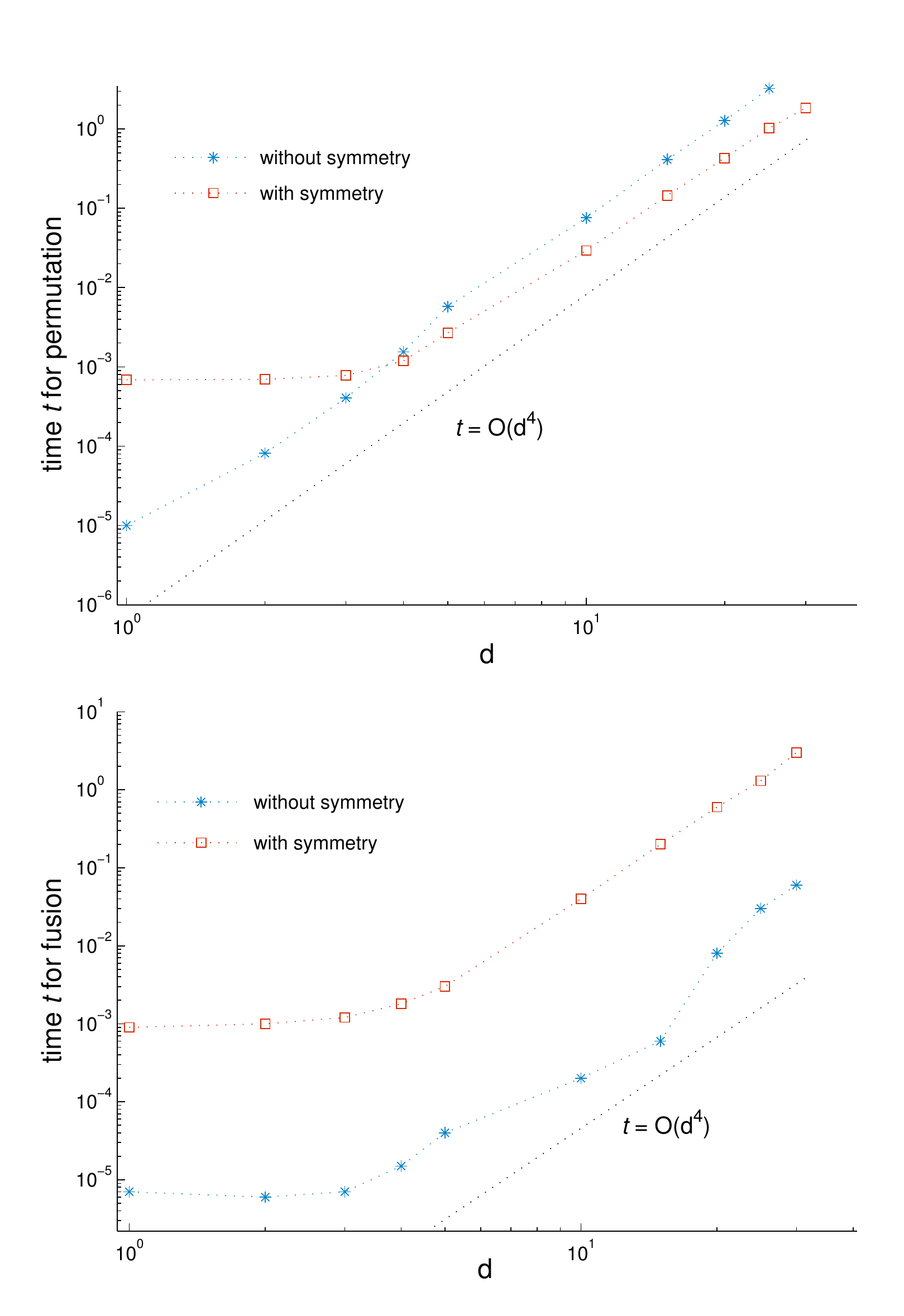}
\caption{
Computation times (in seconds) required to permute and fuse two indices of a rank-four tensor $\hat{T}$, as a function of the size of the indices. All four indices of $\hat{T}$ have the same size $5d$, and therefore the tensor contains $|\hat{T}|=5^4d^4$ coefficients. The figures compare the time required to perform these operations using a regular tensor and a $U(1)$ invariant tensor, where in the second case each index contains 5 different values of the particle number $n$ (each with degeneracy $d$) and the canonical form of Eqs.~\ref{eq:Tcanon}-\ref{eq:Tcanon2} is used. The upper figure shows the time required to permute two indices: For large $d$, exploiting the symmetry of a $U(1)$ invariant tensor by using the canonical form results in shorter computation times. The lower figure shows the time required to fuse two adjacent indices. In this case, maintaining the canonical form requires more computation time. Notice that in both figures the asymptotic cost scales as $O(d^4)$, or the size of $\hat{T}$, since this is the number of coefficients which need to be rearranged.
We note that the fixed-cost overheads associated with symmetric manipulations could potentially vary substantially with choice of programming language, compiler, and machine architecture. The results given here show the performance of the authors' MATLAB implementation of $U(1)$ symmetry.
\label{fig:permFuse}}
\end{figure}

This example illustrates that fusing and splitting indices while maintaining the canonical form of Eqs.~\ref{eq:Tcanon}-\ref{eq:Tcanon2} requires more work than reshaping regular indices. Indeed, after taking indices $b$ and $c$ into $d=b\times c$ by listing all pairs of values $b\times c$, we still need to reorganize the resulting basis elements according to their particle number $n_D$. Although this can be done by following the simple table given by $\fuser$, it may add significantly to the overall computational cost associated with reshaping a tensor. For instance, Fig.~\ref{fig:permFuse} shows that, when using MATLAB, fusing indices of invariant tensors can be more expensive than fusing indices of regular tensors.

\subsection{Multiplication of two matrices\label{sec:symTN:multiply}}


By permuting and reshaping the indices of a $U(1)$ invariant tensor, we can convert it into a $U(1)$ invariant matrix $\hat{T}= \bigoplus_{n n'} \hat{T}_{nn'} \delta_{n,n'}$, or simply
\begin{equation}
	\hat{T} = \bigoplus_{n} \hat{T}_{n},
\label{eq:Mcanon}
\end{equation}
where $\hat{T}_{n} \equiv \hat{T}_{nn}$. In components, matrix $\hat{T}$ reads
\begin{equation}
	(\hat{T})_{ab} = \left(\hat{T}_{n}\right)_{t_n t_n'}, 
	\label{eq:Mcanon2}
\end{equation}
where $a=(n,t_n)$ and $b = (n, t_n')$. In particular, similar to the discussion in Sec.~\ref{sec:tensor:multiply} for regular tensors, the multiplication of two tensors invariant under the action of $U(1)$ can be reduced to the multiplication of two $U(1)$ invariant matrices.

Let $\hat{R}$ and $\hat{S}$ be two $U(1)$ invariant matrices, with canonical forms 
\begin{equation}
	\hat{R} = \bigoplus_n \hat{R}_n, ~~~~ \hat{S} = \bigoplus_n \hat{S}_n.
\end{equation}
Their product $\hat{T} = \hat{R}\cdot \hat{S}$, Eq.~\ref{eq:Mmultiply}, is then another matrix $\hat{T}$ which is also block diagonal,
\begin{equation}
	\hat{T} = \bigoplus_n \hat{T}_n,
\end{equation}
such that each block $\hat{T}_n$ is obtained by multiplying the corresponding blocks $\hat{R}_n$ and $\hat{S}_n$,
\begin{equation}
	\hat{T}_n = \hat{R}_n\cdot \hat{S}_n.
\label{eq:TRSblock}
\end{equation}

Eqs.~\ref{eq:Mcanon} and \ref{eq:TRSblock} make evident the potential reduction of computational costs that can be achieved by manipulating $U(1)$ invariant matrices in their canonical form. First, a reduction in memory space follows from only having to store the diagonal blocks in Eq.~\ref{eq:Mcanon}. Second, a reduction in computational time is implied by just having to multiply blocks in Eq.~\ref{eq:TRSblock}. This is illustrated in the following example

\textbf{Example 9 :} Consider a $U(1)$ invariant matrix $\hat{T}$ which is a linear map in a space $\mathbb{V}$ that decomposes into $q$ irreps $\mathbb{V}_n$, each of which has the same degeneracy $d_n=d$.
That is, $\hat{T}$ is a square matrix of dimensions $dq\times dq$, and with the block-diagonal form of Eq.~\ref{eq:Mcanon}. Since there are $q$ blocks $\hat{T}_n$ and each block has size $d\times d$, the $U(1)$ invariant matrix $\hat{T}$ contains $qd^2$ coefficients. For comparison, a regular matrix of the same size contains $q^2d^2$ coefficients, a number greater by a factor of $q$.

Let us now consider multiplying two such matrices. We use an algorithm that requires $O(l^3)$ computational time to multiply two matrices of size $l\times l$. The cost of performing $q$ multiplications of $d\times d$ blocks in Eq.~\ref{eq:TRSblock} scales as $O(qd^3)$. In contrast the cost of mutiplying two regular matrices of the same size scales as $O(q^3d^3)$, requiring $q^2$ times more computational time. 

\begin{figure}[t]
  \includegraphics[width=8cm]{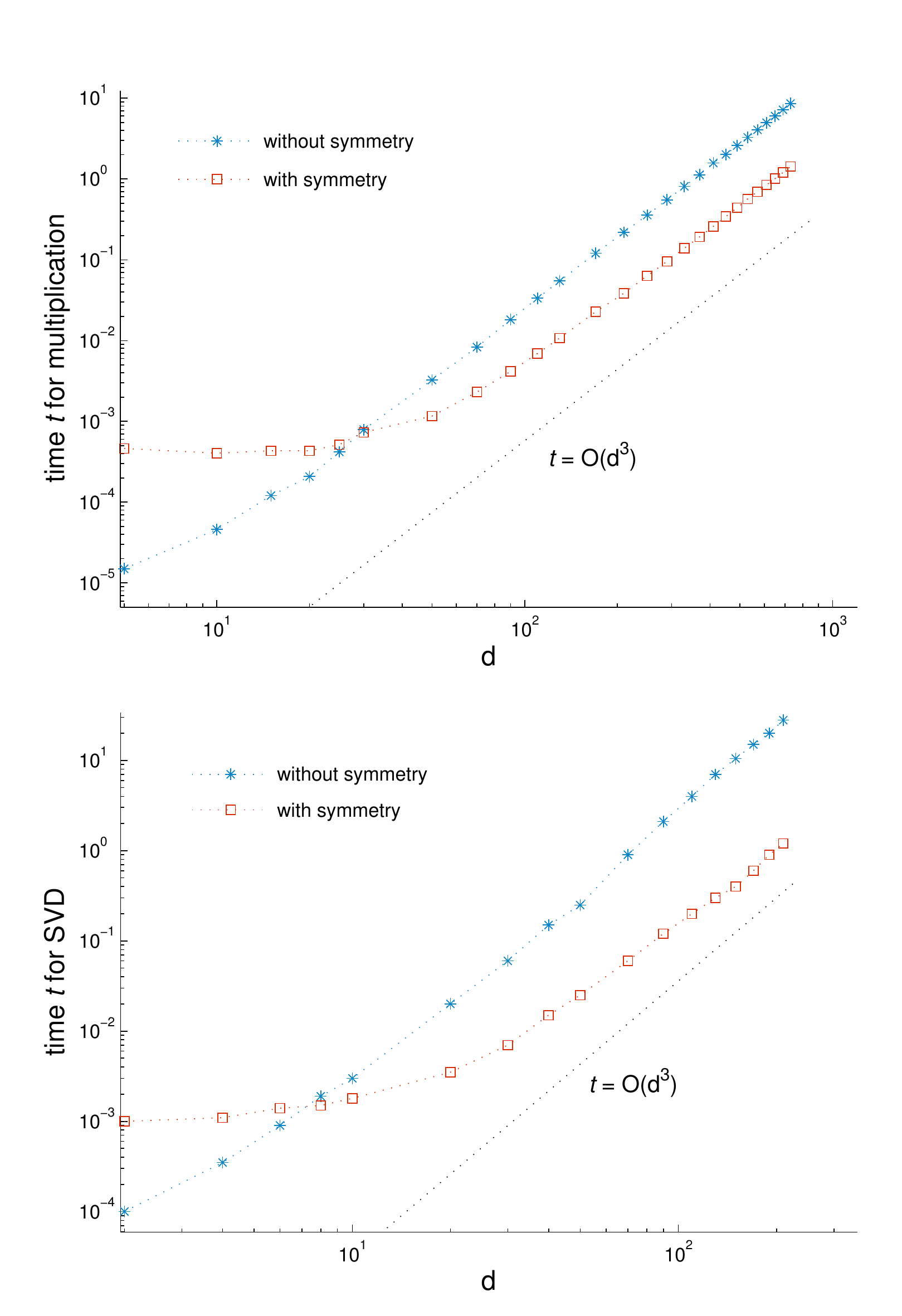}\\
\caption{
Computation times (in seconds) required to multiply two matrices (upper panel) and to perform a singular value decomposition (lower panel), as a function of the size of the indices. Matrices of size $5d \times 5d$ are considered.
The figures compare the time required to perform these operations using regular matrices and $U(1)$ invariant matrices, where for the $U(1)$ matrices each index contains 5 different values of the particle number $n$, each with degeneracy $d$, and the canonical form of Eqs.~\ref{eq:Mcanon}-\ref{eq:Mcanon2} is used. That is, each matrix decomposes into $5$ blocks of size $d\times d$. 
For large $d$, exploiting the block diagonal form of $U(1)$ invariant matrices results in shorter computation time for both multiplication and singular value decomposition. The asymptotic cost scales with $d$ as $O(d^3)$, while the size of the matrices grows as $O(d^2)$.
We note that the fixed-cost overheads associated with symmetric manipulations could potentially vary substantially with choice of programming language, compiler, and machine architecture. The results given here show the performance of the authors' MATLAB implementation of $U(1)$ symmetry.
\label{fig:multSvd}}
\end{figure}

Fig.~\ref{fig:multSvd} shows a comparison of computation times when multiplying two matrices with MATLAB, for both $U(1)$ symmetric and regular matrices.

\subsection{Factorization of a matrix\label{sec:symTN:factorize}}


The factorization of a $U(1)$ invariant matrix $\hat{T}$, Eq.~\ref{eq:Mcanon}, can also benefit from the block-diagonal structure. Consider, for instance, the singular value decomposition $\hat{T} = \hat{U}\hat{S}\hat{V}$ of Eq.~\ref{eq:singular}. In this case we can obtain the matrices
\begin{equation}
	\hat{U} = \bigoplus_n \hat{U}_n 
	~~~~ \hat{S} = \bigoplus_n \hat{S}_n 
	~~~~ \hat{V} = \bigoplus_n \hat{V}_n
\end{equation}
by performing the singular value decomposition of each block $\hat{T}_n$ independently,
\begin{equation}
	\hat{T}_n = \hat{U}_n \hat{S}_n \hat{V}_n.
\end{equation}

The computational savings are analogous to those described in Example 9 above for the multiplication of matrices. Fig.~\ref{fig:multSvd} also shows a comparison of computational times required to perform a singular value decomposition on $U(1)$ invariant and regular matrices using MATLAB.

\subsection{Discussion\label{sec:symTN:discussion}}

In this section we have seen that $U(1)$ invariant tensors can be written in the canonical form of Eqs.~\ref{eq:Tcanon}-\ref{eq:Tcanon2}, and that this canonical form is of interest because it offers a compact description in terms of only those coefficients which are not constrained by the symmetry. We have also seen that maintaining the canonical form during tensor manipulations adds some computational overhead when reshaping (fusing or splitting) indices, but reduces computation time when permuting indices (for sufficiently large tensors) and when multiplying or factorizing matrices (for sufficiently large matrix sizes).

The cost of reshaping and permuting indices is proportional to the size $|\hat{T}|$ of the tensors, whereas the cost of multiplying and factorizing matrices is a larger power of the matrix size, for example $|\hat{T}|^{3/2}$. The use of the canonical form when manipulating large tensors therefore results in an overall reduction in computation time, making it a very attractive option in the context of tensor network algorithms. This is exemplified in the next section, where we apply the MERA to study the ground state of quantum spin models with a $U(1)$ symmetry.

On the other hand, the cost of maintaining invariant tensors in the canonical form becomes more relevant when dealing with smaller tensors. In the next section we will also see that in some situations, this additional cost may significantly reduce, or even offset, the benefits of using the canonical form. In this event, and in the specific context of algorithms where the same tensor manipulations are iterated many times, it is possible to significantly decrease the additional cost by \textit{precomputing} the parts of the tensor manipulations that are repeated on each iteration. Precomputation schemes are described in more detail in the Appendices. Their performance is illustrated in the next section.




\begin{figure}[t]
  \includegraphics[width=8cm]{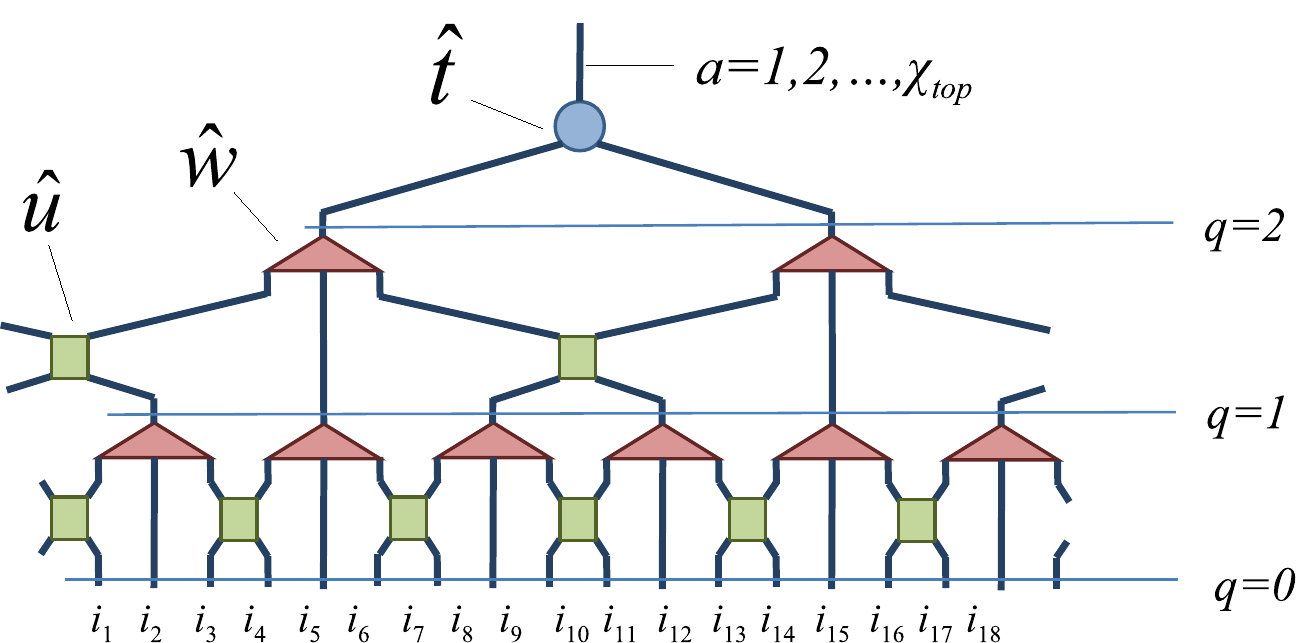}
\caption{
MERA for a system of $L=2\times 3^{2}= 18$ sites, made of two layers of disentanglers $\hat{u}$ and isometries $\hat{w}$ and a top tensor $\hat{t}$.\label{fig:MERA}}
\end{figure}




\section{ Tensor network algorithms with U(1) symmetry: a practical example\label{sec:MERA}}


In previous sections we have described a strategy to incorporate a $U(1)$ symmetry into tensors, tensor networks, and their manipulations. To further illustrate how the strategy works in practice, in this section we consider its implementation in the context of the multi-scale entanglement renormalization ansatz, or MERA.

\subsection{Multi-scale entanglement renormalization ansatz\label{sec:MERA:ansatz}}


Fig.~\ref{fig:MERA} shows a MERA that represent states $\ket{\Psi}\in \mathbb{V}^{\otimes L}$ of a lattice $\mathcal{L}$ made of $L=18$ sites (see Sec.~\ref{sec:tensor:TNstates}). Recall that the MERA is made of layers of isometric tensors, known as disentanglers $\hat{u}$ and isometries $\hat{w}$, that implement a coarse-graining transformation. In this particular scheme, isometries map three sites into one and the coarse-graining transformation reduces the $L=18$ sites of $\mathcal{L}$ into two sites using two layers of tensors. A collection of states on these two sites is then encoded in a top tensor $\hat{t}$, whose upper index $a=1,2,\cdots, \chi_{\tiny \mbox{top}}$ is used to label $\chi_{\tiny \mbox{top}}$ states $\ket{\Psi_a} \in \mathbb{V}^{\otimes L}$. 

In this section we will consider a MERA analogous to that of Fig.~\ref{fig:MERA} but with $Q$ layers of disentanglers and isometries, which we will use to describe states on a lattice $\mathcal{L}$ made of $2\times 3^{Q}$ sites. We will use this variational ansatz to obtain an approximation to the ground state and first excited states of two quantum spin chains that have a global internal $U(1)$ symmetry, namely the spin-$1/2$ quantum XX chain of Eq.~\ref{eq:XX} and the spin-$1/2$ antiferromagnetic quantum Heisenberg chain of Eq.~\ref{eq:XXX}. Each spin-1/2 degree of freedom of the chain is described by a vector space spanned by two orthonormal states $\{\ket{\downarrow}, \ket{\uparrow}\}$. Here we will represent them by the states $\{\ket{0}, \ket{1}\}$ corresponding to zero and one particles, as in Example 1 of Sec.~\ref{sec:symmetry:irreps}. For computational convenience, we will consider a lattice $\mathcal{L}$ where each site contains two spins, or states, $\{\ket{\downarrow \downarrow}, \ket{\downarrow \uparrow}, \ket{\uparrow \downarrow},  \ket{\uparrow \uparrow}\}$. Therefore each site of $\mathcal{L}$ is described by a space $\mathbb{V} \cong \mathbb{V}_0 \oplus \mathbb{V}_1 \oplus \mathbb{V}_2$, where $d_0=d_2=1$ and $d_1=2$, as in Example 2 of Sec.~\ref{sec:symmetry:irreps}. Thus, a lattice $\mathcal{L}$ made of $L$ sites corresponds to a chain of $2L$ spins. In such a system, the total particle number $N$ ranges from $0$ to $2L$. [Equivalently, the $z$-component of the total spin $S_z$ ranges from $-L$ to $L$, with $S_z = N-L$].
 
\subsection{MERA with U(1) symmetry}

A $U(1)$ invariant version of the MERA, or $U(1)$ MERA for short, is obtained by simply considering $U(1)$ invariant versions of each isometric tensors, namely the disentanglers $\hat{u}$, isometries $\hat{w}$, and top tensor $\hat{t}$. This requires assigning a particle number operator to each index of the MERA. Each open index of the first layer of disentanglers corresponds to one site of $\mathcal{L}$. The particle number operator on any such index is therefore given by the quantum spin model under consideration. We can characterize the particle number operator by two vectors $\vec{n}$ and $\vec{d}$---a list of the different values the particle number takes and the degeneracy associated with each such particle number, respectively. In the case of the vector space $\mathbb{V}$ for each site of $\mathcal{L}$ described above, $\vec{n} = [0, 1, 2]$ and $\vec{d}=[1, 2, 1]$. For the open index of the tensor $\hat{t}$ at the very top the MERA, the assignment of charges is also straighforward. For instance, to find an approximation to the ground state and first seven excited states of the quantum spin model with particle number $N$, we choose $\vec{n} = [N]$ and $\vec{d} = [8]$. [In particular, a vanishing $S_z$ corresponds to $N=L$.]

For each of the remaining indices of the MERA, the assignment of the pair $(\vec{n},\vec{d})$ needs careful consideration and a final choice may only be possible after numerically testing several options and selecting the one which produces the lowest expectation value of the energy. Table \ref{table:degdist} shows the assignment of particle numbers and degeneracies made to represent the ground state and several excited states in a system of $L=2\times 3^3 = 54$ sites (that is, $108$ spins) with total particle number $N=L=54$ [or $S_z=0$]. Notice that at level $q$ of the MERA ($q=1,2,3$) each index effectively corresponds to a block of $n_q \equiv 3^q$ sites of $\mathcal{L}$. Therefore having exactly $n_q$ particles in a block of $n_q$ sites corresponds to a density of $1$ particle per site of $\mathcal{L}$. The assigned particle numbers of Table \ref{table:degdist}, namely $[n_q-2,n_q-1,n_q,n_q+1,n_q+2]$ for level $q$, then correspond to allowing for fluctuations of up to two particle with respect to the average density. The sum of corresponding degeneracies $\vec{d}=[d_{n_q-2},d_{n_q-1},d_{n_q},d_{n_q+1},d_{n_q+2}]$ gives the bond dimension $\chi$, which in the example is $\chi=13$.

\begin{table}[ht]
\centering 
\begin{tabular}{c| c| c} 
\hline\hline 
Level $q$ & Particle numbers $\vec{n}$ & Degeneracy $\vec{d}$ \\ [0.5ex] 
\hline
top & $N=54$ & $[\chi_{\tiny \mbox{top}}]$ \\ 
3 & $\left[25, 26, 27, 28, 29\right]$ & $\left[1, 3, 5, 3, 1\right]$  \\
2 & $\left[7, 8, 9, 10, 11\right]$ & $\left[1, 3, 5, 3, 1\right]$  \\
1 & $\left[1, 2, 3, 4, 5\right]$ & $\left[1, 3, 5, 3, 1\right]$  \\
0 & $\left[0, 1, 2\right]$ & $\left[1, 2, 1\right]$  \\ 
[1ex]
\hline 
\hline 
\end{tabular}
\caption{
Example of particle number assignment in a U(1) MERA for $L = 54$ sites (or $108$ spins). The total bond dimension is $\chi = 1+3+5+3+1 = 13$. \label{table:degdist} 
}
\end{table}

In order to find an approximation to the ground state of either $\hat{H}_{XX}$ or $\hat{H}_{XXX}$ in Eqs.~\ref{eq:XX}-\ref{eq:XXX}, we set $\chi_{\tiny\mbox{top}}=1$ and optimize the tensors in the MERA so as to minimize the expectation value 
\begin{equation}
	\bra{\Psi} \hat{H} \ket{\Psi}
\end{equation}
where $\ket{\Psi}\in \mathbb{V}^{\otimes L}$ is the pure state represented by the MERA and $\hat{H}$ is the relevant Hamiltonian. In order to find an approximation to the $\chi_{\tiny\mbox{top}}>1$ eigenstates of $\hat{H}$ with lowest energies, we optimize the tensors in the MERA so as to minimize the expectation value
\begin{equation}
	\sum_{a=1}^{\chi_{\tiny\mbox{top}}}\bra{\Psi_a} \hat{H} \ket{\Psi_a}.
\end{equation}
The optimization is carried out using the MERA algorithm described in Ref.~\onlinecite{Evenbly09}, which requires contracting tensor networks (by sequentially multiplying pairs of tensors) and performing singular value decompositions. In the present example, all of these operations will be performed exploiting the $U(1)$ symmetry.

\begin{figure}[t]
  \includegraphics[width=8cm]{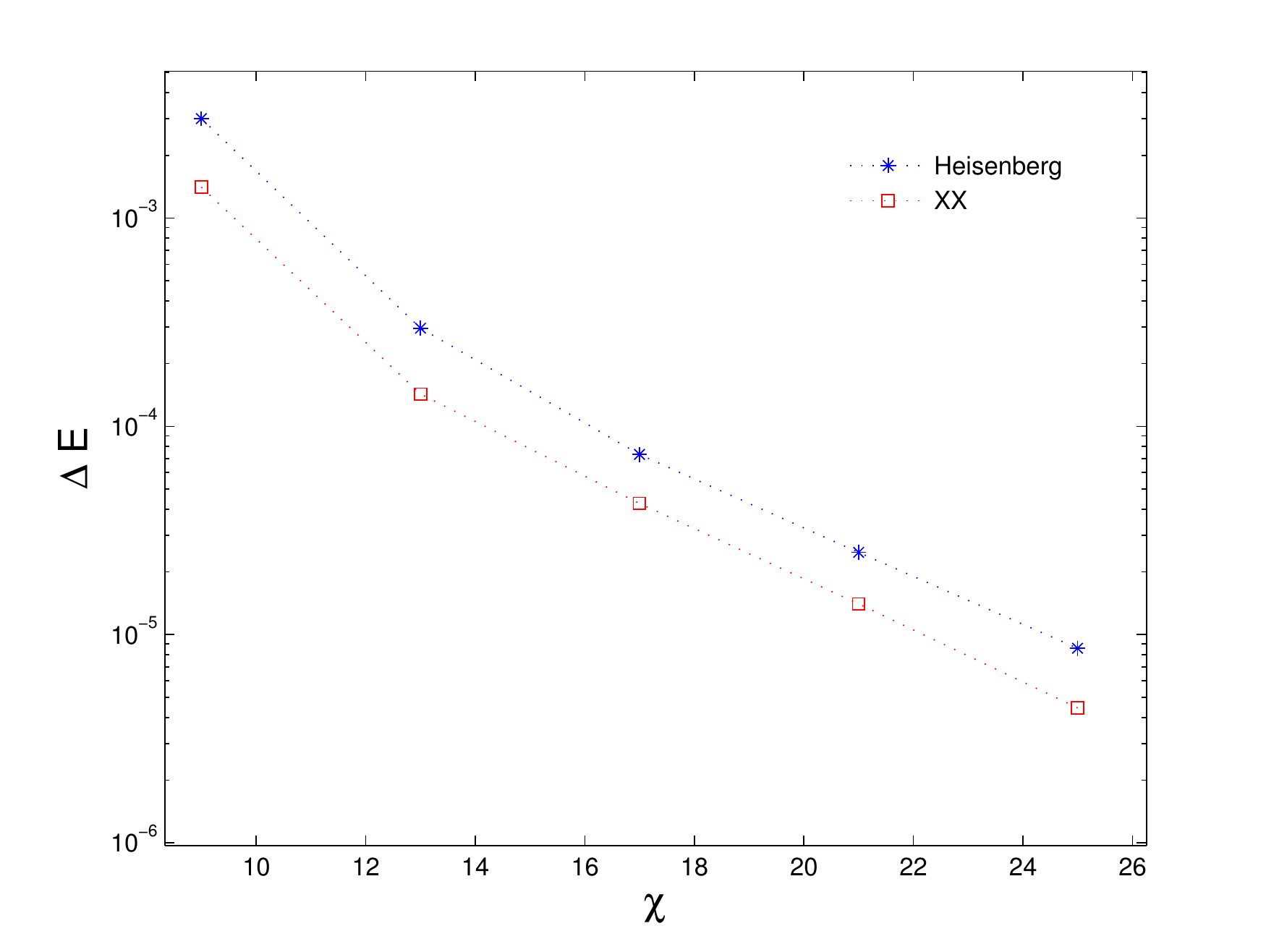}
\caption{Error in ground state energy $\Delta E$ as a function of $\chi$ for the XX and Heisenberg models with $2L=108$ spins and periodic boundary conditions, in the particle number sector $N=L$ (or $S_z=0$). The error is seen to decay exponentially with $\chi$.\label{fig:gserror}}
\end{figure}

Fig.~\ref{fig:gserror} shows the error in the ground state energy as a function of the bond dimension $\chi$, for assignments of degeneracies similar to those in Table \ref{table:speedup}. The error is seen to decay exponentially with increasing $\chi$, indicating increasingly accurate approximations to the ground state.

\begin{table}[t]
\centering 
\begin{tabular}{c|c| r| r| c} 
\hline\hline 
$\chi$ & Degeneracy $\vec{d}$ & no. of coefficients & no. of coefficients & ratio \\ 
       &                      & (regular)~~~~~~~    & (symmetric)~~~~     & \\
[1ex] 
\hline
& & & & \\
~~4~~ & $\left[0, 1, 2, 1, 0\right]$  & 1552~~ & 426~~ & ~~3.6~:~1~~   \\ 
~~8~~ & $\left[0, 2, 4, 2, 0\right]$  & 17216~~ & 4714~~  & ~~3.7~:~1~~   \\ 
~~13~~ & $\left[1, 3, 5, 3, 1\right]$  & 115501~~ & 21969~~ & ~~5.3~:~1~~  \\
~~17~~ & $\left[1, 4, 7, 4, 1\right]$  & 335717~~ & 68469~~ & ~~5.0~:~1~~ \\
~~21~~ & $\left[1, 5, 9, 5, 1\right]$  & 779965~~ & 166901~~ & ~~4.7~:~1~~   \\
~~30~~ & $\left[2, 7, 12, 7, 2\right]$ & 3243076~~ & 639794~~ & ~~5.1~:~1~~   \\ [1ex]
\hline 
\hline 
\end{tabular}
\caption{
Number of coefficients required to specify the tensors of a MERA for $L=54$ as a function of the bond dimension $\chi$, which decomposes into a degeneracy vector $\vec{d}$. A comparison is made between regular tensors and $U(1)$ invariant tensors.\label{table:speedup} 
}
\end{table}

\subsection{Exploiting the symmetry}

We now discuss some of the advantages of using the $U(1)$ MERA.

\subsubsection{Selection of particle number sector}

An important advantage of the $U(1)$ MERA is that it exactly preserves the $U(1)$ symmetry. In other words, the states resulting from a numerical optimization are exact eigenvectors of the total particle number operator $\hat{N}$ \eref{eq:hatN}. In addition, the total particle number $N$ can be pre-selected at the onset of optimization by specifying it in the open index of the top tensor $\hat{t}$. 

\begin{figure}[t]
  \includegraphics[width=8cm]{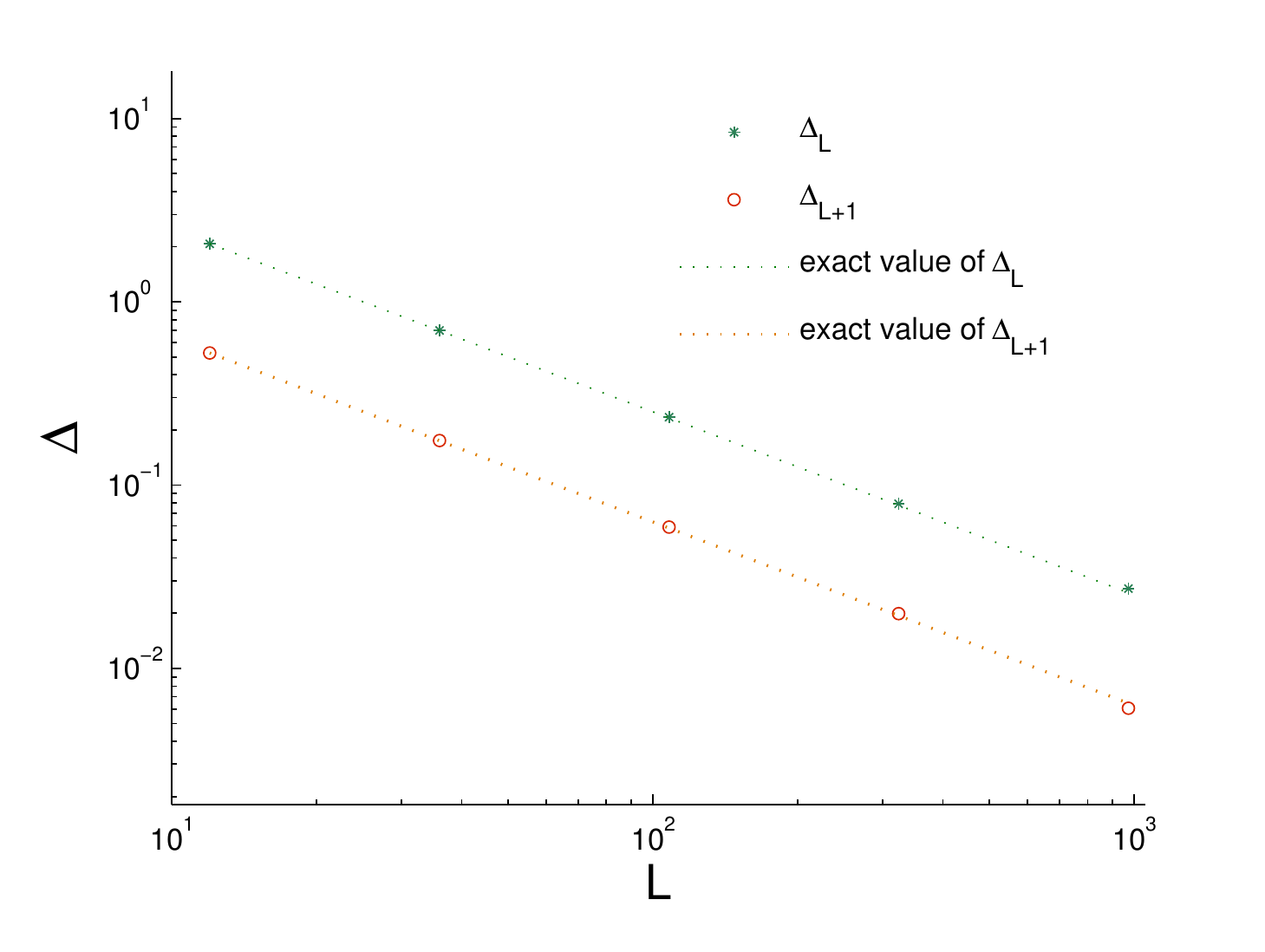}
\caption{Decay of energy gaps $\Delta$ with system size $L$ in the XX model. The upper line corresponds to the energy gap $\Delta_L$ between the ground state and the first excited state in the $N=L$ particle number (or $S_z=0$) sector. The lower line corresponds to the energy gap $\Delta_{L+1}$ between the ground states of the $N=L$ and $N=L+1$ particle number sectors.
\label{fig:XXgaps1}}
\end{figure}

Fig.~\ref{fig:XXgaps1} shows the energy gap between the ground state of an $XX$ chain with $2L$ spins (or $L$ sites), for $N=L$ particles ($S_z=0$) and two excited states. One is the first excited state with also $N=L$ particles. The other is the ground state in the sector with $N=L+1$ particles. The two energy gaps are seen to decay with the system size as $L^{-1}$. The ability to pre-select a given particle number $N$ means that only two optimizations were required: one MERA optimization for $N=L$ with $\chi_{\tiny \mbox{top}}=2$ in order to obtain an approximation to the ground state and first excited state of $\hat{H}_{XX}$ in that particle number sector; and one MERA optimization for $N=L+1$ with $\chi_{\tiny \mbox{top}}=1$ in order to obtain an approximation to the ground state of $\hat{H}_{XX}$ in the particle number sector $N=L+1$. 

Similar results can be obtained with the regular MERA. For instance, one can obtain an approximation to the ground state of a given particle number sector by adding a chemical potential term  $-\mu\sum_{s}\hat{n}^{(s)}$ to the Hamiltonian and carefully tuning the chemical potential term $\mu$ until the expectation value of the particle number $\hat{N}$ is the desired one. However, the regular MERA cannot garantee that the states obtained in this way are exact eigenvectors of $\hat{N}$. Instead, the resulting states are likely to have particle number fluctuations.

\begin{figure}[t]
  \includegraphics[width=8.5cm]{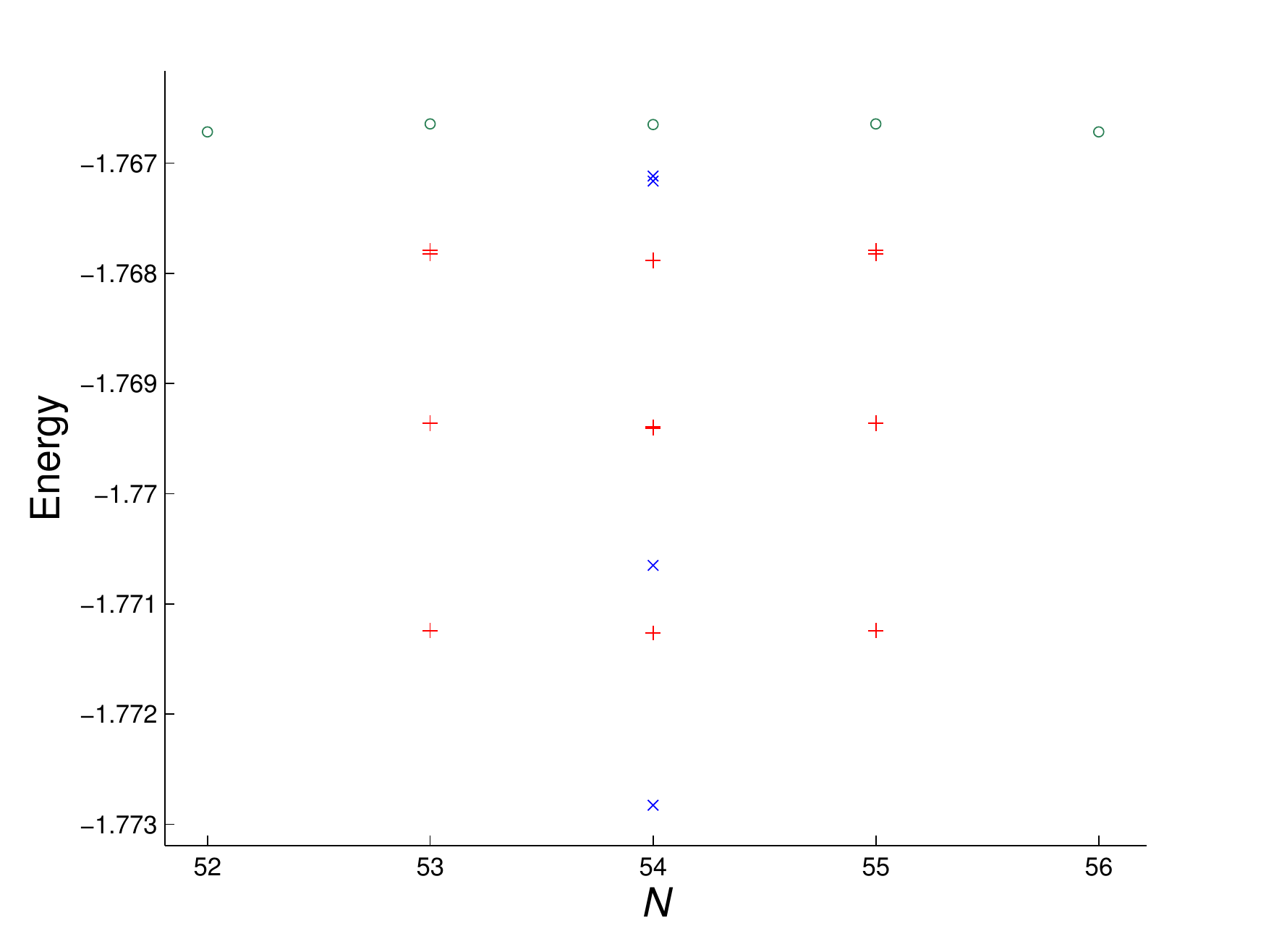}
\caption{
Low energy spectrum of $\hat{H}_{XXX}$ with $L=54$ sites (=108 spins). Depicted states have spins of zero $(\times)$, one (+), or two ($\circ$), and total number of particles ($N$) between 52 and 56. Note that the second and third spin-1 triplets are twofold degenerate.\label{fig:h3spectra}}
\end{figure}

Fig.~\ref{fig:h3spectra} shows the low energy spectrum of the Heisenberg model $\hat{H}_{XXX}$ for a periodic system of $L=54$ sites (or $108$ spins), including the ground state and several excited states in the particle sector $N=54$ (or $S_z=0$) and neighboring particle sectors. Recall that $\hat{H}_{XXX}$ is actually invariant under a global internal \SU(2) symmetry, of which particle number is a $U(1)$ subgroup. Correspondingly the spectrum is organized according to irreps of \SU(2), namely singlets (total spin $0$), triplets (total spin $1$), quintuplets (total spin $2$), etc. Again, using the $U(1)$ MERA, the five particle number sectors $N=52,53,54,55$ and $56$ can be addressed with independent computations. This implies, for instance, that in order to find the gap between the first and fourth singlets, we can simply set $N=54$ and $\chi_{\tiny \mbox{top}} = 9$ on the open index of the top tensor $\hat{t}$. In order to capture the fourth singlet using the regular MERA, we would need to consider at least $\chi_{\tiny\mbox{top}} = 19$ (at a larger computational cost and possibly lower accuracy), since this state has only the $19$\textsuperscript{th} lowest energy overall.

\subsubsection{Reduction of computational costs}

The use of $U(1)$ invariant tensors in the MERA also results in a reduction of computational costs. 

First, $U(1)$ invariant tensors, when written in the canonical form of Eqs.~\ref{eq:Tcanon}-\ref{eq:Tcanon2}, are block diagonal and therefore require less storage space. Table \ref{table:speedup} compares the number of MERA coefficients that need to be stored in the regular and symmetric case, for different choices of particle number assignments relevant to the present examples. 

\begin{figure}[t]
  \includegraphics[width=9.5cm]{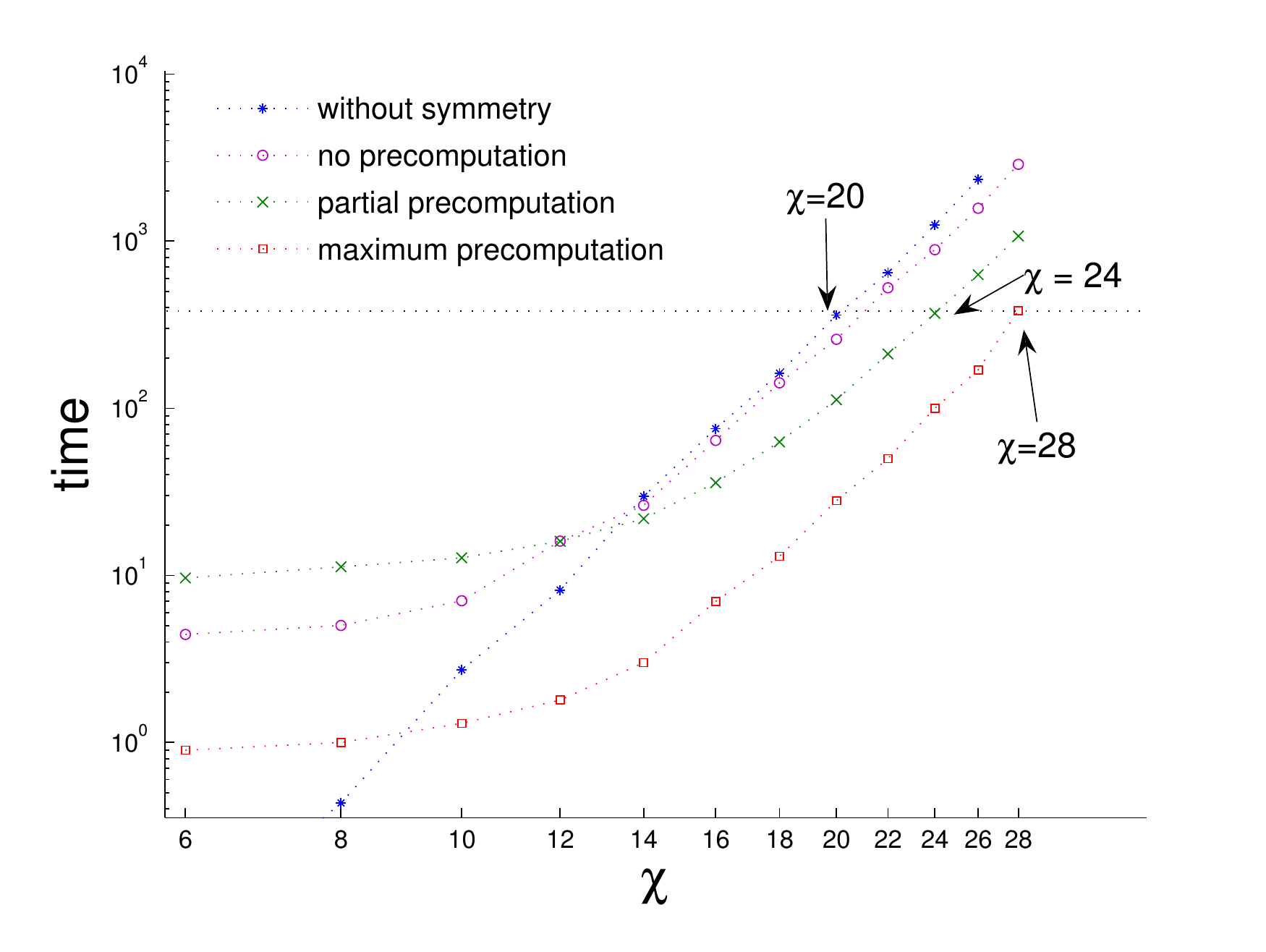}
\caption{ 
Computation time (in seconds) for one iteration of the MERA energy minimization algorithm, as a function of the bond dimension $\chi$. For sufficiently large $\chi$, exploiting the $U(1)$ symmetry leads to reductions in computation time.\label{fig:all1}
The horizontal line on this graph shows how this reduction in computation time equates to the ability to evaluate MERAs with a higher bond dimension $\chi$. For the same cost per iteration incurred when optimising a standard MERA in MATLAB with bond dimension $\chi=20$, one may choose instead to optimise a $U(1)$ symmetric MERA with partial precomputation and $\chi=24$, or with full precomputation and $\chi=28$.}
\end{figure}

Second, the computation time required to manipulate tensors is also reduced when using $U(1)$ invariant tensors in the canonical form. Fig.~\ref{fig:all1} shows the computation time required for one iteration of the energy minimization algorithm of Ref.~\onlinecite{Evenbly09} (during which all tensors in the MERA are updated once), as a function of the total bond dimension $\chi$. The plot compares the time required using regular tensors and $U(1)$ invariant tensors. For $U(1)$ invariant tensors, we display the time per iteration for three different levels of precomputation, as described in the appendix.
The figure shows that for sufficiently large $\chi$, using $U(1)$ invariant tensors always leads to a shorter time per iteration of the optimization algorithm. 

However, in the authors' reference implementation (written in C++ and MATLAB), using the symmetry without precomputation only reduces the computational time by about a factor of two for the largest $\chi$ under consideration. This is due to the fact that maintaining the canonical form for $U(1)$ invariant tensors still imposes a significant overhead for the values of $\chi$ considered (notice that the gap between the cost for regular and symmetric tensors without precomputation is still increasing as a function of $\chi$). While the magnitude of this overhead is necessarily dependent on factors such as programming language and machine architecture, more significant gains can be obtained by making maximum use of precomputation (giving computation times shorter by a factor of ten or more). This option, however, requires a significant amount of additional memory (see appendix), and a more convenient middle ground can be obtained by using a partial precomputation scheme.

\section{CONCLUSIONS\label{sec:conclusions}}


In this paper we have provided a detailed explanation of how a global internal Abelian symmetry may be incorporated into any tensor network algorithm.
Following Ref.~\onlinecite{Singh09} we considered tensor networks constructed from tensors which are invariant under the action of the internal symmetry, and showed how each tensor may be decomposed according to a canonical form 
into \textit{degeneracy} tensors (which contain all the degrees of freedom that are not affected by the symmetry) and \textit{structural} tensors (which are completely determined by the symmetry). We then introduced a set of primitive operations $\mathcal{P}$ which may be used to carry out tensor network algorithms such as MPS, PEPS, and MERA, and showed how each of these operations 
can be implemented in a way such that the canonical form is both preserved and exploited for computational gain. 

We then demonstrated the implementation of this decomposition for tensors with an internal $U(1)$ symmetry, and computed multiple benchmarks demonstrating the computational costs and speed-ups inherent in this approach. We found that although maintaining the canonical form imposed additional costs when combining or splitting tensor indices, for simulations of a sufficiently large scale these costs can be offset by the gains made when performing permutations, matrix multiplications, and matrix decompositions. 

Finally, we implemented the 
MERA on a quantum spin chain with $U(1)$ symmetry and showed that exploitation of this symmetry can lead to a decrease in computational cost by between ten and twenty times.

To demonstrate the practical nature of these gains, we applied $U(1)$ symmetry to an implementation of the Multi-scale Entanglement Renormalization Ansatz on a quantum spin chain, and achieved performance increases by a factor of ten or more. These gains may be used either to reduce overall computation time or to permit substantial increases in the MERA bond dimension $\chi$. 

Although in this paper we have focused 
on an example which is a continuous Abelian group, the formalism presented may equally well 
be applied 
to a finite Abelian group. In particular, let us consider a cyclic group $Z_q$, $q\in\mbb{Z}^+$.
\footnote{The fundamental theorem of Abelian groups states that every finite Abelian group may be expressed as a direct sum of cyclic subgroups of prime-power order.} 
As in the case of $U(1)$, the Hilbert space decomposes under the action of the group
into a direct sum of one dimensional irreps which are each characterized by an integer charge $a$,
and consequently most of the analysis presented in this paper remains 
unchanged. 
In particular, matrices which are invariant under the action of the group will be block diagonal in the basis labeled by the charge $a$ according to Eq.~\ref{eq:Schur}, and symmetric tensors enjoy the canonical decomposition stated in Eqs.~\ref{eq:Tcanon}-\ref{eq:Tcanon2}. The only objects which need modification are the fusion and splitting maps, which need to be altered so that they encode the fusion rules of $Z_q$ instead of $U(1)$. For a cyclic group $Z_q$, the fusion of two charges $a$ and $a'$ gives rise to a charge $a''$ according to $a'' = (a+a')|_q$ where $|_q$ indicates that the addition is performed modulo $q$. 
For example, $Z_3$ has
charges $a=0,1,2$, and the fusion rules for $Z_3$
take the form $a\times a'\rightarrow a''$ where the value of $a''$ is given in the following table:

\begin{table}[h]
\begin{tabular}{cc||ccc}
&&&$a$&\\
&&0&1&2\\
\hline
\hline
&0&0&1&2\\
$a'$&1&~~~1~~~&~~~2~~~&~~~0~~~\\
&2&2&0&1
\end{tabular}
\end{table}
More generally, a generic abelian group 
will be characterised by a set of charges $(a_1, a_2, a_3,\ldots)$. 
When fusing two such sets of charges $(a_1,a_2,a_3,\ldots)$ and $(a'_1,a'_2,a'_3,\ldots)$, each charge $a_i$ 
is combined with its counterpart $a'_i$ 
according to the fusion rule of the relevant subgroup. Once again, 
this behaviour may be encoded in a single fusion map $\fuser$ and its inverse $\splitter$. The formalism presented in this paper 
is therefore directly applicable to any Abelian group.

{\it Acknowledgements:} The authors thank Ian P. McCulloch for fruitful discussions. Support from the Australian Research Council (APA, FF0668731, DP0878830, DP1092513) is acknowledged.

\appendix

\section*{APPENDIX: USE OF PRECOMPUTATION IN ITERATIVE ALGORITHMS}

We have seen that the use of the canonical form given in Eqs.~\ref{eq:Tcanon}-\ref{eq:Tcanon2} to represent $U(1)$ invariant tensors can potentially lead to substantial reductions in memory requirements and in calculation time. We also pointed out, however, that there is an additional cost in maintaining an invariant tensor in its canonical form, and that this is associated with the reshaping (fusing and/or splitting) of its indices. In some situations this additional cost may significantly reduce, or even offset, the benefits of using the canonical form. 

In this appendix we investigate 
techniques for reducing this additional cost in the context of iterative tensor network algorithms. Many of the algorithms discussed in \sref{sec:tensor:TNstates} 
are iterative algorithms, repeating the same sequence of tensor network manipulations many times over. Examples include algorithms which compute tensor network approximations to the ground state by 
minimizing the expectation value of the energy, 
or by
simulating evolution in imaginary time,
with each iteration yielding an
increasingly accurate 
approximation to the ground state of the system. 

The goal of this appendix is to identify calculations which depend only on the symmetry group, and are independent of the variational coefficients of such algorithms.
Where these calculations are repeated in each iteration of the algorithm, we can
effectively eliminate the associated computational cost by performing them 
only once, either during or prior to the first iteration of the algorithm, and then 
storing and reusing these \emph{precomputed} results 
in subsequent iterations.
We will illustrate this procedure by considering the precomputation of a series of operations applied to a single tensor $\hat T$. 

To do this, we begin by revisiting the fusion and splitting tables of \sref{sec:symmetry:tp} and introducing a graphical representation of these objects. We then introduce a convenient decomposition of a symmetric tensor into a matrix accompanied by multiple fusion and/or splitting tensors, and linear maps $\Gamma$ that map one such decomposition into another. These linear maps are independent of the coefficients of the tensor being reorganized, and consequently they are precisely the objects which can be precomputed in order to quicken an iterative algorithm at the expense of additional memory cost. Finally we describe two specific precomputation schemes, differing in what is precomputed and in how the precomputed data is utilized during the execution of the algorithm, in order to illustrate the trade off between the amount of memory needed to store the precomputation data and the associated computational speedup which may be obtained. In practice, the nature of the specific implementation employed will depend on available computational resources.

\subsection{Diagrammatic notation of fusing and splitting tensors}

\begin{figure}[t]
  \includegraphics[width=8cm]{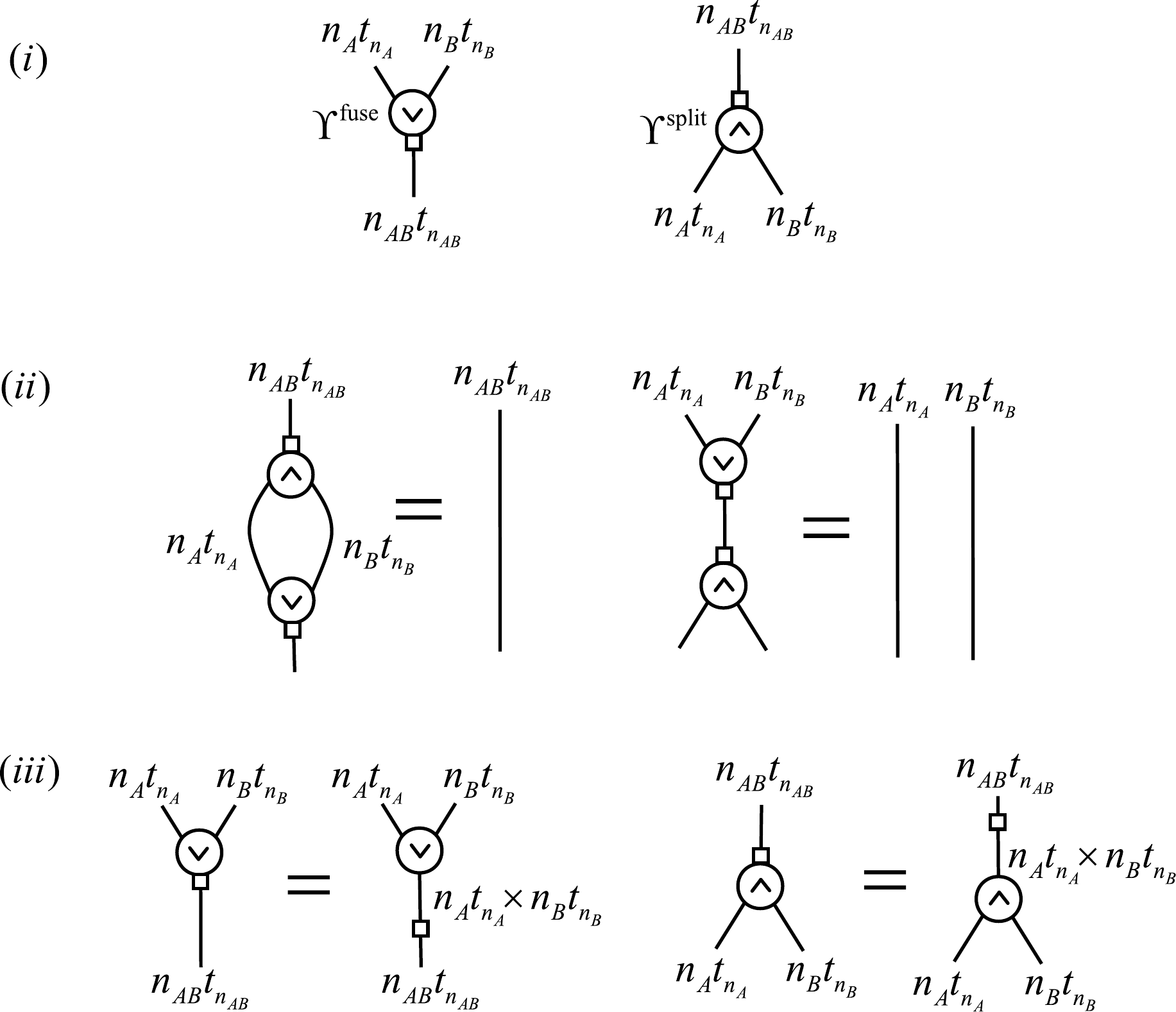}
\caption{
(i) Graphical representation of the fusion tensor $\fuser$ and the splitting tensor $\splitter$. 
(ii) The tensors $\fuser$ and $\splitter$ are unitary, and thus yield the identity when contracted pairwise as shown.
(iii) A fusion tensor decomposed into two parts. The first part (indicated by a circle with an arrow) performs the tensor product of input irreps, $n_A t_A \times n_B t_B$. The result is an index that labels pairs $(n_A t_A, n_B t_B)$. The second part (indicated by a rectangle) is a permutation that associates each pair $(n_A t_A, n_B t_B)$ with a unique $(n_{AB} t_{n_{AB}})$, corresponding to a vector of the coupled basis of $\mathbb{V}^{(AB)}$.}
\label{fig:u1fuse}
\end{figure}

In describing how we can precompute repeated manipulations of this tensor $\hat T$, we will find it useful to employ diagrammatic representations of the fusion and splitting tables $\fuser$ and $\splitter$ introduced in \sref{sec:symmetry:tp}. These tables implement a linear map between a pair of indices and their fusion product, and thus can be understood as trivalent tensors having two input legs and one output leg (or vice versa) in accordance with \sref{sec:tensor:linear}. We choose to represent them graphically as shown in \fref{fig:u1fuse}(i), where the arrow within the circle always points toward the coupled index.
The linear maps $\fuser$ and $\splitter$ are unitary, and consequently we impose that the tensors of \fref{fig:u1fuse}(i) must satisfy the identities given in \fref{fig:u1fuse}(ii), corresponding to unitarity under the action of the conjugation operation employed in diagrammatic tensor network notation (vertical reflection of a tensor and the complex conjugation of its components, typically denoted $^\dagger$).
Our notation also reflects the property
, first noted in section \ref{sec:symmetry:tp}, that $\fuser$ and $\splitter$ may be decomposed into two pieces (Fig. \ref{fig:u1fuse}(iii)). For the fusion tensor, we identify the first piece (represented by a circle containing an arrow) with the creation of a composed index using the manner we would employ in the absence of symmetry (\ref{eq:fuse}). The second piece, represented by the small square, permutes the basis elements of the composed index, reorganizing them according to total particle number. The two components of the splitting tensor are then uniquely defined by consistency with the process of conjugation for the diagrammatic representation of tensors, 
and with the unitarity condition of \fref{fig:u1fuse}(ii).

\begin{figure}[t]
  \includegraphics[width=5.2cm]{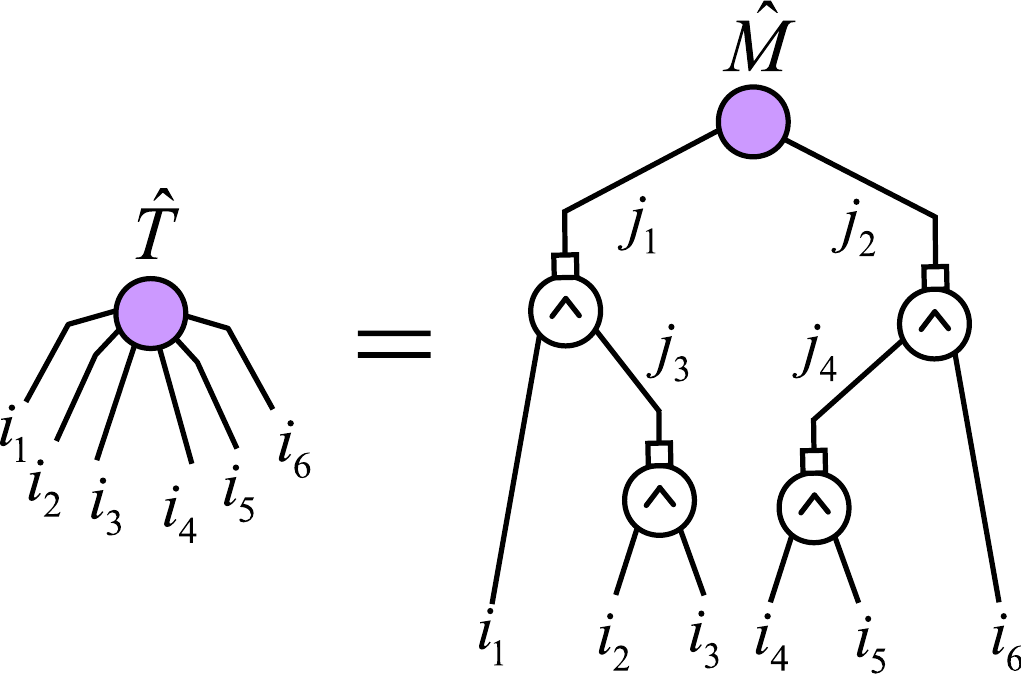}
\caption{
Binary tree decomposition of a symmetric tensor $\hat{T}$ having components $\hat{T}_{i_1 i_2 i_3 i_4 i_5 i_6}$. The tree $\mathcal{T}$ is comprised of a matrix $\hat{M}$ as the root node, 
four splitting tensors as internal nodes, and $i_1, i_2, ..., i_6$ as its leaf indices. No incoming or outgoing arrows are indicated on the indices in the figure, as the decomposition is valid for any such assignment of directional arrows.}
\label{fig:treeDeco}
\end{figure}

These requirements have an important consequence. Suppose the first part of $\fuser$ implements $b\times c\rightarrow d$ by iterating rapidly over the values of $b$ and more slowly over the values of $c$, and $b$ lies clockwise of $c$ on the graphical representation of $\fuser$. This then means that on the graphical representation of $\splitter$ which implements $d\rightarrow b\times c$, index $b$ must lie \emph{counter}clockwise of $c$. It is therefore vitally important to distinguish between the splitting tensor and a rotated depiction of the fusing tensor. To this end we require that when using this diagrammatic notation,
all tensors (with the exception of the fusion and splitting tensors) must be drawn with only downward-going legs, as seen for example in \fref{fig:treeDeco}, though the legs are still free to carry either incoming or outgoing arrows as before.

\begin{figure}[t]
  \includegraphics[width=7.39cm]{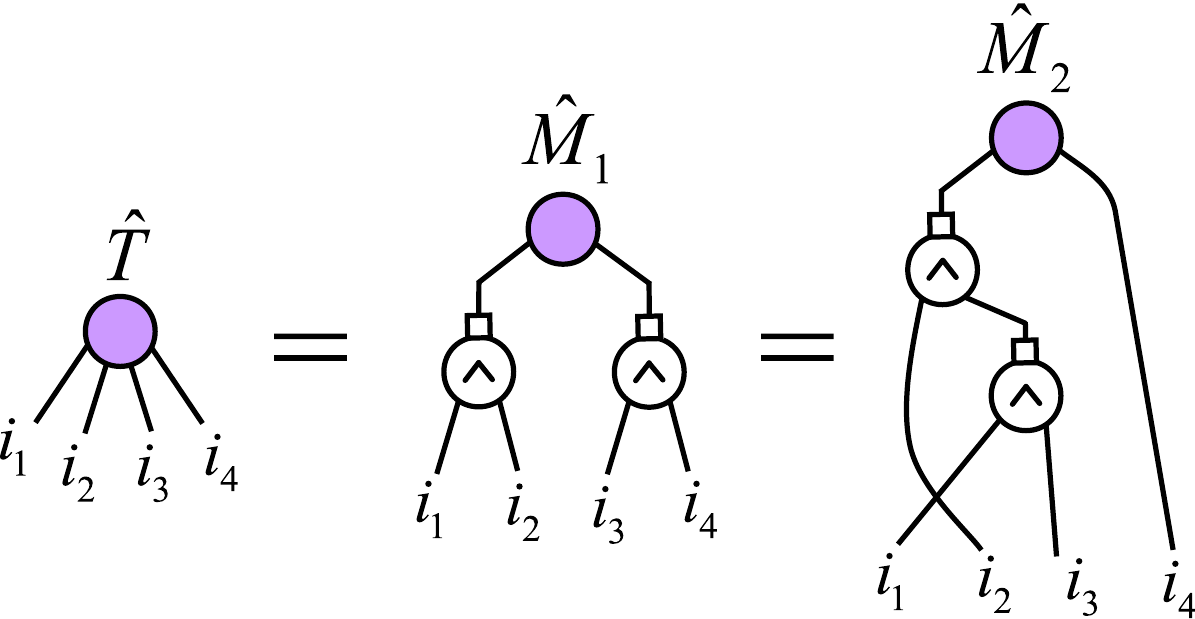}
\caption{
Two possible tree decompositions of a rank-4 tensor $\hat{T}$. Different choices $\mathcal{T}_1, \mathcal{T}_2, \cdots$ of tree decomposition for tensor $\hat{T}$ lead to different matrices $\hat{M}_1, \hat{M}_2, \cdots$ for the same tensor.}
\label{fig:treeDeco2}
\end{figure}

\subsection{Tree decomposition}

We find it convenient to decompose a rank-$k$, $U(1)$ invariant tensor $\hat{T}$, having components $\hat{T}_{i_1i_2\cdots i_k}$, as a binary tree tensor network $\mathcal{T}$ consisting of a matrix $\hat{M}$ which we will call the \textit{root node}, and of $k-2$ splitting tensors $\splitter$ as branching \textit{internal nodes}, with the \textit{leaf} indices of tree $\mathcal{T}$ corresponding to the indices $\{i_1, i_2, \cdots, i_k\}$  of tensor $\hat{T}$. We refer to decomposition $\mathcal{T}$ as a tree decomposition of $\hat{T}$. Fig.~\ref{fig:treeDeco} shows an example of tree decomposition for a rank-6 tensor. It is of the form
\begin{equation}\label{decomposeob}
\hat{T}_{i_1 i_2 i_3 i_4 i_5 i_6} = \sum_{j_1, j_2, j_3, j_4} \hat{M}_{j_1 j_2} \splitt{j_1}{i_1}{j_3} \splitt{j_2}{j_4}{i_6} \splitt{j_3}{i_2}{i_3} \splitt{j_4}{i_4}{i_5},
\end{equation}
where $\{j_1,j_2, j_3, j_4\}$ are the internal indices of the tree.

\begin{figure}[t]
  \includegraphics[width=7cm]{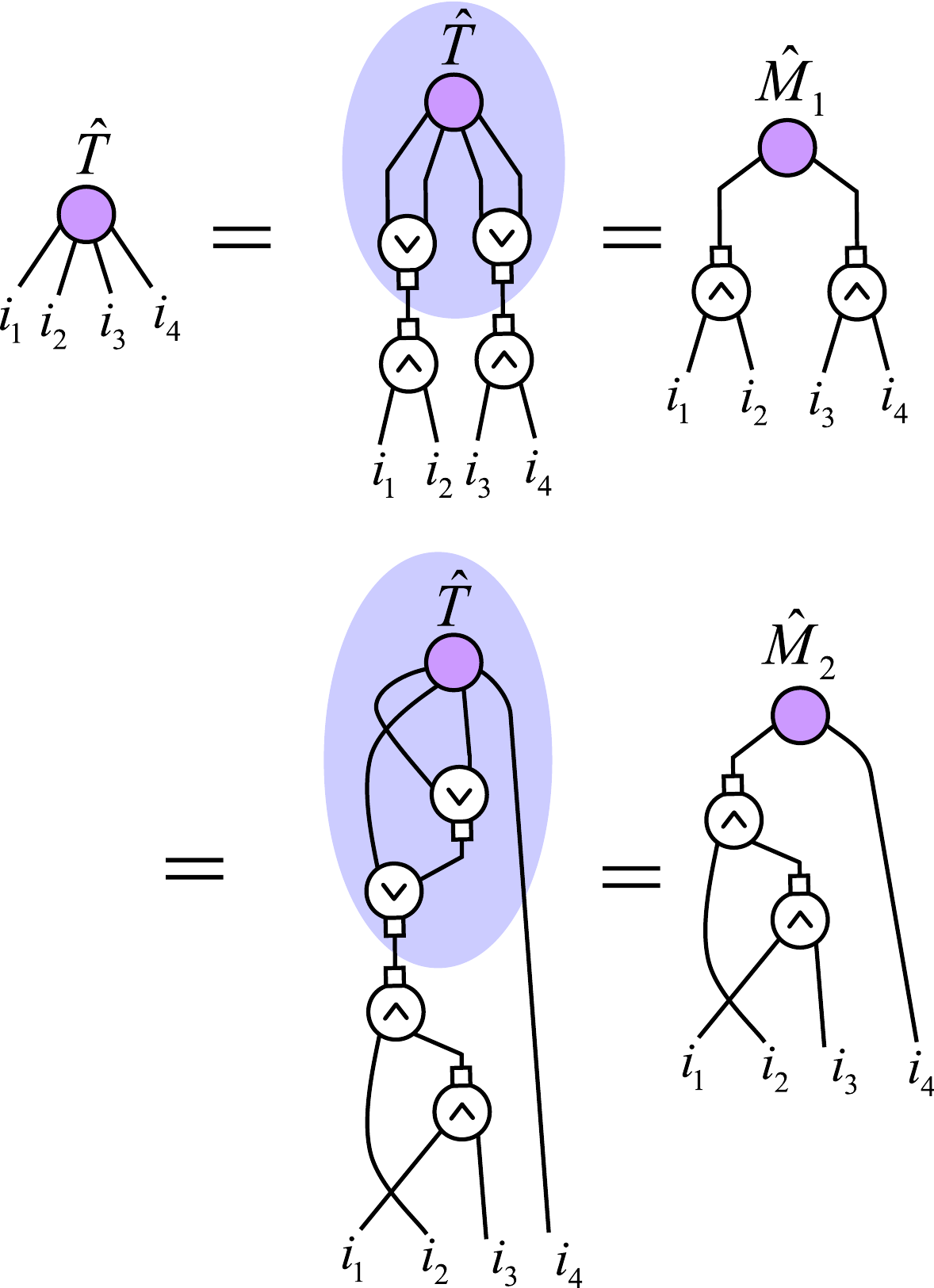}
\caption{Tree decompositions of tensor $\hat{T}$ are obtained by contracting the tensor with an appropriate resolution of the identity on its indices, selected according to the desired choice of the fusion tree $\mathcal{T}$. In each instance, evaluation of the contents of the shaded region yields the appropriate matrix $\hat M$.}
\label{fig:treeDeco3}
\end{figure}

The same tensor $\hat{T}$ may be decomposed as a tree in many different ways, corresponding to different choices of the fusion tree. As an example we show some different, but equivalent, decompositions of a rank-4 tensor in Fig. \ref{fig:treeDeco2}.
Different choices $\mathcal{T}_1, \mathcal{T}_2, \cdots$ of tree decomposition for tensor $\hat{T}$ will lead to different matrices representations $\hat{M}_1, \hat{M}_2, \cdots$ of the same tensor. Finally, Fig.~\ref{fig:treeDeco3} shows how to obtain the tree decompositions from $\hat{T}_{i_1i_2i_3i_4}$ by introducing an appropriate resolution of the identity, constructed from pairs of fusion operators $\fuser$ and splitting operators $\splitter$ in accordance with Fig.~\ref{fig:u1fuse}(ii). 

The representation of a tensor $\hat T$ by means of a tree decomposition is particularly useful because many tensor network algorithms may be understood as a sequence of operations carried out on tensors reduced to matrix form. For example, tensor network algorithms such as MPS, MERA, and PEPS consist primarily of (i) tensor network contractions, and (ii) tensor decompositions. In \sref{sec:tensor:TN}, we argued that all such operations may be reduced to matrix multiplications, matrix decompositions, and a set of primitive operations $\mathcal{P}$. When tensors are updated in these algorithms they are typically created as matrices, to which operations from $\mc{P}$ are then applied, and when they are decomposed or contracted with other tensors, this once again may take place with the tensor in matrix form. Any such matrix form may always be understood as the matrix component of an appropriate tree decomposition $\mathcal{T}$ of tensor $\hat T$, where the sequence of operations required to reshape tensor $\hat T$ to matrix $\hat M$ corresponds to the contents of the shaded area in \fref{fig:treeDeco3}.

\begin{figure}[t]
  \includegraphics[width=5.5cm]{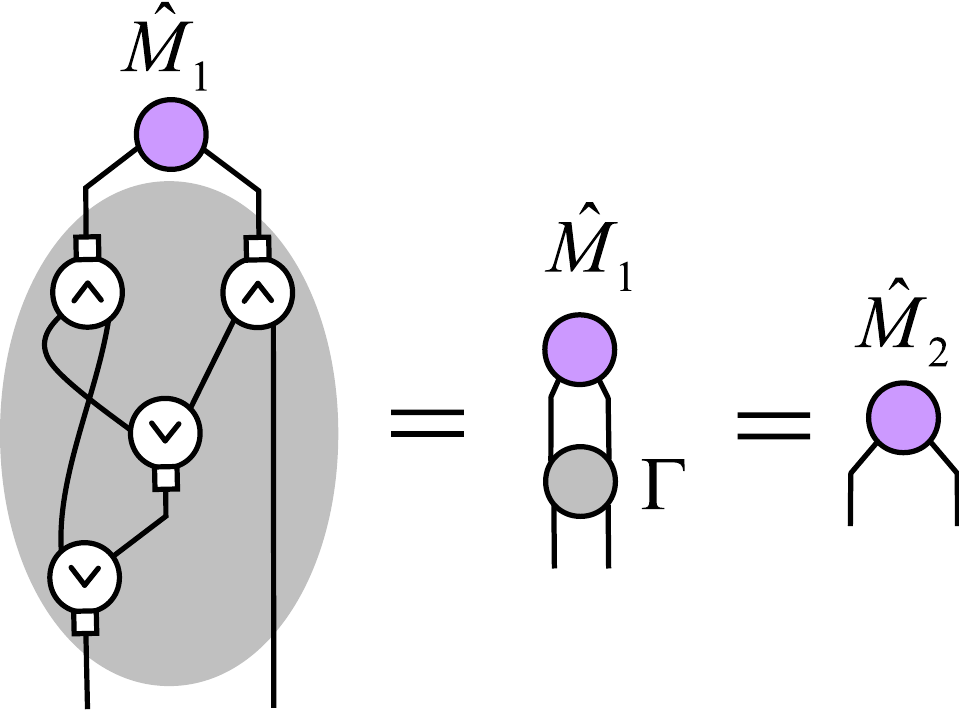}
\caption{
A matrix $\hat{M}_1$ can be reorganized into another matrix $\hat{M}_2$ by means of fusion tensors, splitting tensors, and the permutation of indices. 
These operations define a one to one linear map $\Gamma$ that acts to reorganize the coefficients of $\hat{M}_1$. $\Gamma$ does not depend on the coefficients of $\hat{M}_1$, but solely on the sequence of operations performed. 
}
\label{fig:MtoM}
\end{figure}

\subsection{Mapping between tree decompositions\label{app:gammamap}}

Suppose now that we have a tensor $\hat T$ in matrix form $\hat M_1$, which is associated with
a particular choice of tree decomposition $\mathcal{T}_1$, and we wish to transform it into another matrix form $\hat M_2$, corresponding to another tree decomposition $\mathcal{T}_2$. As indicated, this process may frequently arise during the application of many common tensor network algorithms.
The new matrix $\hat{M}_2$ can be obtained from $\hat{M}_1$ by means of a series of reshaping (splitting/fusing) and permuting operations, as indicated in \fref{fig:MtoM}, and this series of operations may be understood as defining a map $\Gamma$: 
\begin{equation}
	\hat{M}_2 = \Gamma(\hat{M}_1).
	\label{eq:Gamma}
\end{equation}
The map $\Gamma$ is a linear map which depends only on the tree structure of $\mathcal{T}_1$ and $\mathcal{T}_2$, and is independent of 
the coefficients of $\hat{M}_1$. Moreover $\Gamma$ is unitary, and 
it follows from the construction of fusing and splitting tensors and the behaviour of permutation of indices (which serves to relocate the coefficients of a tensor) that $\Gamma$ simply reorganizes the coefficients of $\hat{M}_1$ into the coefficients of $\hat{M}_2$ in a one-to-one fashion.

Therefore, one way to compute the matrix $\hat{M}_2$ from matrix $\hat{M}_1$ is by first computing the linear map $\Gamma$, which is independent of the specific coefficients in tensor $\hat{T}$, and by then applying it to $\hat{M}_1$.

\subsection{Precomputation schemes for iterative tensor network algorithms}

\begin{figure}[t]
  \includegraphics[width=8cm]{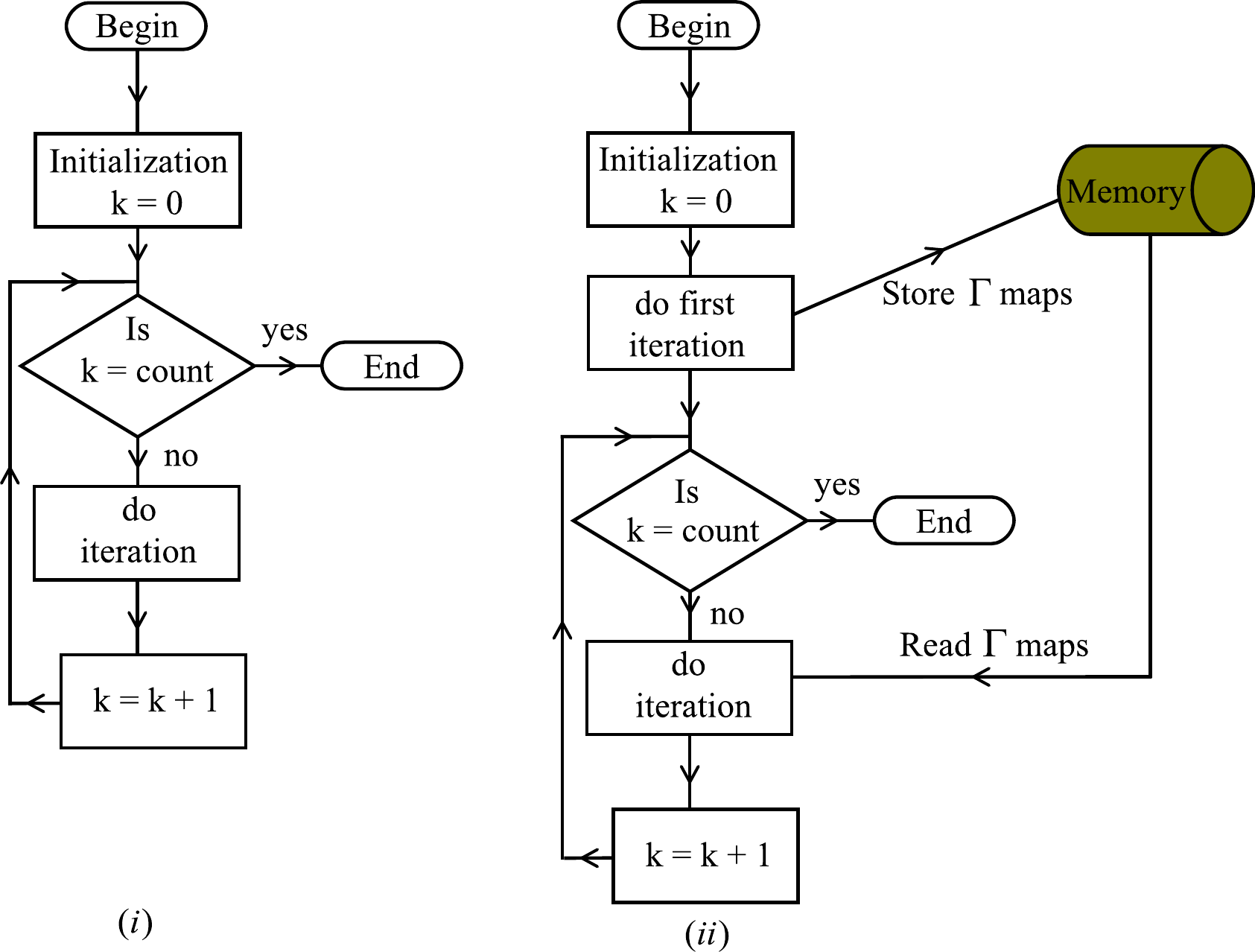}
\caption{Flow diagram for the execution of a predetermined number of iterations of a generic iterative tensor network algorithm (i) without any precomputation and (ii) with precomputation of the operations $\Gamma$.}
\label{fig:preCompute}
\end{figure}

The observation that the map $\Gamma$ is independent of the specific coefficients in $\hat{M}_1$ is particularly useful in the context of iterative tensor network algorithms. It implies that, although the coefficients in $\hat M_1$ will change from iteration to iteration, the linear map $\Gamma$ in Eq.~\ref{eq:Gamma} remains unchanged. It is therefore possible to calculate the map $\Gamma$ once, during the first iteration of the simulation, and then to store it in memory and re-use it during subsequent iterations.
We refer to such a strategy as a \textit{precomputation scheme}. Fig.~\ref{fig:preCompute} contrasts the program flow of a generic iterative tensor network algorithm with and without precomputation of the transformations $\Gamma$.

Using such a precomputation scheme a significant speed-up of 
simulations can be obtained, at the price of storing potentially large amounts of precomputed data (as a single iteration of the algorithm may require the application of many different transformations $\Gamma$). There therefore necessarily exists a trade-off between the amount of speed-up which can be obtained and the memory requirement which this entails.
In this section we describe two different precomputation schemes. The first one fully precomputes and stores all maps $\Gamma$, and is relatively straightforward to implement. This results in the maximal increase in simulation speed, but implementation requires a large amount of memory. The second scheme only partially precomputes the maps $\Gamma$, resulting in a moderate speed-up of simulations, but with memory requirements which are also similarly more modest.

\subsubsection{Maximal precomputation scheme}

As noted in \sref{app:gammamap} of this appendix, applying the map $\Gamma$ to a matrix $\hat{M}_1$ simply reorganizes its coefficients to produce the matrix $\hat{M}_2$. Moreover, if the indices of matrices $\hat M_1$ and $\hat M_2$ are fused to yield vectors $\hat V_1$ and $\hat V_2$ then the map $\Gamma$ may be understood as a permutation matrix, and this in turn may be concisely represented as a string of integers $\Gamma = \gamma_1,\ldots, \gamma_{|\hat M_1|}$ such that entry $i$ of $\hat V_2=\Gamma\hat V_1$ is given by entry $\gamma_i$ of vector $\hat V_1$. Because all of the elements from which $\Gamma$ is composed are sparse, unitary, and composed entirely of zeros and ones, the permutation 
to which $\Gamma$ corresponds may be calculated at a total cost of only $O(|\hat M_1|)$, where $|\hat M_1|$ counts only the 
elements of $\hat M_1$ which are not fixed to be zero by the 
symmetry constraints of Eq.~\ref{eq:Tcanon}. In essence, for each element of the vector $\hat V_1$ one identifies the corresponding number and degeneracy indices $(n^{\hat M_1}_i,t^{\hat M_1}_i)$ on each leg $i\in\{1,2\}$ of matrix $\hat M_1$. One can then read down the figure, applying each table $\fuser$ or $\splitter$ in turn to identify the corresponding labels $(n',t')$ on the intermediate legs, until finally
the corresponding labels on the indices of $\hat M_2$ 
are obtained. 
There is then a further 1:1 mapping from each set of labels $(n^{\hat M_2}_1,t^{\hat M_2}_1)$, $(n^{\hat M_2}_2,t^{\hat M_2}_2)$ on $\hat M_2$ to the corresponding entry in $\hat V_2$, 
completing the definition of $\Gamma$ as a map
from $\hat V_1$ to $\hat V_2$.

Storing the map $\Gamma$ for a transformation such as the one shown in \fref{fig:MtoM} 
imposes a memory cost 
of $O(|\hat M_1|)$. The application of this map also incurs a computational cost of $O(|\hat M_1|)$, but computational overhead is saved in not having to reconstruct the map $\Gamma$ on every iteration of the algorithm.

\begin{figure}[t]
  \includegraphics[width=7cm]{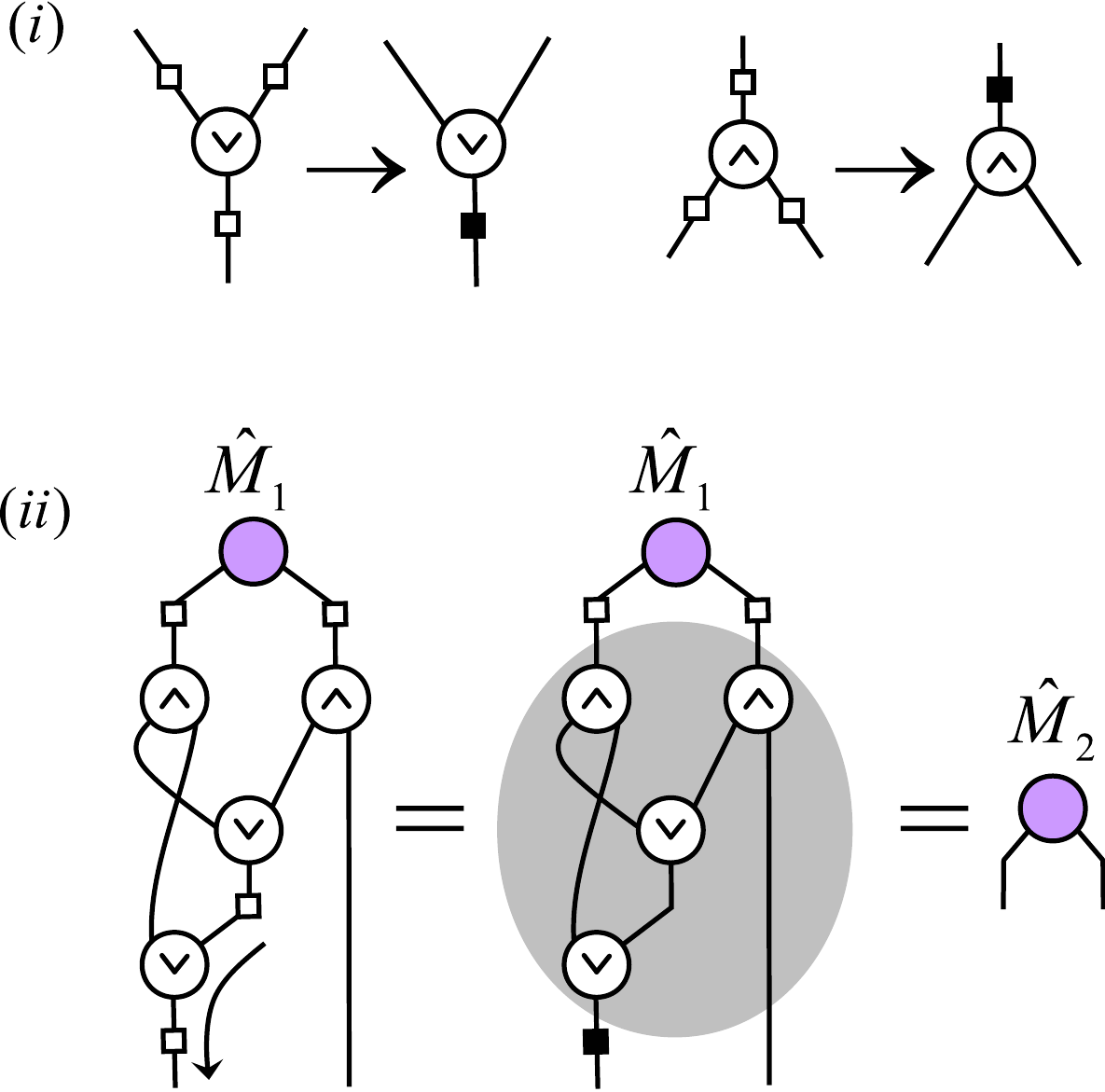}
\caption{(i) Permutations applied to one or more legs of a fusion or splitting tensor can be replaced by an appropriate permutation on the coupled index. This process can be used to replace all permutations applied on internal indices of a diagram such as \protect{\fref{fig:MtoM}} with net permutations on the indices of $\hat{M}_1$ and on the open indices of the network, as in shown in (ii). The residual fusion and splitting operations, depicted as an arrow in a circle, simply perform the basic tensor product operation and its inverse, \eref{eq:fuse}-\eref{eq:split} as described in \protect{\fref{fig:u1fuse}}(iii).}
\label{fig:partial}
\end{figure}

\subsubsection{Partial precomputation scheme}

The $O(|\hat{M}_1|)$ memory cost incurred in the previous scheme can be significant for large matrices. However, we may reduce this cost by replacing the single permutation $\Gamma$ employed in that scheme with multiple smaller operations which may also be precomputed. In this approach $\hat{M}_1$ is retained in matrix form rather than being reshaped into a vector, and we precompute permutations to be performed on its rows and columns. 

First, we decompose all the the fusion and splitting tensors into two pieces in accordance with Fig. \ref{fig:u1fuse}(iii). Next, we recognise that any permutations applied to one or more legs of a fusion or splitting tensor may always be written as a single permutation applied to the coupled index (\fref{fig:partial}(i)). 
We use this to replace all permutations on the intermediate indices of the diagram with equivalent permutations acting only on the indices of $\hat{M}_1$ and the open indices, as shown for a simple example in Fig. \ref{fig:partial}(ii). The residual fusion and splitting operations, depicted by just a circle enclosing an arrow, then simply carry out fusion and splitting of indices as would be performed in the absence of symmetry \eref{eq:fuse}-\eref{eq:split}. These operations are typically far faster than their symmetric counterparts as they do not need to sort the entries of their output indices 
according to particle number.

In subsequent iterations, the matrix $\hat{M}_2$ is obtained from $\hat{M}_1$ by consecutively
\begin{enumerate}
\item Permuting the rows and columns of $\hat{M}_1$ using the precomputed net permutations which act on the legs of $\hat M_1$. 
\item Performing any elementary (non-symmetric) splitting, permuting of indices, and fusing operations, as described by the grey-shaded region in \fref{fig:partial}(ii). 
\item Permuting the rows and columns of the resulting matrix, using the precomputed net permutations which act on the open legs of \fref{fig:partial}(ii).
\end{enumerate}
When matrix $\hat{M}_1$ is defined compactly, as in \eref{eq:Tcanon}, so that elements which are identically zero by symmetry are not explicitly stored, 
a tensor $\hat T$ is constructed from multiple blocks identified by $U(1)$ charge labels on their indices ($\hat T_{n_1n_2\ldots n_k}$ in Eq.~\ref{eq:Tcanon}). Under these conditions the elementary splitting, fusing and permutation operations of step 2 above are applied to each individual block, but some additional computational overhead is incurred in determining the necessary rearrangements of these blocks arising out of the actions performed. This rearrangement may be computed on the fly, or may also be precomputed as a mapping between the arrangement of blocks in $\hat M_1$ and that in $\hat M_2$.

The memory required to store the precomputation data in this scheme is dominated by the size of the net permutations collected on the matrix indices, and is therefore of $O(\sqrt{|\hat{M}_1|})$. The overall cost of obtaining $\hat{M}_2$ from $\hat{M}_1$ is once again of $O(|\hat{M}_1|)$, but is in general higher than the previous scheme as this cost now involves two complete permutations of the matrix coefficients, as well as 
a reorganisation of the block structure of $\hat M_1$ which may possibly be computed at runtime.
Nevertheless, in situations where memory constraints are significant, partial precomputation schemes of this sort may be preferred.



\begin{thebibliography}{73}
\bibitem{Fannes92} M. Fannes, B. Nachtergaele, and R. Werner, Commun. Math. Phys., \textbf{144}, 443 (1992).
\bibitem{Ostlund95} S. Ostlund and S. Rommer, Phys. Rev. Lett., \textbf{75}, 3537 (1995). 
\bibitem{Perez-Garcia07} D. Perez-Garcia, F. Verstraete, M.M. Wolf, and J.I. Cirac, Quantum Inf. Comput., \textbf{7}, 401 (2007). 
%
\bibitem{Wilson75} K.G. Wilson, Rev. Mod. Phys., \textbf{47}, 773 (1975).
%
\bibitem{White92} S.R. White, Phys. Rev. Lett., \textbf{69}, 2863 (1992).
\bibitem{White93} S.R. White, Phys. Rev. B, \textbf{48}, 10345 (1993).
\bibitem{Schollwoeck05} U. Schollw\"ock, Rev. Mod. Phys., \textbf{77}, 259 (2005). 
\bibitem{McCulloch08} I.P. McCulloch, arXiv:0804.2509v1 [cond-mat.str-el] (2008).
%
\bibitem{Vidal03} G. Vidal, Phys. Rev. Lett., \textbf{91}, 147902 (2003). 
\bibitem{Vidal04} G. Vidal, Phys. Rev. Lett., \textbf{93}, 040502 (2004). 
\bibitem{Daley04} A. J. Daley, C. Kollath, U. Schollw\"ock, and G. Vidal, J. Stat. Mech. Theor. Exp., P04005 (2004). 
\bibitem{White04} S. R. White and A. E. Feiguin, Phys. Rev. Lett., \textbf{93}, 076401 (2004).
\bibitem{Schollwoeck05b} U. Schollw\"ock, J. Phys. Soc. Jpn., \textbf{74S}, 246 (2005). 
\bibitem{Vidal07} G. Vidal, Phys. Rev. Lett., \textbf{98}, 070201 (2007).
%
\bibitem{Shi06} Y. Shi, L.-M. Duan and G. Vidal, Phys. Rev. A, \textbf{74}, 022320 (2006). 
%
\bibitem{Vidal07b} G. Vidal, Phys. Rev. Lett., \textbf{99}, 220405 (2007). 
\bibitem{Vidal08} G. Vidal, Phys. Rev. Lett., \textbf{101}, 110501 (2008).  
\bibitem{Evenbly09} G. Evenbly and G. Vidal, Phys. Rev. B, \textbf{79}, 144108 (2009).
\bibitem{Giovannetti08} V. Giovannetti, S. Montangero, and R. Fazio, Phys. Rev. Lett., \textbf{101},
180503 (2008).
\bibitem{Pfeifer09} R.N.C. Pfeifer, G. Evenbly, and G. Vidal, Phys. Rev. A, \textbf{79},
040301(R) (2009).
\bibitem{Vidal10} G. Vidal, in \textit{Understanding Quantum Phase Transitions}, edited by L. D. Carr (Taylor \& Francis, Boca Raton, 2010) (in press). 
%
\bibitem{Verstraete04} F. Verstraete, and J. I. Cirac, arXiv:cond-mat/0407066v1 (2004).
\bibitem{Sierra98} G. Sierra and M.A. Martin-Delgado, arXiv:cond-mat/9811170v3 (1998).
\bibitem{Nishino98} T. Nishino and K. Okunishi, J. Phys. Soc. Jpn., \textbf{67}, 3066, 1998. 
\bibitem{Nishio04} Y. Nishio, N. Maeshima, A. Gendiar, and T. Nishino, arXiv:cond-mat/0401115.  
\bibitem{Murg07} V. Murg, F. Verstraete, and J. I. Cirac, Phys. Rev. A, \textbf{75}, 033605 (2007).
\bibitem{Jordan08} J. Jordan, R. Orus, G. Vidal, F. Verstraete, and J. I. Cirac, Phys. Rev. Lett., \textbf{101}, 250602 (2008).
\bibitem{Gu08} Z.-C. Gu, M. Levin, and X.-G. Wen, Phys. Rev. B, \textbf{78}, 205116 (2008). 
\bibitem{Jiang08} H. C. Jiang, Z. Y. Weng, and T. Xiang, Phys. Rev. Lett., \textbf{101}, 090603 (2008). 
\bibitem{Xie09} Z. Y. Xie, H. C. Jiang, Q. N. Chen, Z. Y. Weng, and T. Xiang, Phys. Rev. Lett., \textbf{103}, 160601 (2009). 
\bibitem{Murg09} V. Murg, F. Verstraete, and J. I. Cirac, Phys. Rev. B, \textbf{79}, 195119 (2009).
%
\bibitem{Tagliacozzo09} L. Tagliacozzo, G. Evenbly, and G. Vidal, Phys. Rev. B, \textbf{80}, 235127 (2009). 
\bibitem{Murg10} V. Murg, O. Legeza, R. M. Noack, and F. Verstraete, arXiv:1006.3095v1 [cond-mat.str-el] (2006).
%
\bibitem{Evenbly10f} G. Evenbly and G. Vidal, Phys. Rev. B, \textbf{81}, 235102 (2010). 
\bibitem{Evenbly10b} G. Evenbly and G. Vidal, New J. Phys., \textbf{12}, 025007 (2010). 
\bibitem{Aguado08} M. Aguado and G. Vidal, Phys. Rev. Lett., \textbf{100}, 070404 (2008).
\bibitem{Cincio08} L. Cincio, J. Dziarmaga, and M. M. Rams Phys. Rev. Lett., \textbf{100}, 240603 (2008).
\bibitem{Evenbly09b} G. Evenbly and G. Vidal, Phys. Rev. Lett., \textbf{102}, 180406 (2009).
\bibitem{Koenig09} R. Koenig, B.W. Reichardt, and G. Vidal, Phys. Rev. B, \textbf{79}, 195123
(2009).
\bibitem{Evenbly10} G. Evenbly and G. Vidal, Phys. Rev. Lett., \textbf{104}, 187203 (2010).
%
\bibitem{Corboz09} P. Corboz, G. Evenbly, F. Verstraete, and G. Vidal, Phys. Rev. A, \textbf{81}, 010303(R) (2010). 
\bibitem{Kraus09} C. V. Kraus, N. Schuch, F. Verstraete, and J. I. Cirac, Phys. Rev. A, \textbf{81}, 052338 (2010). 
\bibitem{Pineda09} C. Pineda, T. Barthel, and J. Eisert, Phys. Rev. A, \textbf{81}, 050303(R) (2010).
\bibitem{Corboz09b} P. Corboz and G. Vidal, Phys. Rev. B, \textbf{80}, 165129 (2009). 
\bibitem{Barthel09} T. Barthel, C. Pineda, and J. Eisert, Phys. Rev. A, \textbf{80}, 042333 (2009).
\bibitem{Shi09} Q.-Q. Shi, S.-H. Li, J.-H. Zhao, and H.-Q. Zhou, arXiv:0907.5520v1 [cond-mat.str-el] (2009). S.-H. Li, Q.-Q. Shi, H.-Q. Zhou, arXiv:1001.3343v1 [cond-mat.supr-con] (2010).
\bibitem{Corboz10b}  P. Corboz, R. Orus, B. Bauer, and G. Vidal, Phys. Rev. B, \textbf{81}, 165104 (2010). 
\bibitem{Pizorn10} I. Pizorn and F. Verstraete, Phys. Rev. B, \textbf{81}, 245110 (2010).
\bibitem{Gu10} Z.-C. Gu, F. Verstraete, and X.-G. Wen, arXiv:1004.2563v1 [cond-mat.str-el] (2010).
%
\bibitem{Cornwell97} J. F. Cornwell, {\it Group Theory in Physics} (Academic Press, San Diego, 1997).
%
\bibitem{Ramasesha96} S. Ramasesha, S. K. Pati, H. R. Krishnamurthy, Z. Shuai, and J. L. Bredas, Phys.Rev. B, \textbf{54}, 7598 (1996).
\bibitem{Sierra97} G. Sierra and T. Nishino, Nucl. Phys., \textbf{B495}, 505 (1997). 
\bibitem{Tatsuaki00} W. Tatsuaki, Phys. Rev. E, \textbf{61}, 3199 (2000).
\bibitem{McCulloch02} I. P. McCulloch and M. Gulacsi, Europhys. Lett., \textbf{57}, 852 (2002). 
\bibitem{Daley05} A.J. Daley, S. R. Clark, D. Jaksch, and P. Zoller, Phys. Rev. A, \textbf{72}, 043618(2005).
\bibitem{Bergkvist06} S. Bergkvist, I. McCulloch, and A. Rosengren, Phys. Rev. A, \textbf{74}, 053419 (2006). 
\bibitem{Pittel06} S. Pittel and N. Sandulescu, Phys. Rev. C, \textbf{73}, 014301 (2006).
\bibitem{McCulloch07} I. McCulloch, J. Stat. Mech., P10014 (2007). 
\bibitem{Danshita07} I. Danshita, J. E. Williams, C. A. R. S\'a de Melo, and C. W. Clark, Phys. Rev. A, \textbf{76}, 043606(2007).
\bibitem{Perez-Garcia08} D. Perez-Garcia, M. M. Wolf, M. Sanz, F. Verstraete, and J. I. Cirac, Phys. Rev. Lett., \textbf{100}, 167202 (2008). 
\bibitem{Sanz09} M. Sanz, M. M. Wolf, D. Perez-Garcia, and J. I. Cirac, Phys. Rev. A, \textbf{79}, 042308 (2009). 
\bibitem{Muth09} D. Muth, B. Schmidt, and M. Fleischhauer, arXiv:0910.1749v3 [quant-ph] (2010).
\bibitem{Mishmash09} R. V. Mishmash and L. D. Carr, Math. Comput. Simul., \textbf{80}, 732 (2009). 
\bibitem{Singh10} S. Singh, H.-Q. Zhou, and G. Vidal, New J. Phys., \textbf{12}, 033029 (2010).
\bibitem{Cai10} Z. Cai, L. Wang, X.C. Xie, and Y. Wang, Phys. Rev. A, \textbf{81}, 043602 (2010).
%
\bibitem{Singh09} S. Singh, R.N.C. Pfeifer, and G. Vidal, arXiv:0907.2994v1 [cond-mat.str-el] (2009).
\bibitem{Perez-Garcia10} D. Perez-Garcia, M. Sanz, C.E. Gonzalez-Guillen, M.M. Wolf, and J.I. Cirac, New J. Phys., \textbf{12}, 025010 (2010). 
\bibitem{Zhao10} H.H. Zhao, Z.Y. Xie, Q.N. Chen, Z.C. Wei, J.W. Cai, and T. Xiang, Phys. Rev. B, \textbf{81}, 174411 (2010).
%
\bibitem{Schuch10} N. Schuch, I. Cirac, and D. Perez-Garcia, arXiv:1001.3807v2 [quant-ph] (2010).
\bibitem{Swingle10} B. Swingle and X.-G. Wen, arXiv:1001.4517v1 [cond-mat.str-el] (2010).
\bibitem{Chen10} X. Chen, B. Zeng, Z.-C. Gu, I. L. Chuang, and X.-G. Wen, arXiv:1003.1774v1 [cond-mat.str-el] (2010).
\bibitem{Tagliacozzo10} L. Tagliacozzo and G. Vidal, arXiv:1007.4145v1 [cond-mat.str-el] (2010).
%
\bibitem{Singh11} S. Singh et al., in preparation. 
\end{thebibliography}
\end{document}